\documentclass{aa}
\usepackage{graphicx}
\usepackage{txfonts}

\usepackage[colorlinks=true,citecolor=blue,linkcolor=blue,urlcolor=blue,breaklinks=true]{hyperref}
\usepackage{xcolor,float}
\usepackage[ruled,vlined]{algorithm2e}
\usepackage{multirow}
\usepackage{natbib}
\bibpunct{(}{)}{;}{a}{}{,}

\newcommand{\XMM}{\textit{XMM-Newton}} 
\newcommand{\Vbox}{$\mathcal{V}_{\text{box}}$}
\newcommand{\Source}[3]{\noindent\textbf{#1, OBS #2, Source #3}\\}
\newcommand{\errors}[2]{$\pm^{#1}_{#2}$}
\newcommand{\ctss}{\,cts\,s$^{-1}$} 
\newcommand{\fcgs}{\,erg\,cm$^{-2}$\,s$^{-1}$} 
\newcommand{\ergs}{\,erg\,s$^{-1}$} 

\begin{document}

\title{The {\em EXOD} search for faint transients in \textit{XMM-Newton} observations: \\
Method and discovery of four extragalactic Type I X-ray Bursters}
\titlerunning{The {\em EXOD} search for faint XMM transients}

   \author{I. Pastor-Marazuela
          \inst{1,2,3}
          \and
          N.A. Webb\inst{3}
          \and
          D.T. Wojtowicz\inst{3,4}
          \and
          J. van Leeuwen\inst{2,1}
          }

   \institute{Anton Pannekoek Institute for Astronomy, University of Amsterdam, Science Park 904, 1098 XH Amsterdam, The Netherlands\\
              \email{i.pastormarazuela@uva.nl, }
         \and
             ASTRON, Netherlands Institute for Radio Astronomy, Oude Hoogeveensedijk 4, 7991 PD Dwingeloo, The Netherlands
         \and
             Universit\'e de Toulouse; UPS-OMP; IRAP, Toulouse, France; CNRS; IRAP, 9 Av. colonel Roche, BP 44346, 31028, Toulouse CEDEX 4, France
	 \and
	     IRIT, Universit\'e Toulouse III - Paul Sabatier, 118 route de Narbonne, 31062 Toulouse CEDEX 9, France}

   \date{}


\abstract
{\XMM\ has produced a very extensive X-ray source catalogue in which the standard pipeline determines the variability of sufficiently bright sources through $\chi^2$ and fractional variability tests. Faint sources, however, are not automatically checked for variability. This means that faint, short timescale transients are overlooked. From dedicated X-ray searches, and optical and radio archive searches, we know that some such dim sources can still be identified with high confidence.
   }
{ Our goal is to find new faint, fast transients in \XMM\ EPIC-pn observations. To that end we have created the EPIC-pn \XMM\ Outburst Detector (EXOD) algorithm, which we run on the EPIC-pn full-frame data available in the 3XMM-DR8 catalogue.
   }
{In EXOD, we compute the variability of the whole field of view (FoV) by first binning in time the counts detected in each pixel of the detector. We next compute the difference between the median and maximal number of counts in each time bin and pixel to detect variability. We applied EXOD to 5,751 observations in the full frame mode, and compared the variability of the detected sources to the standard $\chi^2$ and Kolmogorov-Smirnov (KS) variability tests.
   }
{The algorithm is able to detect periodic and aperiodic variability, short and long flares. Of the sources detected by EXOD, 60$-$95\% are also shown to be variable by the standard $\chi^2$ and KS tests.
EXOD computes the variability over the entire field of view faster than the light curve generation takes for all the individual sources.
We detect a total of 2,961 X-ray variable sources. After removing the spurious detections, we obtain a net number of 2,536 variable sources.
Of these we investigate the nature of 35 sources with no previously confirmed classification. 
Amongst the new sources, we find stellar flares and AGNs; but also four extragalactic type I X-ray bursters that double the known neutron-star population in M31.
   }
{This algorithm is a powerful tool to promptly detect interesting variable sources in \XMM\ observations.
EXOD also detects fast transients that other variability tests would classify as non-variable due to their short
duration and low number of counts. This is of increasing importance for the multi-messenger detection of transient sources.
Finally, EXOD allows us to identify the nature of compact objects through their variability, and to detect rare compact
objects. We demonstrate this through the discovery of four extragalactic neutron-star low mass X-ray binaries, doubling the number of known neutron stars in M31.
   }


   \maketitle
%
\section{Introduction}\label{sec:intro}
\XMM\ has been inspecting the X-ray sky since its launch in 1999.
Its three on-board telescopes, each with a geometric effective area of $\sim 1550$\,cm$^2$ at 1.5\,keV, combine to provide the largest total collecting area of any X-ray telescope launched \citep{jansen_xmm-newton_2001}. 
This has allowed the compilation of a very large catalogue of X-ray detections.
We used the eighth data release of the \XMM\ catalogue, 3XMM-DR8, which contains 775,153 detections \citep{rosen_xmm-newton_2016}\footnote{3XMM-DR8 website: \url{http://xmmssc.irap.omp.eu/Catalogue/3XMM-DR8/3XMM\_DR8.html}}.

The large majority of these detections are of steady, unvarying sources. Compared to this static sky, the dynamic sky remains relatively unexplored.
Many violent and rare variable phenomena such as tidal disruption events 
and X-ray bursters  are observed, and can be used for studying (astro-)physics including strong gravity and thermonuclear explosions, respectively.
The X-ray regime is one part  of multi-messenger time-domain astronomy, which focuses on this dynamical sky.
The detection of cosmic rays and neutrinos \citep[e.g.][]{hirata_observation_1987}, of extragalactic bursts in radio \citep{lorimer_bright_2007}, type Ia supernovae in the optical \citep[e.g.][]{riess_observational_1998} and of double neutron star mergers in gravitational waves \citep{abbott_gravitational_2017} show that transient surveys are of vital importance to understand the physics behind the explosive universe.

\XMM\ was not built as a transient detector. However, the high sensitivity and high time resolution of its European
Photon Imaging Cameras (EPIC) allow it to record fast X-ray transients. Nevertheless, the detection of these transients and potential follow-up with other instruments depends on the sensitivity and performance of the variability detection pipeline.
The pipeline processing for the 3XMM catalogue includes such a variability study,  for sources with total EPIC counts
(including the detectors pn, MOS1 and MOS2) exceeding 100.

The variability of these brightest sources is studied through two complementary tests; the first is a simple $\chi^2$ test where the time
series are fitted to a constant model. The time bin width is the lowest integer multiple of 10\,s for which the average
number of counts is $\geq 18$ counts/bin. Sources with a $\chi^2$ probability of constancy P($\chi^2$)$\leq 10^{-5}$ are
flagged as variable \citep{watson_xmm-newton_2009}. The second method is the study of the fractional variability
amplitude, $F_{\text{var}}$, as described in \cite{rosen_xmm-newton_2016} and references therein. $F_{\text{var}}$ is
given by the square root of the excess variance normalised by the mean count rate. This provides the scale of the
variability. Although $F_{\text{var}}$ is given in the catalogue, it is not used to flag sources as variable.

The $\chi^2$ statistic can be applied to binned data sets where the observed number of counts deviates from expectation
approximately following a Gaussian distribution. \cite{cash_parameter_1979} showed that when the number of counts per bin falls below $\sim$10-20, the $\chi^2$ statistic becomes inaccurate. This justifies the requirements for the light curve generation in \XMM's pipeline.

There are, however, some drawbacks to this technique. Faint, variable sources can easily go unnoticed, since their time series are not generated. Additionally, sources that are variable on timescales shorter than the duration of the time bins will not be flagged as variable, since one single data point with a very high count rate is likely to be disregarded by the $\chi^2$ variability test.

A number of objects can display variability within a single \XMM\ observation. These can be distinguished in multiple ways. In a sample containing stars, AGN and different compact object systems, the compact objects show a higher X-ray flux variation factor \citep{lin_classification_2012}. Furthermore, the features in the light curves are usually specific to a particular type of object. Thus, a number of source classes have previously been detected and identified:

Type I X-ray bursts are outbursts lasting from seconds to minutes, emitted by low mass X-ray binaries (LMXB) with an
accreting neutron star \citep[NS; e.g.][]{parikh_nucleosynthesis_2013}. One subclass of High mass X-ray binaries (HMXB) called supergiant fast X-ray transients (SFXT) are HMXB with a neutron star accreting the winds of its companion. These sources can present multiple X-ray flares that are visible for a few hours \citep[see][for a review]{sidoli_supergiant_2013}. Cool late-type stars can show flares, due to magnetic reconnection. These flares can last from minutes to hours and show an exponential rise and decay \citep[e.g.][]{pye_survey_2015,favata_bright_2005,imanishi_systematic_2003}. Cataclysmic variables (CVs) show modulated light curves with periods between $\sim 0.2$ and $\sim 1.5$ h \citep[e.g.][]{kuulkers_x-rays_2003,southworth_orbital_2011}.

While some sources from the list above have been previously detected with \XMM, a more sensitive algorithm could detect a higher number; improving population statistics, and increasing the chance of detection.

We also aim to detect rarer transients such as short gamma ray bursts (SGRBs), the electromagnetic counterpart to NS mergers, as well as long gamma ray bursts (LGRBs), believed to be emitted from the jets formed during the death of massive stars in core-collapse supernovae. Whilst GRBs generally have spectra that peak in the gamma-rays, the spectra are broadband and can also be observed in X-rays. For distant GRBs, the peak luminosity will be redshifted into the \XMM\ energy range.
If a GRB with a small angle between the jet and the line of sight is within the field of view (FoV) of an \XMM\ observation, it could be detected as a very faint, short transient. A handful of distant LGRBs has been detected by the \textit{Swift} mission \citep[][and references therein]{salvaterra_high_2015}.

Short X-ray bursts may accompany Fast Radio Bursts (FRBs),
the rare, extragalactic, millisecond radio transients.
Such a connection is implied by the behaviour of Galactic magnetar 
SGR~1935+2154.
That source is capable of emitting short FRB-like radio bursts with fluences as high as a MJy\,ms 
\defcitealias{scholz_atel_2020}{Scholz et al. 2020}
(\citealt{2020ATel13684....1B}, \citetalias{scholz_atel_2020}).
These radio bursts are accompanied by bright, short X-ray bursts
\citep{mereghetti_atel_2020, tavani_atel_2020, zhang_atel_2020, ridnaia_atel_2020}.
Extragalactic FRBs may thus potentially also be accompanied by  a transient X-ray counterpart
\citep[e.g.][]{kaspi_magnetars_2017,metzger_fast_2019}. 
Previous follow-up efforts and archival data searches have so far not
found convincing X-ray counterparts to extragalactic FRBs 
\citep{scholz_repeating_2016,bhandari_survey_2018,delaunay_discovery_2016}. 
More observations, more sensitive instruments or perhaps more sensitive algorithms applied to existing instruments could offer the step needed to detect a counterpart, or set more stringent limits on their existence.

Variability is thus a feature that can be used as a diagnosis tool to shed light upon the nature of a source, and potentially allow the identification of new and rare compact objects. Hence, there is a need for new algorithms computationally cheaper than the generation of individual light curves that can be applied to faint sources.

Previous attempts to detect faint transients in \XMM\ data include the \textit{EXTraS}
project\footnote{\url{http://www.extras-fp7.eu/index.php}}, which searched for transients and periodic variability \citep{de_luca_overview_2015,novara_supernova_2020}. The transient detection method employed there combines a source search on short time intervals with an imaging technique. 
The time intervals are defined through a modified Bayesian block algorithm \citep{scargle_studies_1998, scargle_studies_2013} that finds count-rate changes in the EPIC observations and runs a source detection algorithm on the time-resolved images.

In this paper, we present EXOD, a new algorithm that aims to detect short, faint transients by computing the variability
of the whole FoV, then by detecting variable sources through an imaging technique. This can be used to promptly spot
interesting sources that vary on timescales as short as 3\,s.
Due to the difference in the methodologies between \textit{EXTraS} and EXOD (see Section~\ref{sec:method}), particularly in the way the time windows are defined and the stage at which the source detection algorithm is run, we expect to find different sources.

The remainder of this paper is organised as follows: Section~\ref{sec:data} explains the data we used and how it was reduced; section~\ref{sec:method} details the EXOD algorithm and the variability detection technique as well as the comparison with other variability tests; section~\ref{sec:results} discusses the results and reports the detection of new variable sources. In section~\ref{sec:discussion} we discuss our work and potential applications and we give a conclusion in section~\ref{sec:conclusion}.

\section{Data}\label{sec:data}
\subsection{Observation mode}
We analysed the set of EPIC-pn observations contained in the 3XMM-DR8 catalogue performed in the \textit{full frame} mode. The full frame mode was chosen for the following reasons: (1) The FoV covers a large surface of $27.2\arcmin\times 26.2\arcmin$ and thus contains a higher number of sources than other EPIC-pn observation modes. (2) A  time resolution of 73.3 ms allows the study of short timescale variability. (3) EPIC-pn receives about twice the number of photons that fall on EPIC-MOS \citep{struder_european_2001,turner_european_2001}.
These properties are key to finding new variable, faint sources.
We analysed a total number of 5,751 observations with these characteristics.

\subsection{Filtering observations}\label{sec:filt-obs}
We filtered the EPIC FITS pn IMAGING mode event list (PIEVLI) files using the Science Analysis System (SAS) version 16.1.0\footnote{\url{https://xmm-tools.cosmos.esa.int/external/xmm\_user\_support/documentation/sas\_usg/USG/}} with the following criteria:

We limited the energy range to 0.5--12\,keV.
We filtered out the 0.2 -- 0.5\,keV energy range to
avoid the spurious detections caused by low-energy noise \citep{watson_xmm-newton_2009}.

In order to ignore high background count rates from soft proton flares we removed periods in  which the 10 --
12\,keV count rate exceeds a certain threshold. We tested the \XMM\ Science Analysis System (SAS) task
\texttt{bkgoptrate} for this. It computes the rate maximizing the signal to noise ratio for a given source. However, we found that the rate computed by this function is not optimal to detect faint variable sources. Although this rate increases the number of counts per source, the sources present an extrinsic variability due to the proton flares.

\begin{figure}
\centering
\includegraphics[width=\hsize]{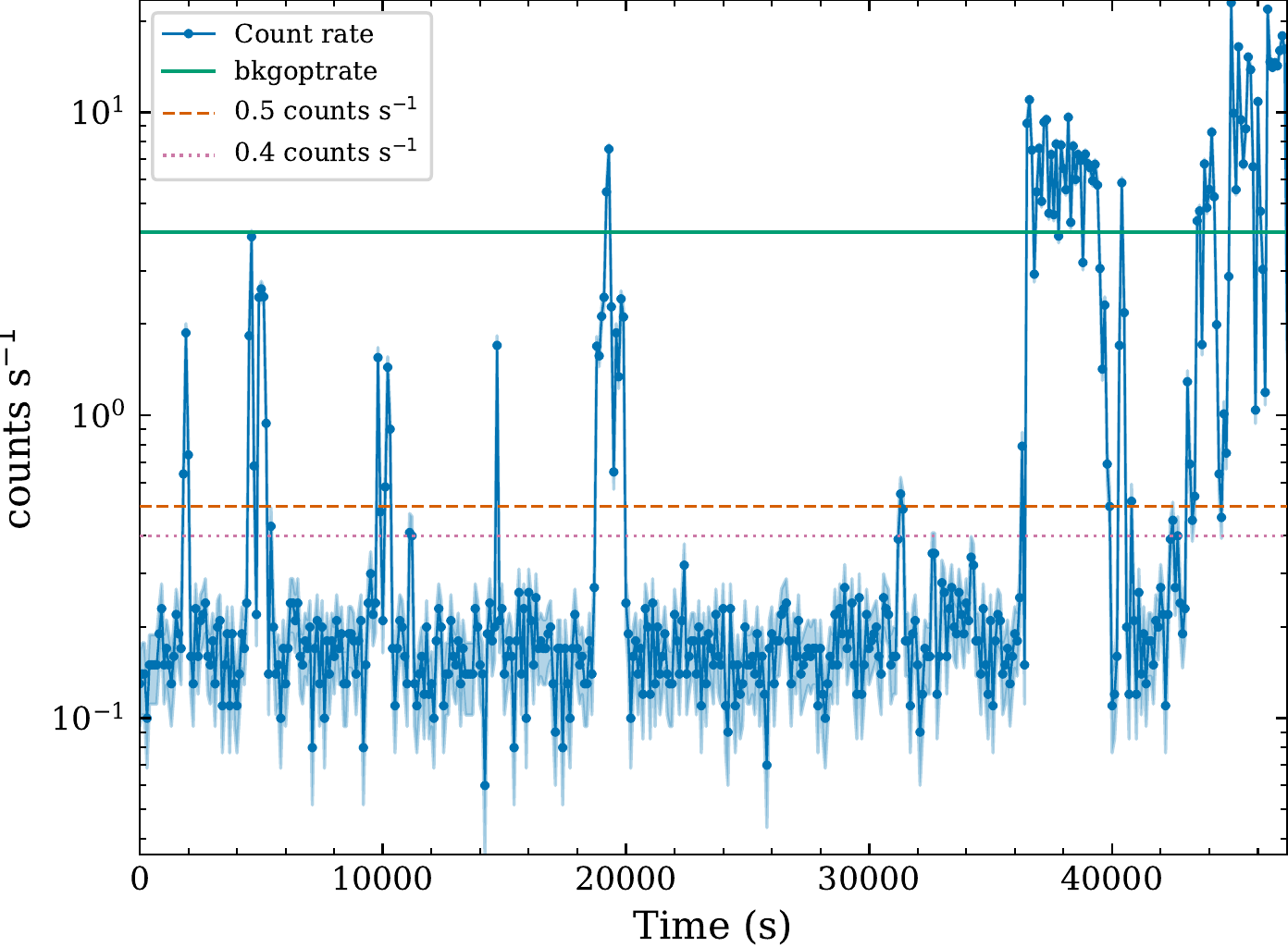}
\caption{Count rate of observation 0112370801 between 10 -- 12\,keV as a function of time since the beginning of the observation. The vertical axis is in logarithmic scale and we use a time binning of 100\,s. \textit{Green} solid line: optimal rate as computed by \texttt{bkgoptrate}, in this case $\sim4$\ctss. \textit{Orange} dashed line: 0.5\ctss. \textit{Pink} dotted line: 0.4\ctss.}
\label{fig:bkgrate}
\end{figure}

Fig.~\ref{fig:bkgrate} is an example, showing the background count rate in observation 0112370801. One can see that the
\texttt{bkgoptrate} optimal background rate (\textit{green} solid line) is too high to filter out several background
flares. Choosing this rate would thus imply keeping a contaminated observation, affecting the variability detection. The count rate recommended by the SAS for filtering EPIC-pn observations, 0.4\ctss\ (\textit{pink} dotted line) filters all the high background periods of this observation. However, this value could filter too many photons in the case of an observation with a higher low background rate or a source showing an extremely bright outburst. We thus chose a threshold of 0.5\ctss\ (\textit{orange} dashed line). This threshold filters most of the background flares that are also filtered with 0.4\ctss. The steepness of the light curve at the beginning and the end of the background flares implies that the difference of the good time intervals between count rates of 0.4 and 0.5 will be small. 
The threshold of 0.5\ctss\ was adopted after testing it on $\sim$700 observations. Whilst a slightly refined value may be better for some observations, this value was found to be appropriate for the large majority of the observations studied.
We used the extra precaution of removing all the time windows (see Section~\ref{sec:var-comp} for definition) where a background flare starts or ends to avoid contamination.

We used the SAS task \texttt{tabgtigen} with a count rate of 0.5\ctss\ and time bins of 100\,s to generate the Good Time Interval (GTI) file and filtered the bad time intervals from the observations. Additionally, we applied the standard filters \texttt{\#XMMEA\_EP} and \texttt{PATTERN<=4}.

\section{Method}\label{sec:method}

\subsection{The algorithm}
Since the light curve generation is necessary for the variability computation of the sources in \XMM's pipeline, only the variability of the brightest sources is studied. In order to detect fast, faint transients coming from fainter sources within \XMM\ observations, the whole FoV should be explored.

We have developed \textit{EXOD}, the EPIC-pn \XMM\ Outburst Detector, an algorithm that detects fast and faint transients by computing the variability of every single EPIC-pn pixel from each observation, as we present below. The code is publicly available online\footnote{\url{https://github.com/InesPM/EXOD}}.

The algorithm computes the variability as explained in the following section using the previously generated filtered events file and GTI file.
The required input parameters used for the variability detection can be found in Table~\ref{tab:input-parameters}.
\begin{table*}
\caption{Input parameters of the variability computation.}
\label{tab:input-parameters}
\centering
\begin{tabular}{p{1cm} p{1.5cm} p{1.5cm} p{1.2cm} p{10cm}}
\hline\hline
	Symbol & Parameter & Accepted Values & Default Value & Function \\
\hline
	$DL$ & Detection level & $\Re^+$ & 10 & Level above which an area is considered as being variable. A lower value allows the detection of more sources, but a higher proportion of the detected sources will be non-variable.\\
	$TW$ & Time window & $\Re^+$ & 100\,s & Duration of the time windows. It is also the minimal timescale of the variability that can be explored.\\
	$b$ & Box size & [3,5,...,63] & 5 & Length in pixels of the detection box, limited by the size of the CCDs. It determines the extension of the variable area that we want to detect.\\
	$r_{GT}$ & Good time ratio & [0;1] & 1.0 & Critical (good time)/TW above which the time window will be taken into account. Choosing 1.0 will remove all time windows that have been partially or entirely contaminated by soft proton flares.\\
\hline
\end{tabular}
\end{table*}

\subsubsection{Variability computation}\label{sec:var-comp}
The variability computation can be divided in the steps given below. Fig.~\ref{fig:variability_computation} depicts the algorithm with the same step enumeration. The algorithmic notation is given in Appendix~\ref{app:algo}.

\begin{enumerate}
	\item We extract the time and pixel of arrival of every event detected during the observation from the events file.
	\item Photons detected in a $3\times3$ square around each pixel are added together to increase the signal-to-noise ratio and to reduce stochastic variability.
	\item We bin the events per pixel into time windows ($TW$), whose duration we give as an input.
	\item We extract the data from the GTI file.
	\item We compute the good time ratio of each time window, that is the time belonging to the GTI (good time) of the considered time window divided by the duration of the time window.
	\item We divide the number of counts per time window by the good time ratio. Only those time windows with the good time ratio above a critical value, $r_{GT}$, will be considered. This will normalize the number of photons that have been detected during a time window that has been shortened by the bad time periods.\\ Note that the $r_{GT}$ default value is 1 (see Table~\ref{tab:input-parameters}). This removes every time window that has been partially or totally contaminated by background flares, avoiding detecting these flares as variable sources.
	\item The variability $\mathcal{V}$ of each pixel is given by:

	\begin{eqnarray} \label{eq:variability}
	\mathcal{V} =
	\begin{cases}
		max(\mathcal{C}_{max} - \tilde{\mathcal{C}}, |\mathcal{C}_{min} -\tilde{\mathcal{C}}|)/\tilde{\mathcal{C}} & \quad \text{if } \tilde{\mathcal{C}} \neq 0\\
		\mathcal{C}_{max} & \quad \text{if } \tilde{\mathcal{C}} = 0
	\end{cases}
	\end{eqnarray}
	Where $\mathcal{C}_{max}$ and $\mathcal{C}_{min}$ are respectively the maximal and minimal number of counts per time window for that pixel and $\tilde{\mathcal{C}}$ is the median number of counts over the time windows for the pixel.\\
	The expression $\mathcal{C}_{max} - \tilde{\mathcal{C}}$ targets sources presenting outbursts, while $|\mathcal{C}_{min} -\tilde{\mathcal{C}}|$ points to those sources with a period of lower flux. Considering the maximum between the two allows the detection of a wider variety of phenomena.\\
	The division by the median $\tilde{\mathcal{C}}$ gives the variability relative to the flux.
\end{enumerate}

\begin{figure*}
	\centering
	\includegraphics[width=17cm]{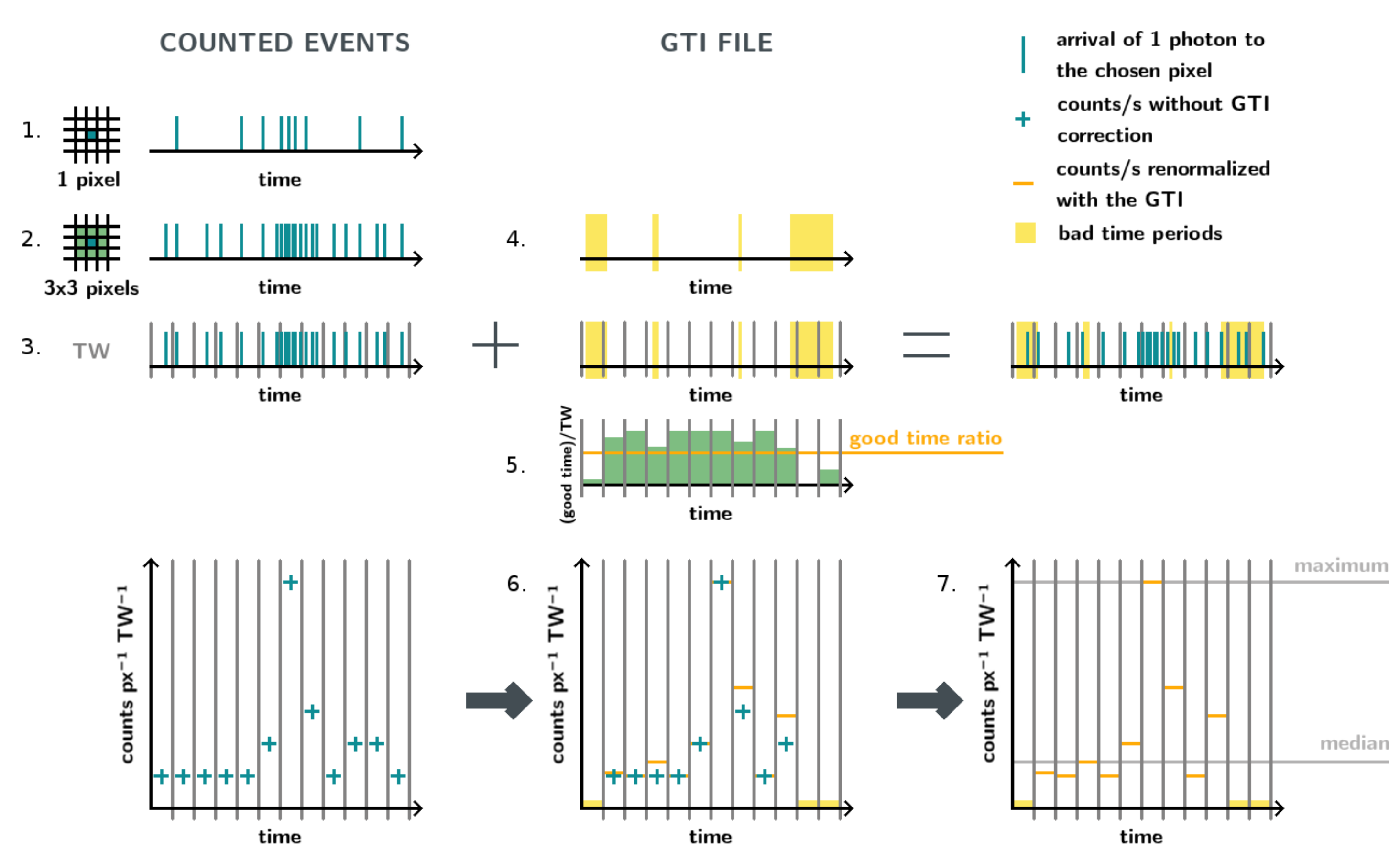}
   	\caption{Diagram of the variability computation with the different stages of the algorithm. The vertical \textit{blue} lines indicate the arrival time of the photons during the observation. The vertical \textit{gray} lines represent the time bins. The \textit{blue} crosses indicate the counts per time window per pixel without filtering for the GTI. The \textit{yellow} shaded regions indicate the time periods contaminated by soft proton flares. The \textit{green} rectangles show the good time over the duration of the time window for each time window. The \textit{orange} horizontal lines show the counts per time window per pixel after renormalizing with the GTI, and the yellow horizontal lines show the time windows that will not be taken into account because they are below the good time ratio. Finally, the \textit{gray} horizontal lines indicate the maximum $\mathcal{C}_{max}$ and the median $\tilde{\mathcal{C}}$ counts that will be used to compute the variability.}
	\label{fig:variability_computation}
\end{figure*}

\subsubsection{Variable source detection}

The variable areas are detected with the sliding box technique, where a box of size $|b|^2$ pixels will move through all the pixels of the observation, with $b$ an odd number.

 This procedure has been broadly used in X-ray observatories, for instance the \textit{Einstein observatory} \citep{gioia_einstein_1990}, \textit{ROSAT}\footnote{\url{https://heasarc.gsfc.nasa.gov/wgacat/}}, and currently in \textit{Chandra} \citep{calderwood_sliding-cell_2001} and \XMM\ \citep{watson_xmm-newton_2009}.

We first compute the median variability value $\tilde{\mathcal{V}}$ (Eq.~\ref{eq:variability}) for all detector
pixels. 
We next calculate the sum of the variability in a box of size $|b|^2$ to obtain \Vbox. The central position of the box is then shifted to the contiguous pixel and the process is repeated until \Vbox\ has been calculated for every available position within the detector (based on  \citealt{watson_xmm-newton_2009}).
The variability of a box \Vbox\ centered on the pixels x,y is given by the following equation:
\begin{equation}
\mathcal{V}_{\text{\text{box (x,y)}}} = \sum_{i = x-(b-1)/2}^{x+(b-1)/2}\sum_{j = y-(b-1)/2}^{y+(b-1)/2} \mathcal{V}_{i,j}
\end{equation}
Where $b$ is the length of the box in raw pixel units. 
\Vbox\ will not be computed at the border of the CCDs, since the size of the box would be smaller than $|b|^2$.

If for a certain box \Vbox\ is above a chosen threshold, we consider the pixels contained in the box as variable. The value of this threshold is given by the following expression:
\begin{equation}
\mathcal{V}_{\text{box}} > DL \times |b|^2 \times \tilde{\mathcal{V}}
\end{equation}
Where $DL$ is the detection level. When two consecutive boxes are variable, the pixels of both boxes are joined into a single variable area.

The variable sources are located at the center of the variable areas. The position of the (X,Y) coordinates in raw pixels is the mean value of the position of the pixels belonging to the variable areas. The raw pixel position and the number of the CCD of the variable sources are given as an output, as well as the distance from the center to the outermost pixel belonging to the variable area, in pixels.
We do not consider sources detected in CCD=4, RAWX=12; CCD=5, RAWX=12; CCD=10, RAWX=28 since these pixels are known to be damaged \citep{struder_evidence_2001} and produce a high number of spurious detections.

\subsection{Detection parameters}\label{sec:used-parameters}
We applied the algorithm to the observations defined in Section \ref{sec:data} with different time windows, detection
levels and box sizes. We thus empirically determined the optimal parameters for finding most variable
sources, while minimizing  spurious detections.
The different values of the time windows check for variability on different timescales. We chose $TW = 3, 10, 30, 100$\,s to optimise time variability with computation time.

We performed these variability tests with the parameters given in Table~\ref{tab:parameters} and $r_{GT} = 1.0$ on a
subset of 2,284 observations. These are the observations that were included in the incremental releases of the 3XMM-DR7
and 3XMM-DR8 catalogs.
We used the optimal parameters obtained to analyse the light curves of the sources detected in the remaining observations.

\begin{table}
	\centering
	\caption{Parameters used for the variability computation, with Good Time Ratio $r_{GT} = 1.0$.}
	\label{tab:parameters}
	\begin{tabular}{|l|ccc|ccc|ccc|ccc|}
	\hline\hline
	$TW$ (s)    & \multicolumn{3}{c|}{3} & \multicolumn{3}{c|}{10} & \multicolumn{3}{c|}{30} & \multicolumn{3}{c|}{100} \\
	\hline
	$DL$     & 5& 6 & 7 & 6 & 7 & 8 & 7 & 8 & 9 & 8 & 9 & 10 \\
\hline
	$b$ (pixels)    & \multicolumn{6}{c}{3} & \multicolumn{6}{c|}{5} \\
	\hline
	\end{tabular}
	\tablefoot{For each Time Window $TW$, the three respective Detection Levels $DL$ were used, whereas the two given box sizes $b$ were used for all the $TW$-$DL$ combinations}
\end{table}

\subsection{Comparison with other variability tests}\label{sec:light curve}
In order to determine the robustness of the algorithm, we needed a comparison with the existing variability tests. To do
so, we determined the fraction of false positives as well as the fraction of false negatives. 

To measure the fraction of false negatives, we performed a positional cross-match of the EXOD sources with the 3XMM-DR8 sources flagged as variable. For this we used the catalogue manipulation tool \textsc{Topcat} \citep{taylor_topcat_2005}.

To measure the fraction of false positives, we first recreated for reference the standard variability measure of the EXOD-detected sources. We generated the light curve of the detected sources with the SAS function \texttt{evselect} and subsequently computed the probability of constancy with the $\chi^{2}$ and Kolmogorov-Smirnov tests, P($\chi^2$) and P(KS) respectively, with the \textsc{FTOOLS Xronos} function \texttt{lcstats}. Comparisons with other test are found in \citet{pastor-marazuela_searching_2018}\footnote{Searching for fast transients in \XMM\ data. Master thesis: \url{https://www.uva.nl/en/profile/p/a/i.pastormarazuela/i.pastor-marazuela.html}}.
We used the HEASOFT version 6.22.1 \citep{blackburn_ftools:_1995}.

The $\chi^2$ test can only be applied to data with a Gaussian distribution, which occurs when the number of counts per
bin is $\gtrsim 18$. As we generated light curves regardless of the number of counts, not every light curve satisfies this criterion. We thus additionally apply the KS test, which makes no assumptions about the binning of the data and can be safely applied to sources with low counts\footnote{Kirkman, T.W. (1996) Statistics to Use. \url{http://www.physics.csbsju.edu/stats/}}.

If P($\chi^2$)$\leq 10^{-4}$ or P(KS)$\leq 10^{-4}$, we consider the source as being variable. P($\chi^2$)$\leq 10^{-4}$ gives a $>4\sigma$ confidence level on the variability detection with the $\chi^2$ test. We define the robustness of the algorithm as the ratio of sources detected with EXOD that are variable according to at least one of the variability tests $\chi^2$ or KS.

The light curves were extracted automatically by giving the position in raw pixel units of the detected variable source, then using a set of SAS tasks described below that optimise the position and radius of the source. We used the $TW$ duration as the bin size of the light curves. For these tasks, the SAS summary file and the EPIC FITS global background time series (FBKTSR) are needed additionally to the filtered observation and the GTI file.

Since we noted that extended sources, bright sources, or Out of Time (OoT) events can trigger spurious detections of variability within an observation, we only applied the automatic light curve generation when the number of detections per observation was less than 6.

We determined this six-source threshold empirically, by noting that less than $\sim2.6$\% of the observations have more than 5 variability detections, and that these are usually spurious detections. However, we visually inspected the variability plots (such as the one shown in Fig.~\ref{fig:example}) of the fields of view with more than 5 detections in order to avoid missing interesting candidates.

\paragraph{Source region:}
From the raw pixel and CCD of the detected variable sources, one can determine the position in sky pixel space with the SAS task \texttt{ecoordconv}. The position and the radius of the source are corrected with the function \texttt{eregionanalyse} with \texttt{backval=0.1}. This function returns a circle containing 90\% of the energy of the source.

It is known that the shape of the Point Spread Function (PSF) in the EPIC-pn detector depends on its angle with respect to the center of the detector \citep{read_new_2011}. Whereas a centered (on-axis) source is well approximated by a circular region, a non-centered (off-axis) source is elongated due to off-axis aberration (astigmatism), giving a larger, energy-dependent radius. 
Although elliptical regions increase the signal to noise ratio of the light curves, we chose circular extraction regions since this option is a sufficiently good approximation, especially when dealing with faint sources for which only a small number of pixels will be above the background level. We also find it to be more reliable in such an extensive study to automatically extract the counts of all the sources and find nearby background extraction regions. The selection of elliptical regions becomes of greater importance when performing spectral analyses.

\paragraph{Background region:}
To subtract the background from the source light curve, we use background regions determined with the SAS task \texttt{ebkgreg}.
This task searches the optimal background region following only geometrical criteria. To avoid selecting a region containing sources, we extract the background from a filtered events file where we have removed the sources. We obtain the positions of all the sources of the observation from the FBKTSR file. We choose the same radius as for the source region.

\subsection{M31}

There are some fields observed by \XMM\ where we expect to detect more than five variable sources. In this work we have decided to focus on one specific field, M31.
Because of its crowded field, M31 required analysis that was more manual than described above.

M31 is located at $\sim0.78$ Mpc\footnote{M31 properties obtained from NASA/IPAC Extragalactic Database (NED): \url{https://ned.ipac.caltech.edu/byname?objname=M31}. The given distance is an average of the published distances.}, and is the nearest spiral galaxy to the Milky Way. With an angular size of $\sim200\arcmin\times80\arcmin$, EPIC-pn's FoV covers its central region. This makes the study of M31 very interesting, since one observation contains a high number of extragalactic sources.

There is a total of 48 \XMM\ EPIC-pn observations of M31 that have not been strongly polluted by high background
flares. The high density of sources required selecting the extraction regions manually. In most observations, the number of detected sources exceeds the limit of 5. We have thus made an exception for this limit for M31, where we expect to detect a higher number of variable sources.

\subsection{Fast Radio Bursts}
We know EXOD is sensitive to bursts lasting up to 100s (see, Sect.~\ref{sec:vary}). 
There are now both theories and observations linking Fast Radio Bursts (FRBs) with magnetars; 
and the latter are known to show X-ray flares lasting up to a few seconds.
FRBs theories such as those put forward by \citet{beloborodov_flaring_2017} and \citet{metzger_millisecond_2017} require young magnetars as the power source.
Meanwhile, observations of SGR~1935+2154 show that Galactic magnetars  
can emit energetic flashes that show as \emph{both} 
FRB-like radio bursts \citep{2020ATel13684....1B}
and X-ray bursts \citep{zhang_atel_2020}, simultaneously.

We thus tested whether EXOD could again find variability missed by standard tools. 
We reanalysed observations 0792382801 and 0792382901 of FRB 121102, the first known repeating FRB \citep{scholz_simultaneous_2017}.
FRB 121102 was highly active in radio during these \XMM\ observations \citep{chatterjee_direct_2017,law_multi-telescope_2017}, but \citet{scholz_simultaneous_2017} found no transient X-ray counterparts.
Using our same optimal parameters defined in Section~\ref{sec:used-parameters}, we also found no sign of variability.

The energy upper limit estimates that \cite{scholz_simultaneous_2017} set on the X-ray emission of $10^{45}-10^{47}$\,erg at a distance of 972\,Mpc \citep{tendulkar_host_2017}, remains several orders of magnitude lower than that of GRBs. 
The detection of an FRB afterglow with \XMM\ could be possible only if its host galaxy was located at a lower redshift. 
%
%
Follow up of non-repeating low-DM FRBs, such as FRB 110214 at 169\,pc\,cm$^{-3}$ \citep{petroff_fast_2018}, could help to achieve a detection, especially if possible near real-time \citep[cf.][]{maan_real-time_2017}.


\section{Results}\label{sec:results}

\subsection{Detected sources}
In this section, we present the sources EXOD detected in our set of observations.
This includes the detected variable sources for which the light curves were automatically generated (less than six detections per observation) as well as those with more than five detections that we determined as non-spurious through a visual inspection.

There were 221 observations with more than five variability detections, with only 20 containing non-spurious detections. Among these non-spurious detections, we generated the light curve of 83 sources. We include these sources in the following analyses.

\subsubsection{Variability}
\label{sec:variability}
\begin{figure}
\centering
\includegraphics[width=\hsize]{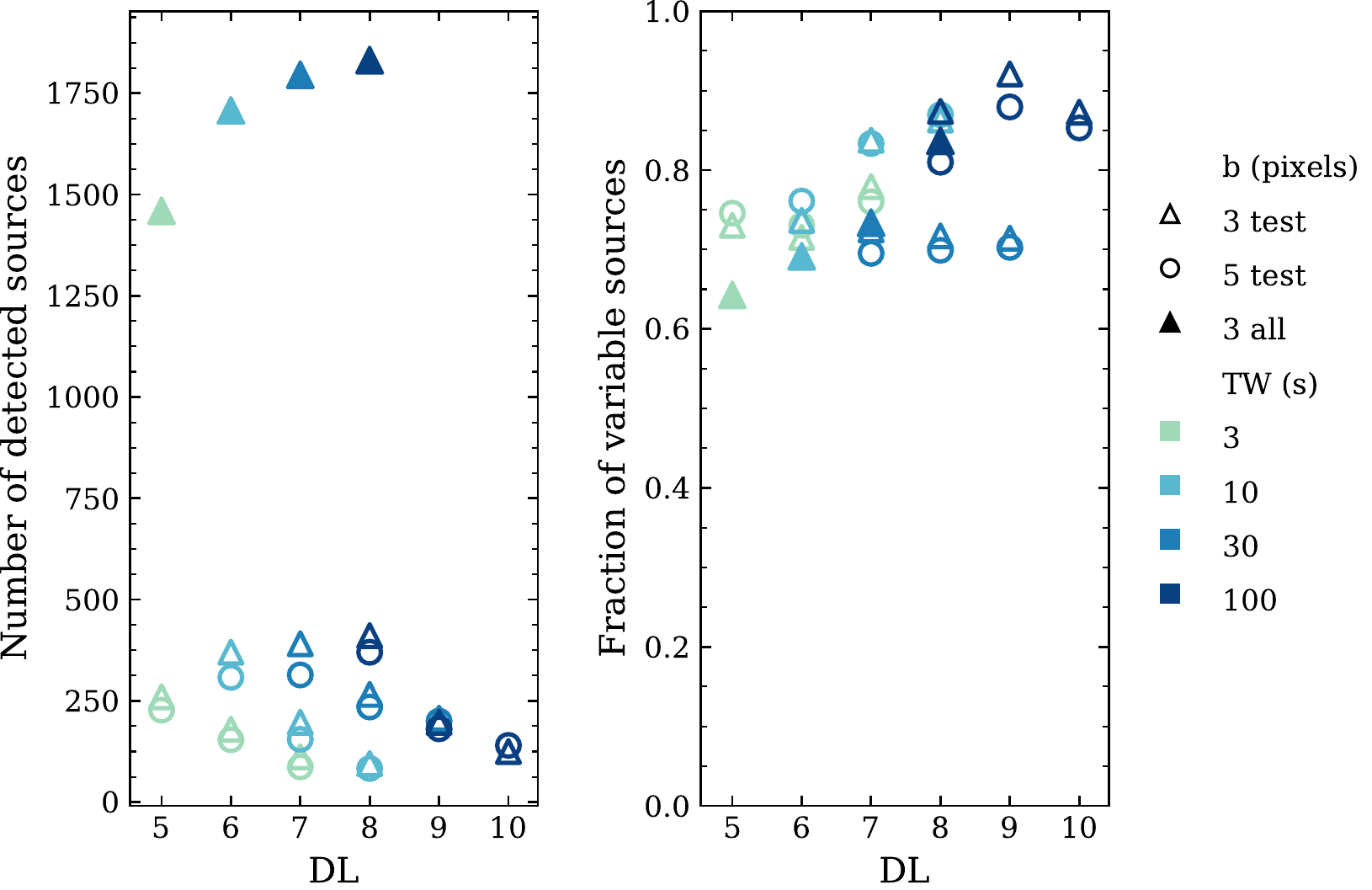}
\caption{Properties of the sources detected by EXOD. Test observations with the detection parameters listed in Table~\ref{tab:parameters}: empty markers. All observations with the parameters in Table~\ref{tab:final_param}: filled markers. \textbf{Left:} Number of detected sources as a function of $DL$. \textbf{Right:} Fraction of sources confirmed to be variable through the $\chi^2$ and KS tests among the detected ones as a function of $DL$. The colors correspond to different $TW$: light \textit{green} for $TW=3$\,s, light \textit{blue} for $TW=10$\,s, medium \textit{blue} for $TW=30$\,s and dark \textit{blue} for $TW=100$\,s. The shapes represent the box size: triangles for $b=3\times3$ pixels and circles for $b=5\times5$ pixels.}
\label{fig:probabilities}
\end{figure}

In Fig.~\ref{fig:probabilities} we show the results obtained after applying EXOD as described in the previous sections.
By looking at the test parameters, one can see that the number of detected sources increases with decreasing detection level $DL$.
Decreasing $DL$ also increases the fraction of non-variable sources according to the $\chi^2$ and KS tests, although this is accompanied by an increase of the net number of variable sources. 
In most cases a $3\times3$ pixel box results in a slightly higher fraction of variable sources than a $5\times5$ box.

In order to detect the highest possible number of previously unknown variable sources, we chose a low detection level
with a box size of $3\times3$ pixels to analyse the full data set, accepting the accompanying increase in false positives. These final optimal parameters are given in Table~\ref{tab:final_param}, together with the number of sources detected among the 5,751 observations. The results are plotted in Fig.~\ref{fig:probabilities} as filled symbols, with the same color code as the test observations.

From these results, we can determine that the robustness of the algorithm as defined in Section~\ref{sec:light curve} is
between $\sim$60 and 95\% and depends on the chosen parameters. For the optimal parameters, $\sim$84\% of the sources are deemed to be variable using the $\chi^2$ and KS tests for a time window $TW=100$s. This percentage decreases when $TW$ decreases, and reaches $\sim$64\% for
$TW=3$s. EXOD is thus more robust for longer time windows.

The remaining fraction of sources are usually detected in short or highly polluted observations, notably when the net exposure time is below 5000\,s, and can be clearly identified. 
We thus advise visually inspecting the results when using EXOD for such short exposure observations. 
 
Although we perform a large number of independent trials (especially the number of TWs, in 4 trials $\times$ number of detection boxes), statistical fluctuations of sources that are not intrinsically variable are not a significant cause of false positives, given the high minimum detection levels we require.

When applying the optimal parameters to the whole set of observations, we find the fraction of non-variable sources to be higher than that in the test observations. We are investigating the cause for this but we suspect the later date of the DR7 and DR8 observations to play a role, as the average duration of the later observations has increased.  Given the longer duration, the number of flares that can be present in an observation increases. In a shorter observation, the lower number of counts may give rise to a higher number of spurious detections due to the aforementioned outliers of the Poissonian fluctuations.

The sum of the sources detected with the optimal parameters is 6,649. Since some of the sources are detected with different time windows, we performed an auto-correlation on the position of the sources with \textsc{Topcat}. We searched for internal matches in a radius of $15\arcsec$ around the location of each source. This radius corresponds to the Half Energy Width (HEW) of EPIC-pn at 1.5\,keV \citep{struder_european_2001}.
After removing the duplicated selections, we obtain a grouped list  with a total number of 2,961 sources detected among the 5,751 observations.

The \XMM\ pipeline finds slightly more variable sources because it searches longer timescales than EXOD.
Sources varying more slowly than our  longest timescale of 100\,s might not be detected. 

Although EXOD may miss some  variables  found by the standard pipeline, it \emph{does} find variable sources this
pipeline missed. We investigate these in Sect.~\ref{sec:vary}.

\begin{table}
	\centering
	\caption{Optimal parameters and detected sources.}
	\label{tab:final_param}
	\begin{tabular*}{\hsize}{@{\extracolsep{\fill} } lrrrr}
	\hline\hline
	Property & \multicolumn{4}{c}{Value}\\
	\hline
	$TW$ (s)    	& 3 & 10 & 30 & 100 \\
	$DL$        	& 5 & 6  & 7  & 8   \\
	$b$ (pixels)    & 3 & 3 & 3 & 3\\
	\hline
	Detected sources     & 1459 & 1707 & 1795 & 1831 \\
	Variable sources (\%) & 64.3 & 69.2 & 73.4 & 83.7 \\
	\hline\hline
	\end{tabular*}
	\begin{tabular*}{\hsize}{@{\extracolsep{\fill} } lclc}
	Sum & 6792 & Total & 2961 \\
	\hline
	\end{tabular*}
	\tablefoot{Parameters used to generate the light curves of all the observations, number of sources detected and percentage of variable sources according to the $\chi^2$ and KS tests.  $r_{GT} = 1.0$. Bottom line: sum of the detected sources and total number of sources after auto-correlation.}
\end{table}

\subsubsection{Cross-correlation with other catalogs}
\label{sub:crosscor}
To determine the nature of the EXOD-detected sources, we
queried them by position in the \textsc{Simbad} database\footnote{\textsc{Simbad}: \url{http://simbad.u-strasbg.fr/simbad/}}
\citep{wenger_simbad_2000}, using \texttt{astroquery}.
This returns the name, location and source type of the nearest catalogued source. We allowed a $15\arcsec$  radius around the detection  coordinates. We then placed the sources in one of the categories listed in Table~\ref{tab:class}.
These include 6 physical classes, one group of known extended sources, and two categories of unknowns: the \textit{no identification} class for previously detected sources whose type is not known; and the \textit{without counterpart} class where no detection has previously been made at all. If two sources were identified with the same \textsc{Simbad} entry, only one is counted. For that reason, the number of classified sources in Table~\ref{tab:parameters} (2,961) differs from the total number of sources in \ref{tab:class} (2,907).

In a preliminary cross-correlation with \textsc{Simbad}, we found $\sim 1,200$ \textit{without counterpart}. A visual inspection of these sources
revealed that most of these belonged to one of the following categories: (1) Out of Time events, (2) detections in the PSF of a bright source, (3) detections in an extended source and (4) hot pixels. 
We thus cross-matched these with the 3XMM-DR8 catalogue with \textsc{Topcat} in a $15\arcsec$ radius around the detection position and the radius of the sources given in the catalogue to take into account their extension. The sources matching a 3XMM-DR8 source were added to the \textit{no identification} (point-like sources) or the \textit{extended sources} class ($> 6\arcsec$ radius). This reduced the number of \textit{without counterpart} to $\sim400$. A visual inspection of the remaining \textit{without counterpart} showed that these were spurious detections.

Given our search criteria, we could have expected to find new sources through the variability search if they were faint and short enough to be drowned in the background noise and would thus not be included in the regular XMM catalogue.
Whilst no completely new sources were identified here, further modifications to the detection parameters could still reveal such objects.

\begin{figure}
\centering
\includegraphics[width=\hsize]{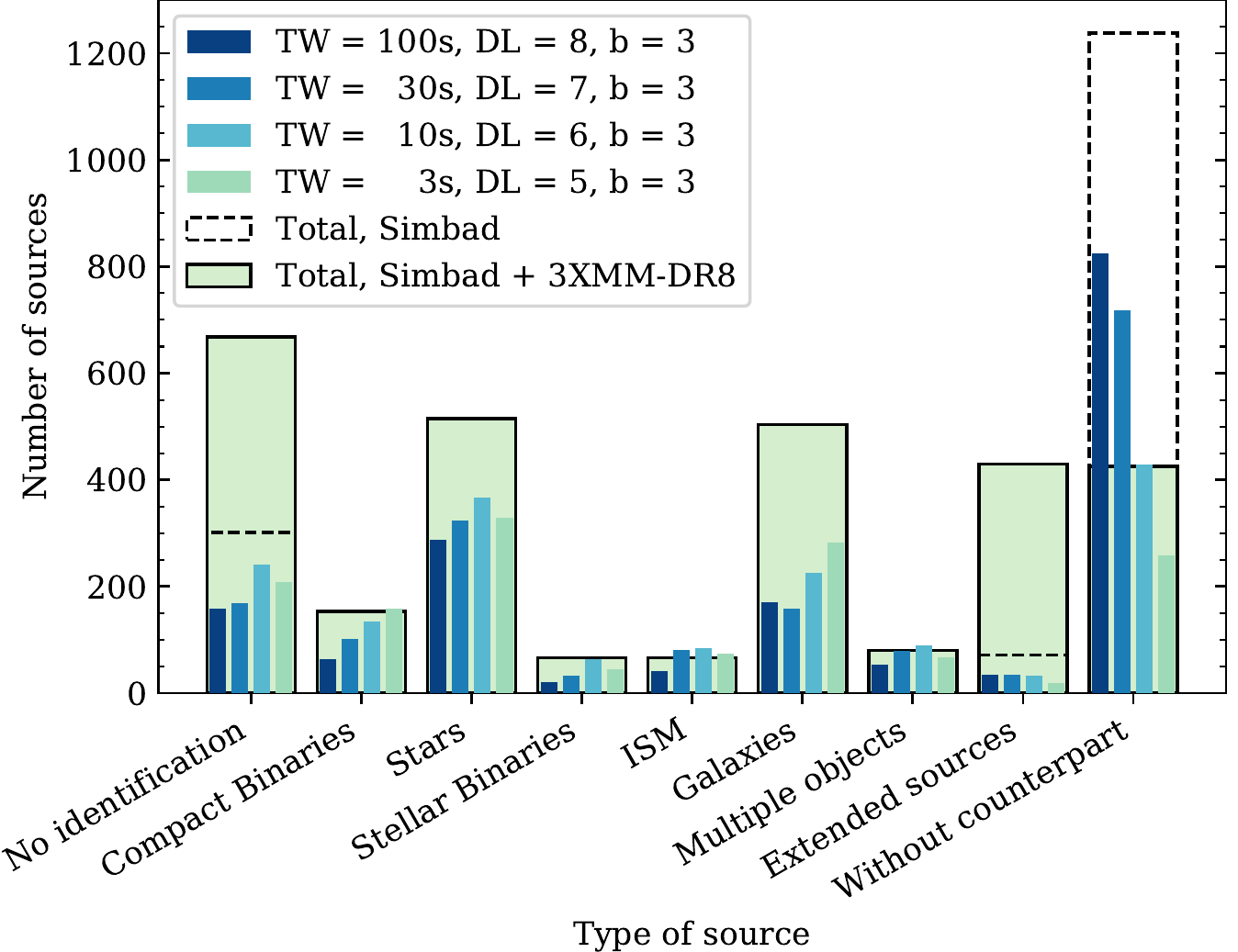}
\caption{Classification of the sources detected by EXOD in source types. The thin bars show the \textsc{Simbad} association of the sources detected, from darker to lighter, with 100\,s, 30\,s, 10\,s and 3\,s respectively. The thick bars with a black dashed contour show the \textsc{Simbad} association of the grouped sources. The thick light \textit{green} bars with a solid black contour show the \textsc{Simbad} association with the 3XMM-DR8 correction.}
\label{fig:otype}
\end{figure}

In Fig.~\ref{fig:otype}, we show the result of the cross-correlation with \textsc{Simbad} for the four sets of parameters as thin bars. The cross-correlation of the grouped sources with \textsc{Simbad} and 3XMM-DR8 is shown as thick bars with a solid black contour. The cross-correlation only with \textsc{Simbad} is plotted as dashed black contours for comparison.

If we remove the number of spurious detections from our set of detected sources, we obtain a net count of 2,536 variable sources detected with EXOD.

We can compare this number to the number of variable sources in the catalogue. 3XMM-DR8 contains 775,153 sources. 438,342 of these sources were detected in observations that we have analysed. 102,498 of these sources have generated time series. Finally, 3,418 sources are catalogued as variable, compared to the net 2,536 sources in this work.

However, the number of sources that are classified as variable in 3XMM-DR8 and EXOD simultaneously is 777. This corresponds to a false negative rate of 77.3\%. The high rate of false negatives was expected, since we are targeting a very specific kind of short-term variability, and sources varying on longer timescales are not detected. Nevertheless, the light curve of 688 (27.1\%) of the EXOD variable sources were not generated in 3XMM-DR8, so there was no previous available information about their variability. Lastly, Table~\ref{tab:final_param} shows that EXOD detects a high number of variable sources according to the $\chi^2$ or the KS tests, indicating that we are sensitive to a variability that the tests applied in \XMM's pipeline are not always adapted to detect.

\begin{table*}
	\centering
	\caption{Source category classification.}
	\label{tab:class}
	\begin{tabular*}{\hsize}{lp{11.5cm}r}
	\hline\hline
	Category & Description & Number of sources\\
	\hline
	Compact binaries & Cataclysmic variables, X-ray binaries, ULXs, novae. & 153\\
	Stars & Stars, pulsars & 515\\
	Stellar binaries & Binary stellar systems & 66\\
	ISM & Interstellar medium & 66\\
	Galaxies & Galaxies, AGNs, QSOs & 504\\
	Multiple objects & Galaxy clusters, groups of galaxies, stellar clusters or associations. & 80\\
	Extended sources & The angular distance $d$ to the catalogued source satisfies $10\arcsec \leq d \leq 15\arcsec$ and it belongs to the category ISM, galaxies or multiple objects. & 430\\
	No identification & Previously known source of unknown type. & 668\\
	Without counterpart & No association has been found within $15\arcsec$. & 425\\
	\hline
	\end{tabular*}
	\tablefoot{Naming of the categories used in Fig.~\ref{fig:otype}, with a description of the objects included in each category and the number of sources detected for each.}
\end{table*}

\subsubsection{Detected variability}

\begin{figure*}
\centering
\includegraphics[width=0.325\textwidth]{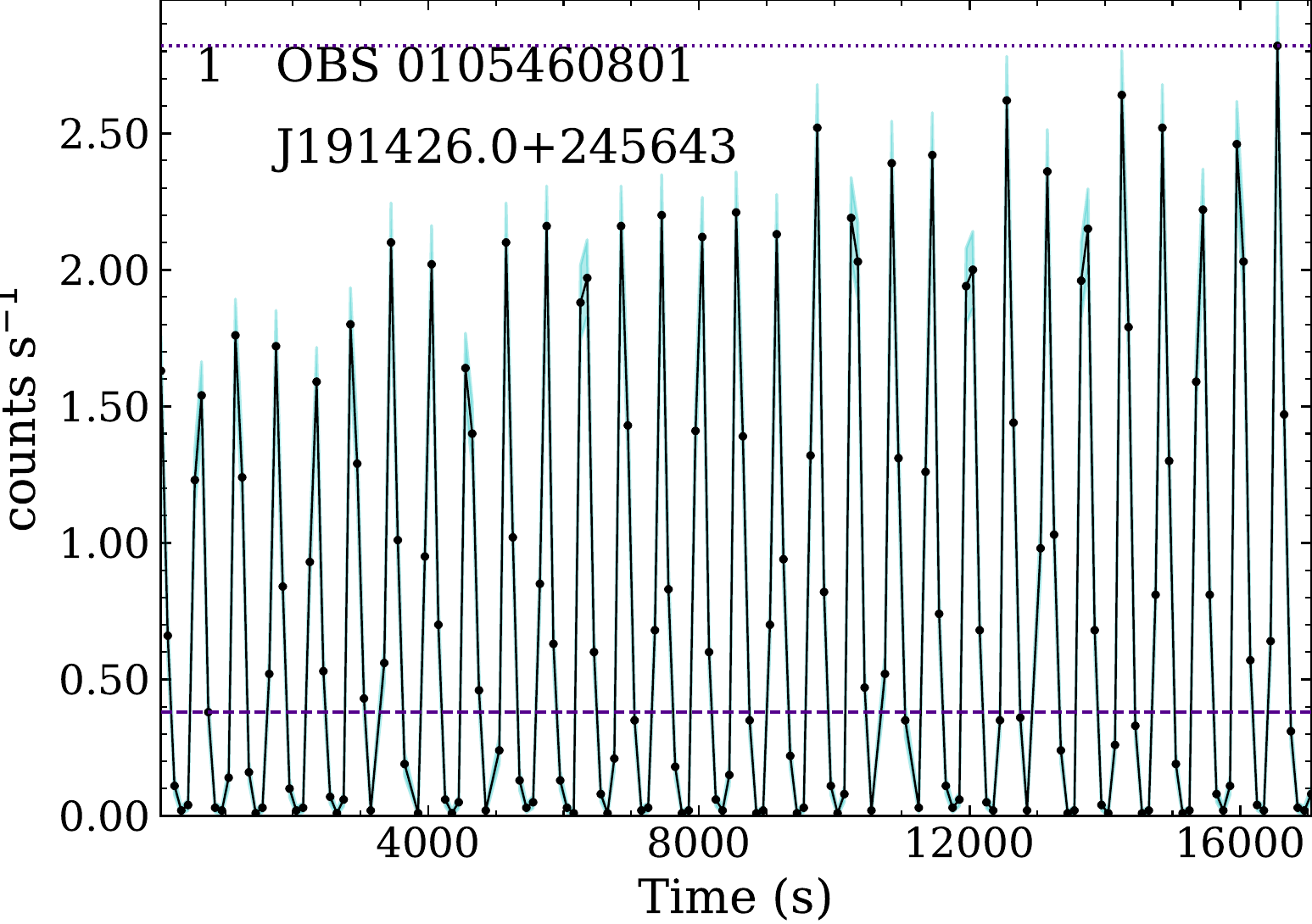} 
\includegraphics[width=0.325\textwidth]{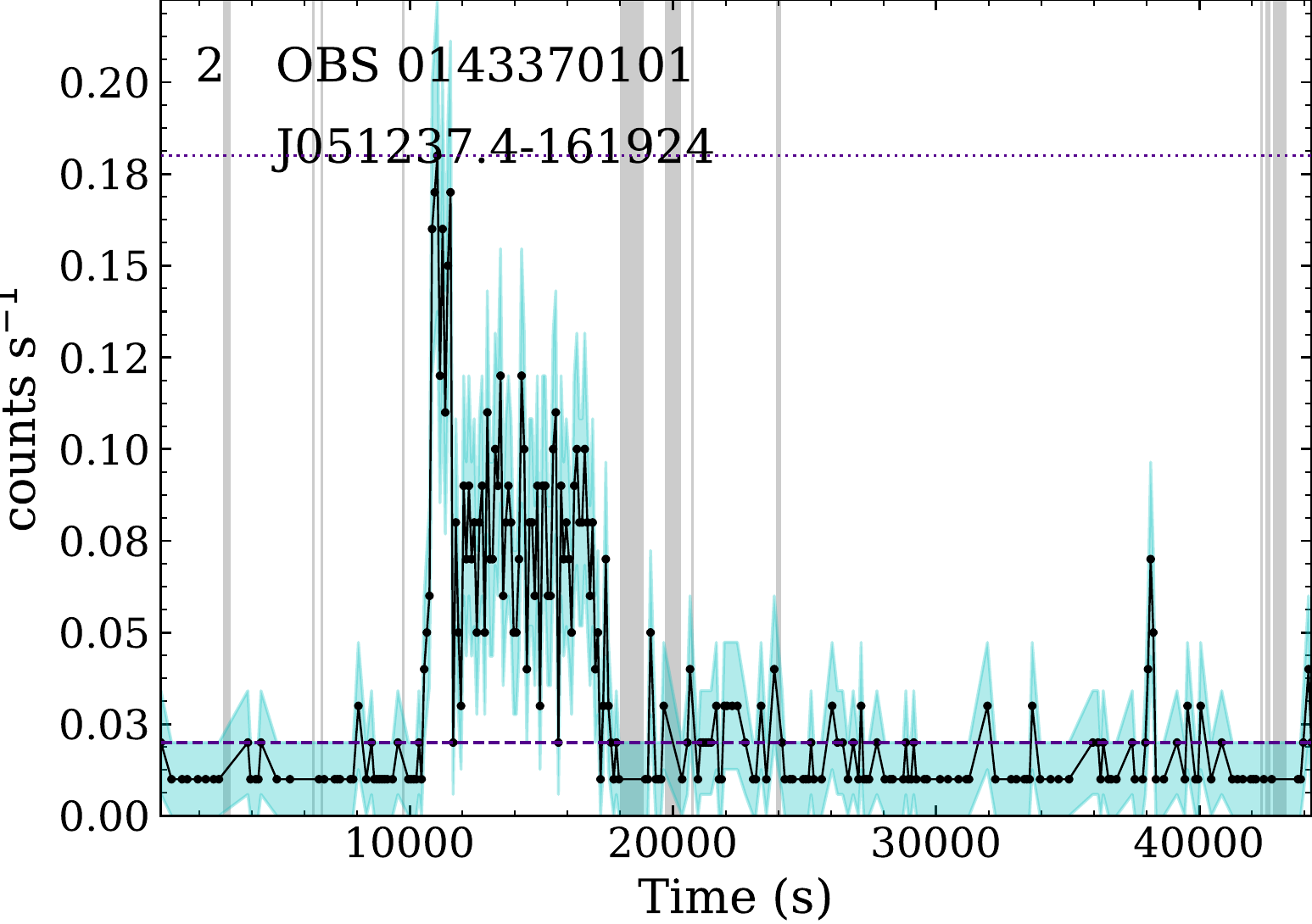} 
\includegraphics[width=0.325\textwidth]{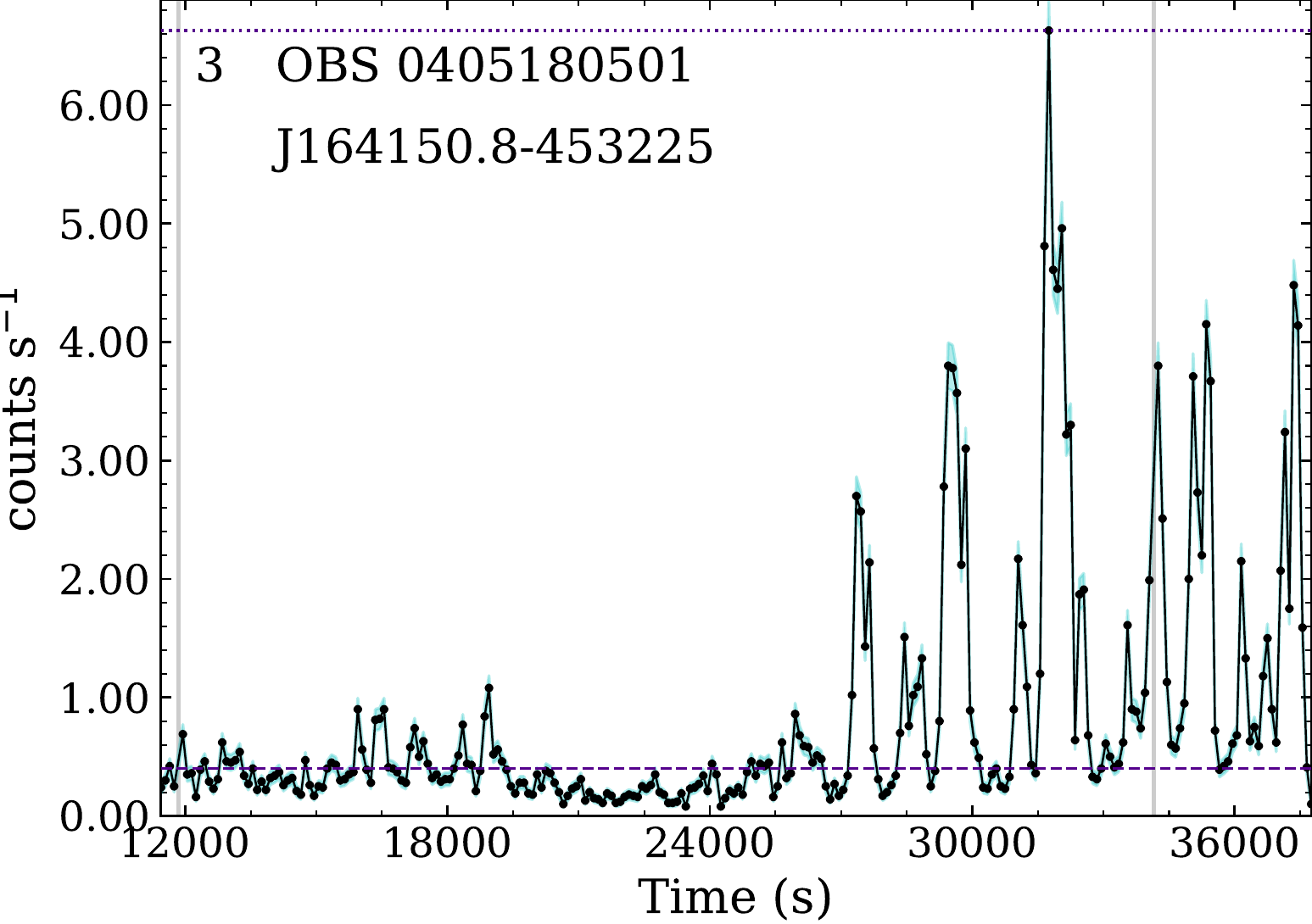} 
\includegraphics[width=0.325\textwidth]{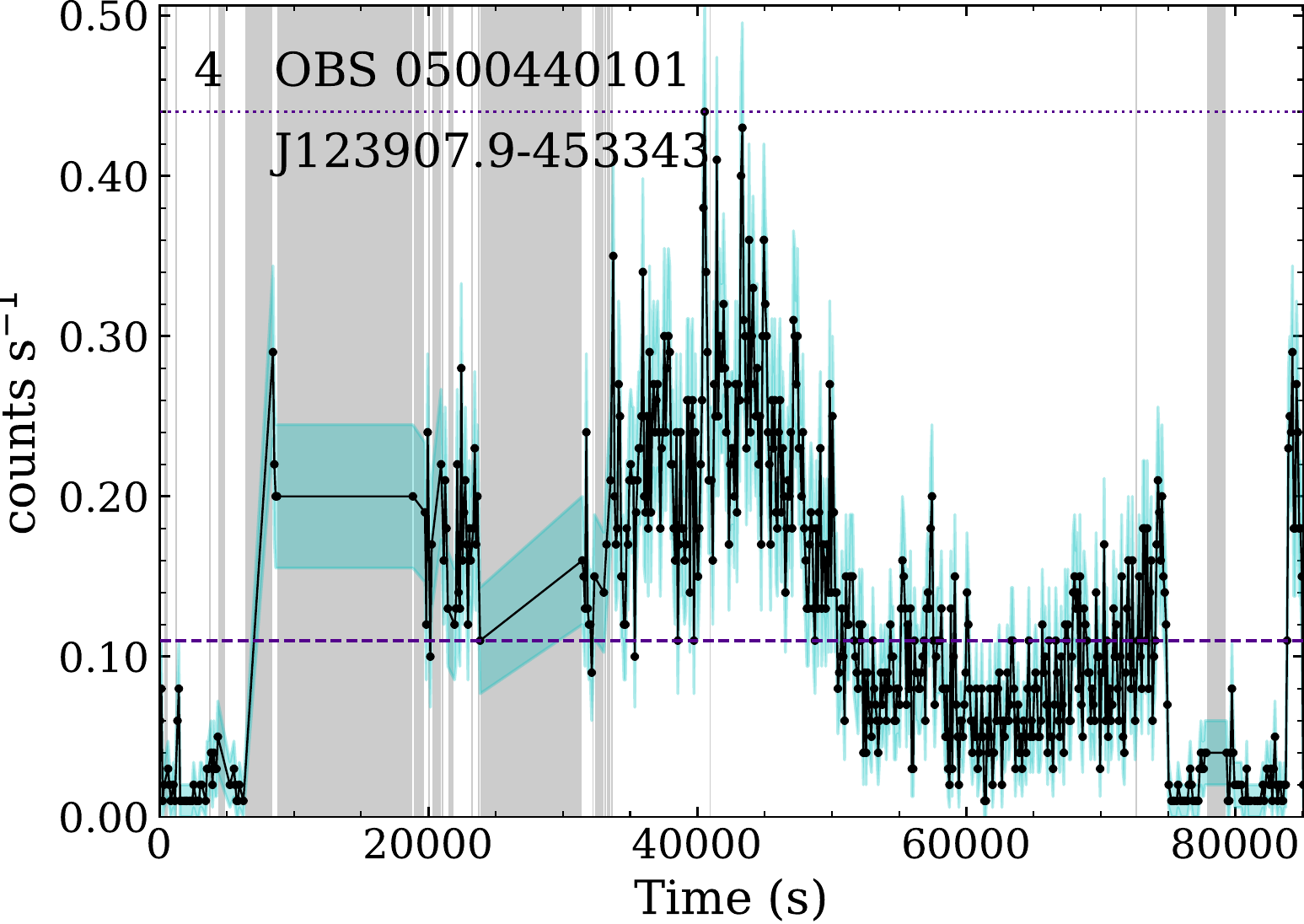} 
\includegraphics[width=0.325\textwidth]{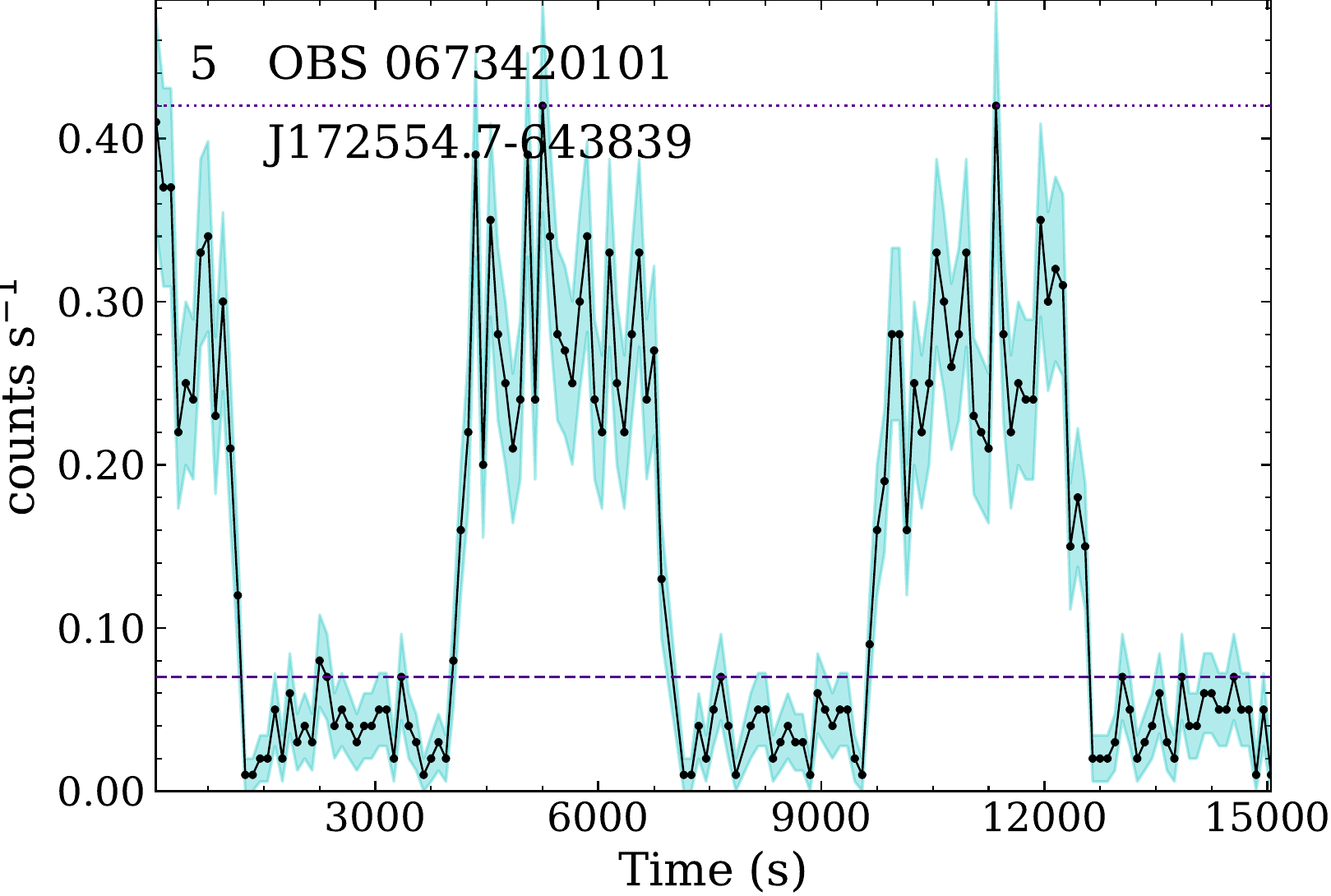} 
\includegraphics[width=0.325\textwidth]{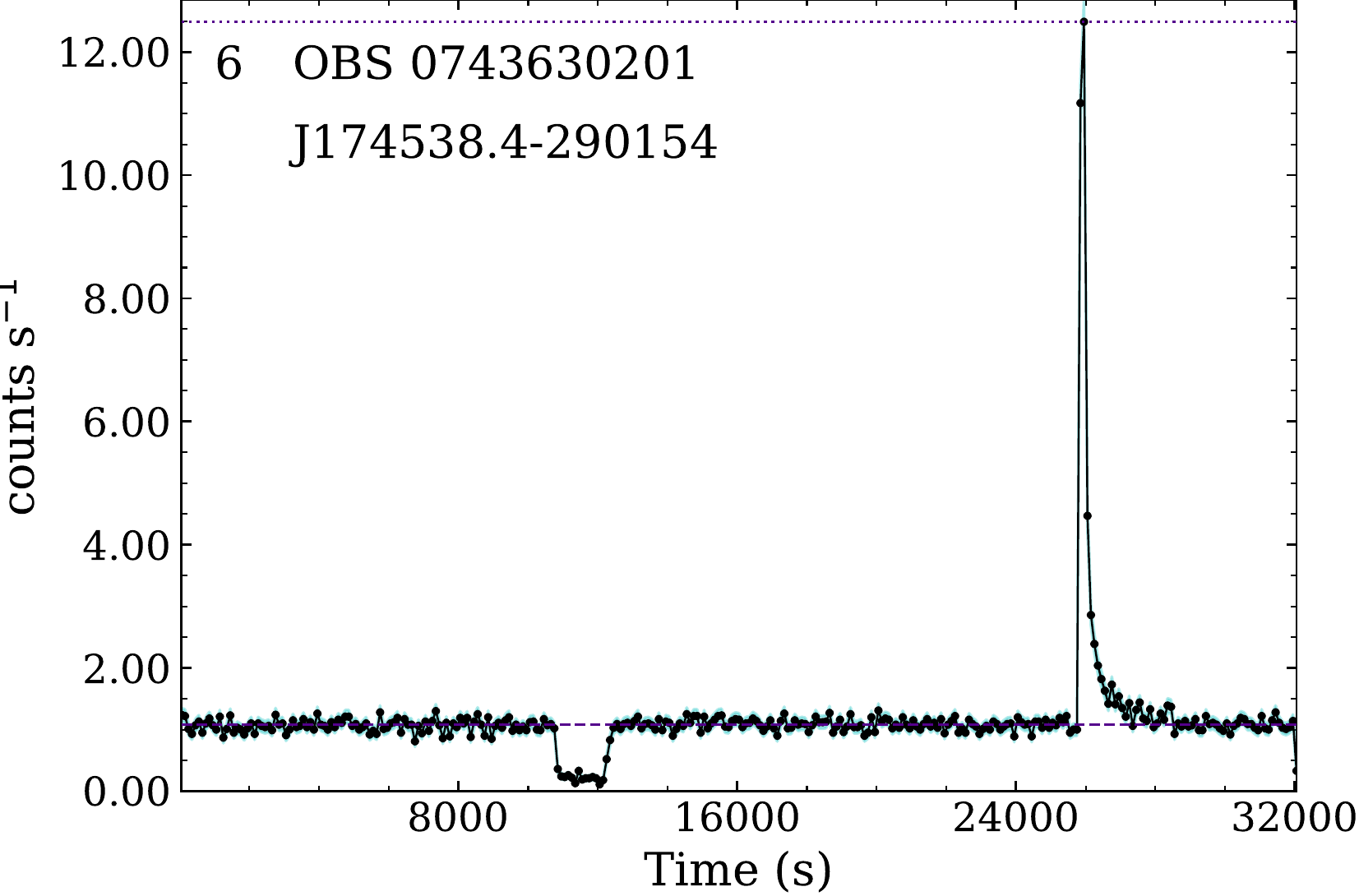} 
\caption{Example of light curves of sources with different types of variability detected by EXOD. The light curves are plotted in black with \textit{cyan} shaded regions representing the 1$\sigma$ error bars. In each plot we give the OBSID where it was detected and the 3XMM name of the source. The dashed \textit{purple} line represents $\tilde{\mathcal{C}}$, the median number of counts. The dotted \textit{purple} line represents $\mathcal{C}_{max}$, the maximal number of counts. The gray vertical shaded regions represent the bad time intervals.
1) RX J1914.4+2456, ultra compact binary showing periodic oscillations \citep{steeghs_gemini_2006}. 2) WISEA J051237.57-161925.2, high proper motion star showing a stellar flare \citep{kirkpatrick_allwise_2016}. 3) IGR J16418-4532, flares from a supergiant fast X-ray transient \citep{sidoli_xmm-newton_2012}. 4) V* V1129 Cen, eclipsing binary of beta Lyr type \citep{bruch_orbital_2017}. 5) LSQ J172554.8-643839, magnetic cataclysmic binary showing periodic eclipses \citep{fuchs_magnetic_2016} 6) AX J1745.6-2901, LMXB showing a type I X-ray burst \citep{in_t_zand_searching_2019}.}
\label{fig:types_var}
\end{figure*}

Although EXOD was specifically designed to detect faint, short outbursts, the inspection of the detected sources showed that it is able to detect a large diversity of variable phenomena. In Fig.~\ref{fig:types_var} we show a selection of six light curves from different variable sources. These light curves were extracted automatically. We also plot $\tilde{\mathcal{C}}$ and $\mathcal{C}_{max}$ to compare the light curves to the value of the variability obtained with EXOD.

These sources differ in their variability timescales (short and long flares), periodicity or aperiodicity, and speed of
decline and rise. They also span a range of physical classes. In Fig.~\ref{fig:types_var}, we present the following sources:
in (1), an ultra compact binary showing periodic oscillations whose maximal values seem to increase during the observation \citep{steeghs_gemini_2006}.
In (2), a high proper motion star showing a flare during the observation \citep{kirkpatrick_allwise_2016}.
In (3), we can see flares from a supergiant fast X-ray transient, aperiodic with varying maxima \citep{sidoli_xmm-newton_2012}.
In (4), an eclipsing binary of beta Lyr type \citep{bruch_orbital_2017}.
In (5), we see a magnetic cataclysmic variable whose periodic variability is due to variation in the opacity of the accretion curtain that comes into view as the system rotates \citep{mason_x-ray_1985,fuchs_magnetic_2016}.
In (6), an LMXB presents first an eclipse during which the flux decreases, then a type-I X-ray burst \citep{in_t_zand_searching_2019}. Note that we refiltered the last observation since the 0.5\ctss\ \texttt{tabgtigen} threshold cut a part of the flare.

\subsection{Computational performance of the algorithm}

\begin{figure}
\centering
\includegraphics[width=\hsize]{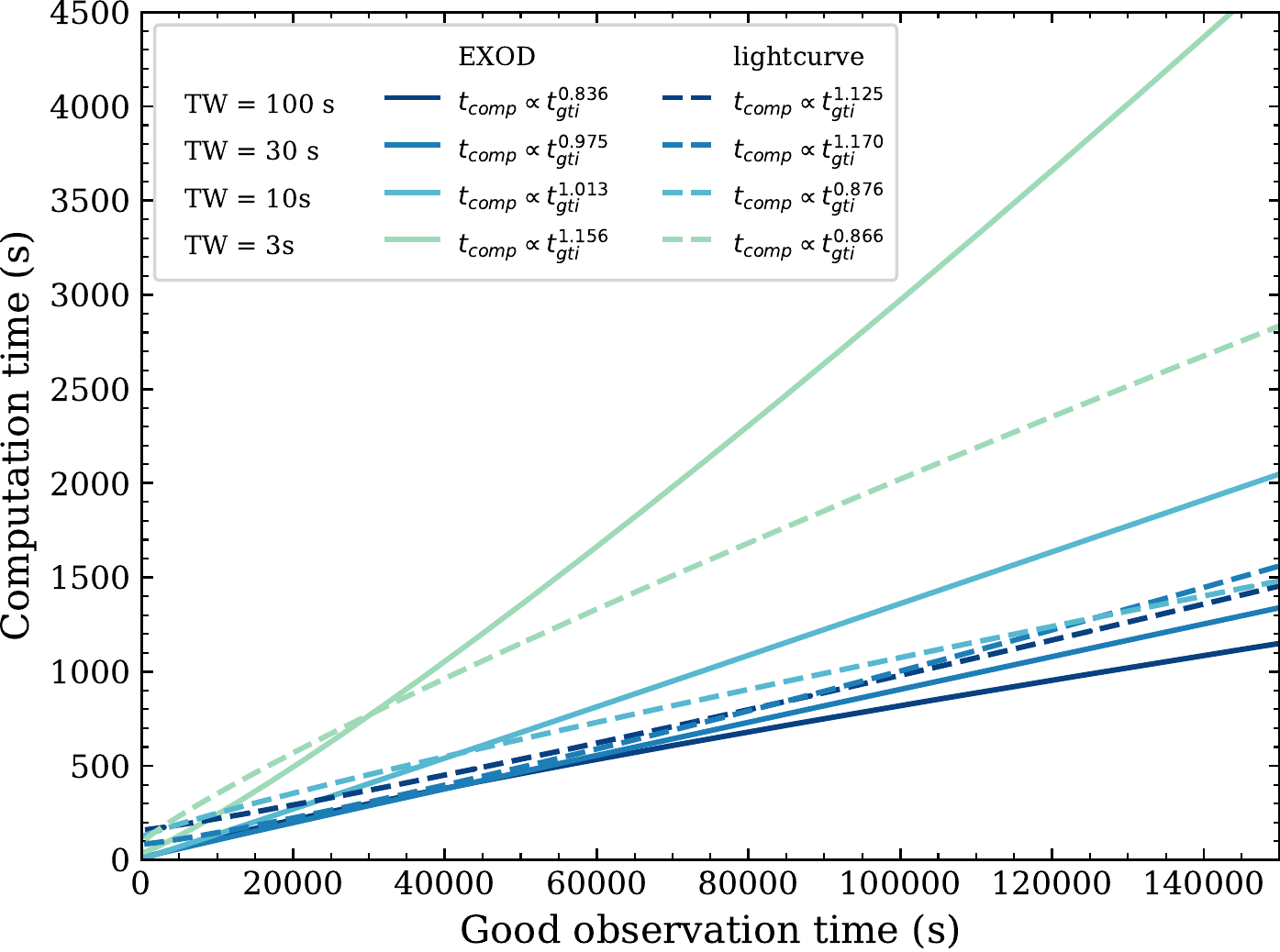}
\caption{Computation time as a function of the observation time that has not been polluted  by background flares (good time) fitted to a power law. The solid lines indicate the EXOD detector computation time. The dashed lines indicate the light curve computation time for one source. Blue (dark) is the computation time for a time window of 100\,s, green (medium) for 30\,s, orange (medium) for 10\,s and pink (light) for 3s.}
\label{fig:time}
\end{figure}

We measured the absolute computation time $t_{comp}$ taken by EXOD to analyse each observation. Additionally, we measured the computation time for the generation of a single light curve for each detected variable source. Experiments were conducted on a virtual machine with 38 cores at 2.3~GHz and 320~GB of RAM. We fitted these values to a power law with the expression given in Eq.~\ref{eq:time} as a function of the total duration of the GTI, $t_{gti}$. The result is plotted in Fig.~\ref{fig:time},
\begin{equation}\label{eq:time}
t_{comp} = a + b \times t_{gti}^{c}
\end{equation}
where $a$, $b$ and $c$ are the parameters determined by the fit. We computed these parameters by considering the computation time for all of the analysed observations with time windows of 3s, 10\,s, 30\,s and 100\,s. These values are shown in Table~\ref{tab:times}.

The computation time of the algorithm includes the variability computation and the source detection time, whereas the computation time of the light curve includes some SAS and FTOOLS tasks that are required for the light curve generation: \texttt{cifbuild}, \texttt{eregionanalyse}, \texttt{ebkgreg}, \texttt{evselect}, \texttt{epiclccorr}, \texttt{lcstats} and \texttt{lcurve}.

The EXOD computation time increases for shorter $TW$, as expected given the increased number of data points. The light curve computation time is similar for 10, 30 and 100\,s $TW$, and increases for $TW = 3$\,s.
For 100 and 30\,s $TW$, EXOD can compute the variability of \emph{the whole observation} before even a single light curve
is generated. While for a shorter $TW$, the computation is slower than generating a single light curve,
EXOD is faster than light curve generation when more than one source is present in the FoV.

\begin{table}
	\centering
	\caption{Computation time fit parameters.}
	\label{tab:times}
	\begin{tabular}{r|rrr|rrr}
	\hline\hline
	& \multicolumn{3}{c|}{EXOD} & \multicolumn{3}{c}{light curve}\\
	\hline
	$TW$(s) & $a$ & $b$ & $c$ & $a$ & $b$ & $c$ \\
	\hline
	100 & 0.151 & 0.054 & 0.836 & 158.204 & 0.002 & 1.125\\
	30 & 12.614 & 0.012 & 0.975 & 83.171 & 0.001 & 1.170 \\
	10 & 5.210 & 0.012 & 1.013  & 121.960 & 0.040 & 0.876\\
	3 & 36.822 & 0.005 & 1.156 & 89.109 & 0.091 & 0.866\\
	\hline
	\end{tabular}
	\tablefoot{Parameters of the fit to the computation time for the expression \ref{eq:time} for EXOD and the light curve generation.}
\end{table}

\subsection{Previously steady sources found to be variable}
\label{sec:vary}

\begin{figure*}
	\centering
	\includegraphics[width=0.325\textwidth]{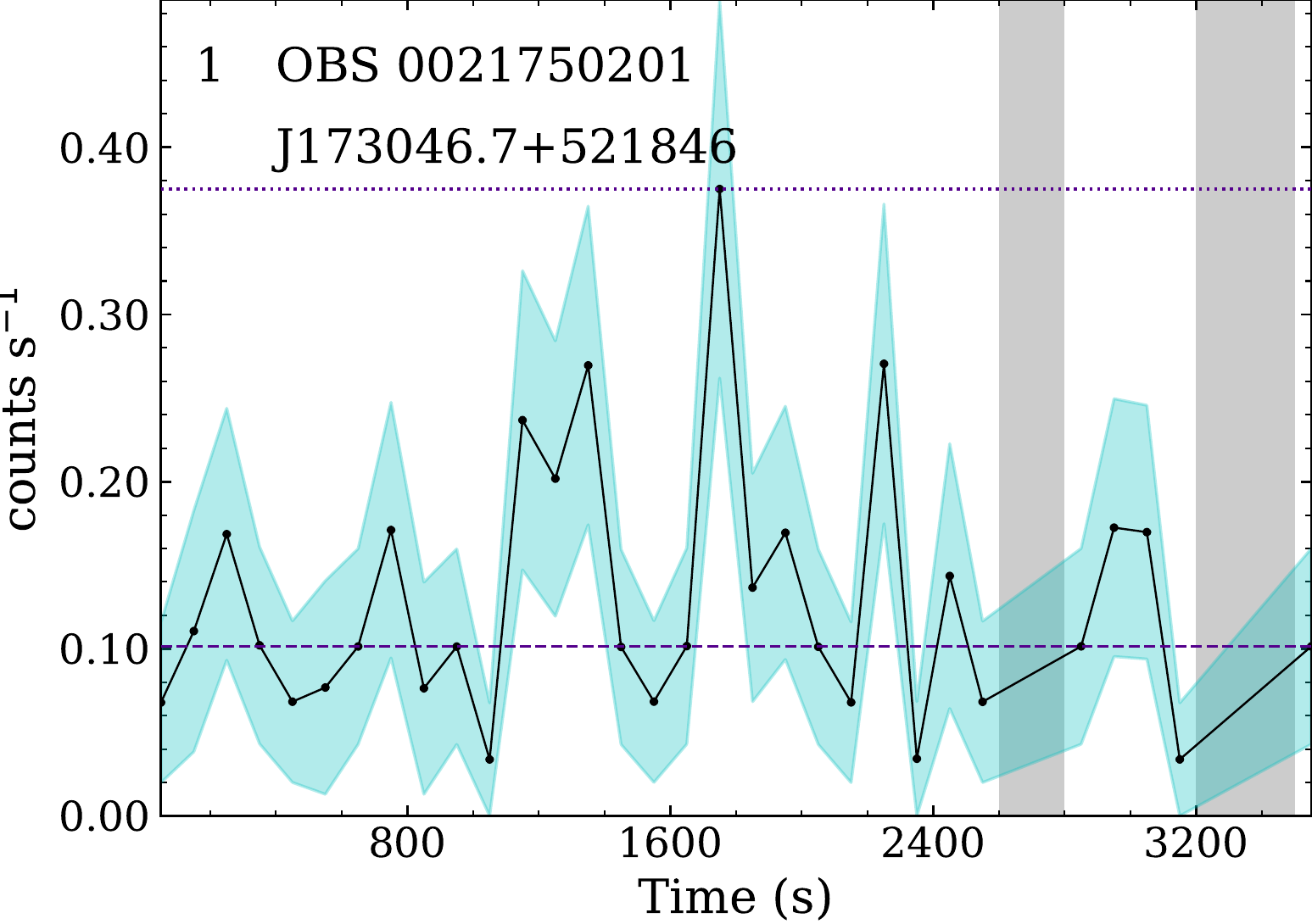}
	\includegraphics[width=0.325\textwidth]{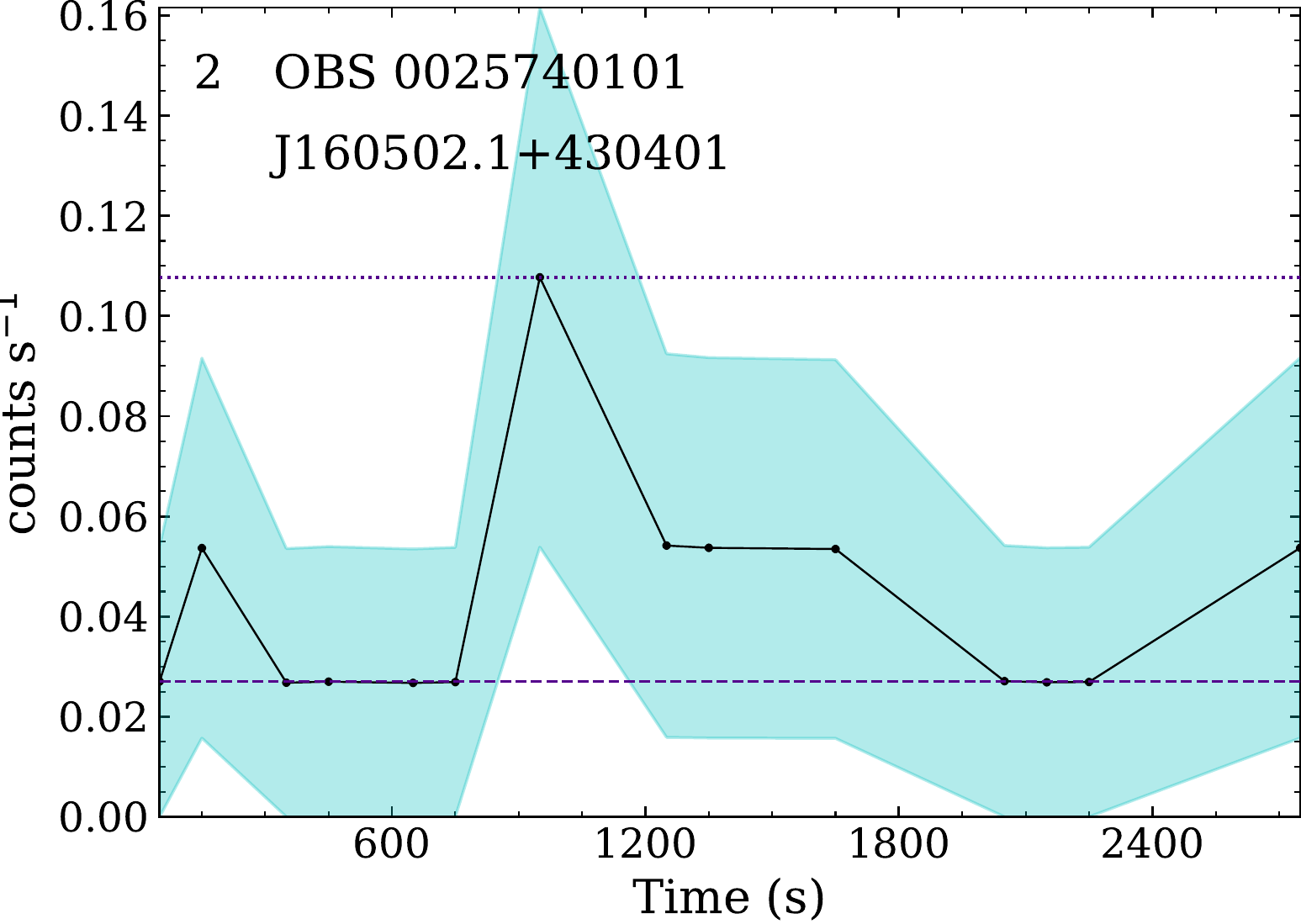}
	\includegraphics[width=0.325\textwidth]{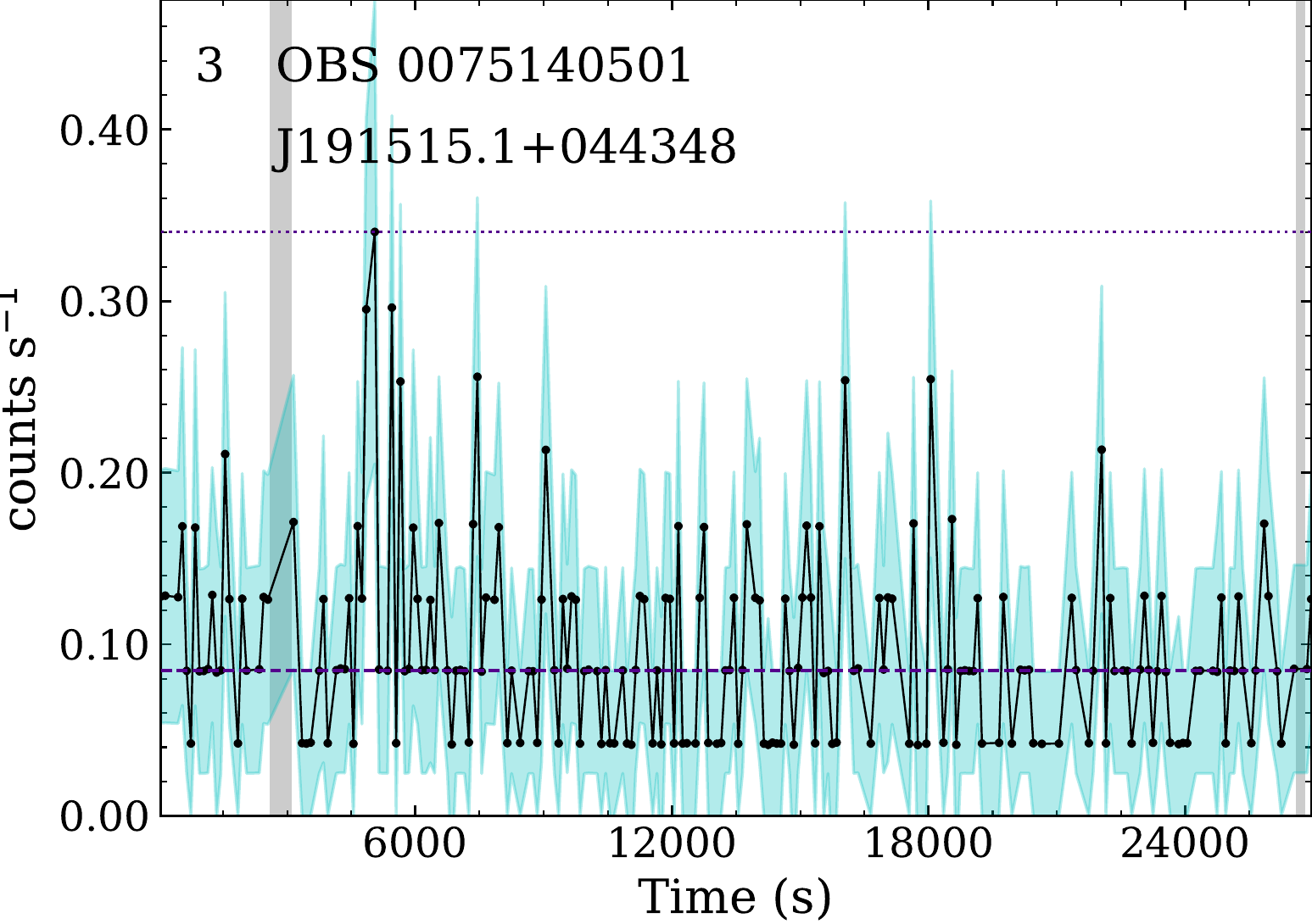}
	\includegraphics[width=0.325\textwidth]{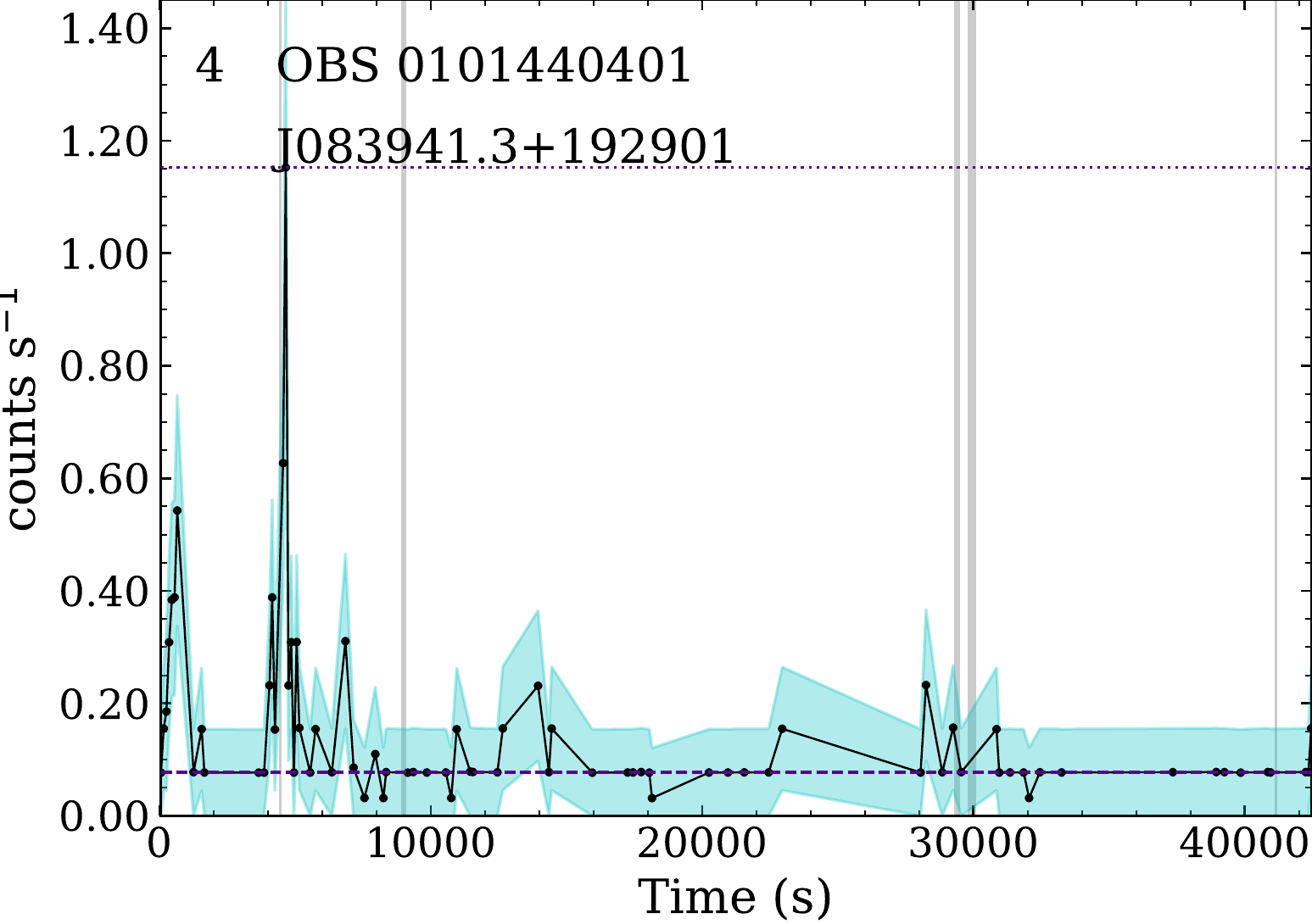}
	\includegraphics[width=0.325\textwidth]{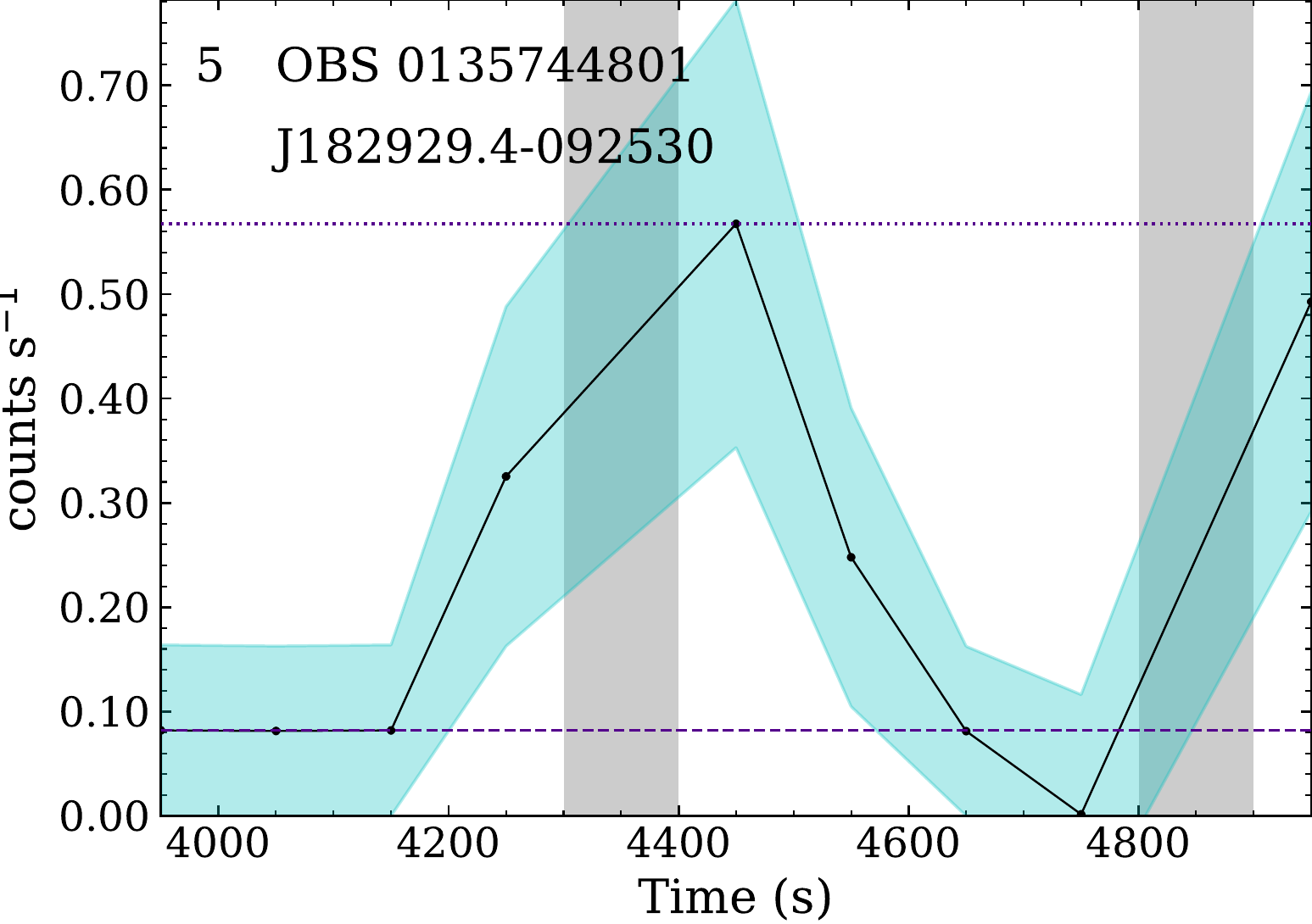}
	\includegraphics[width=0.325\textwidth]{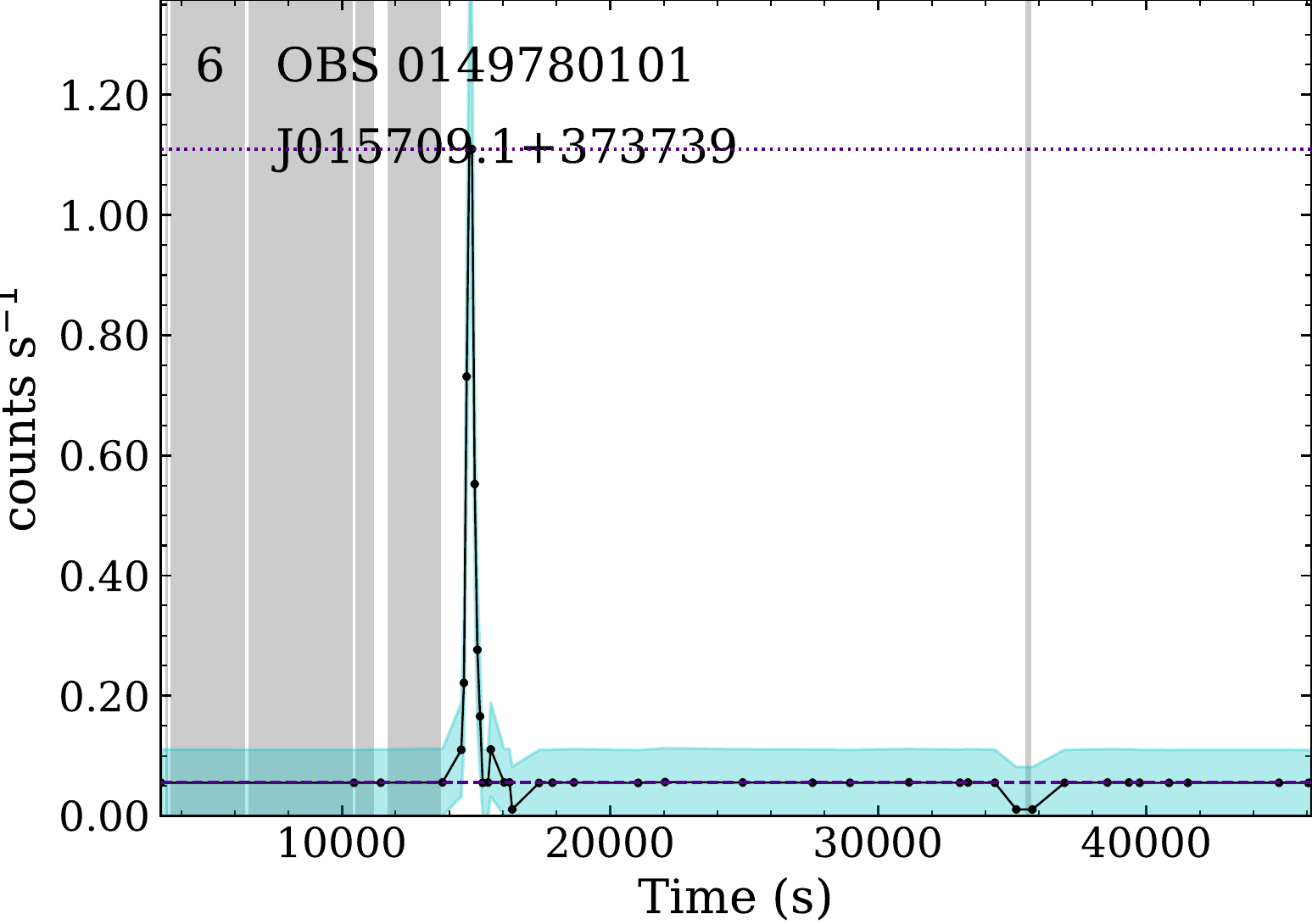}
	\includegraphics[width=0.325\textwidth]{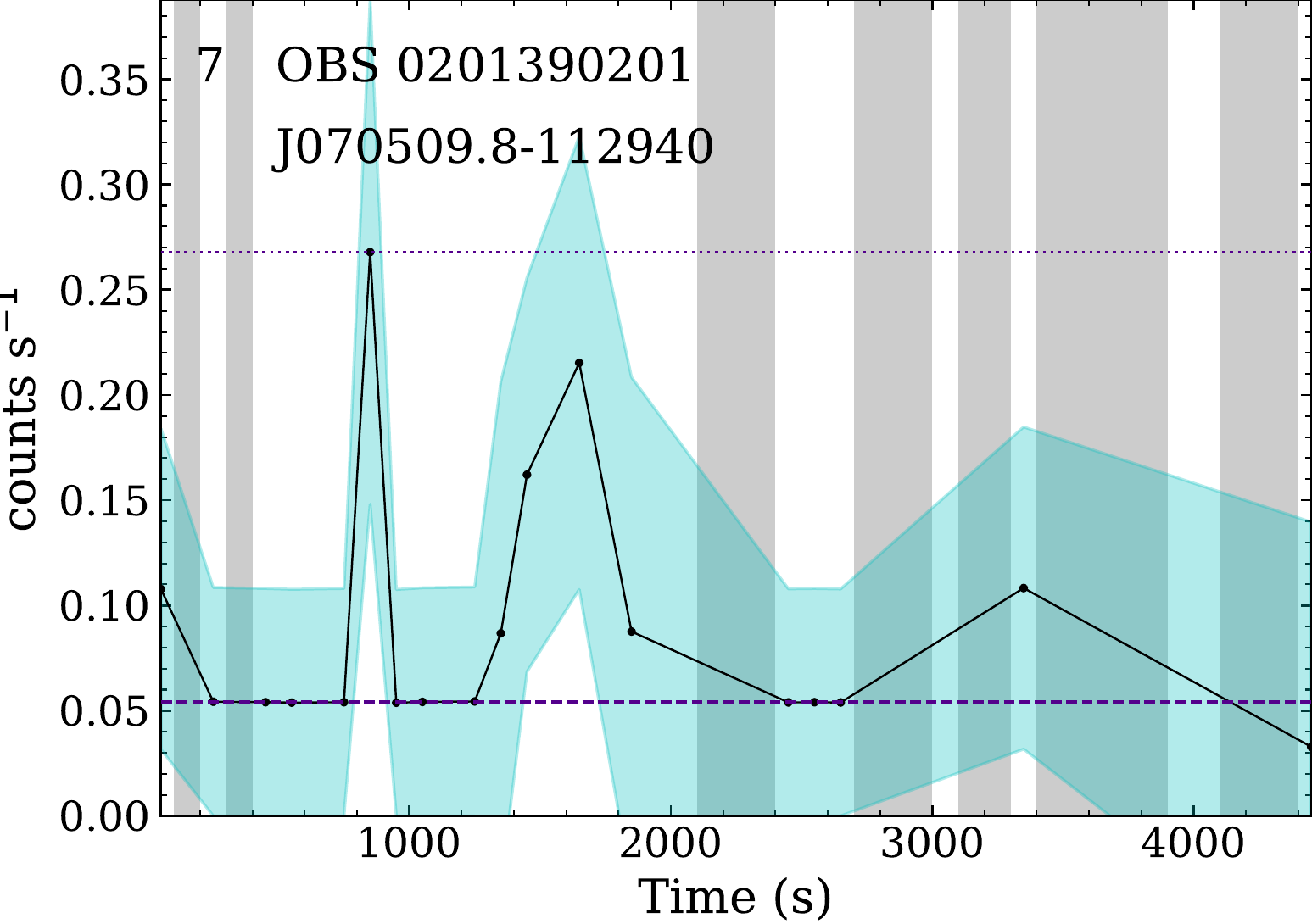}
	\includegraphics[width=0.325\textwidth]{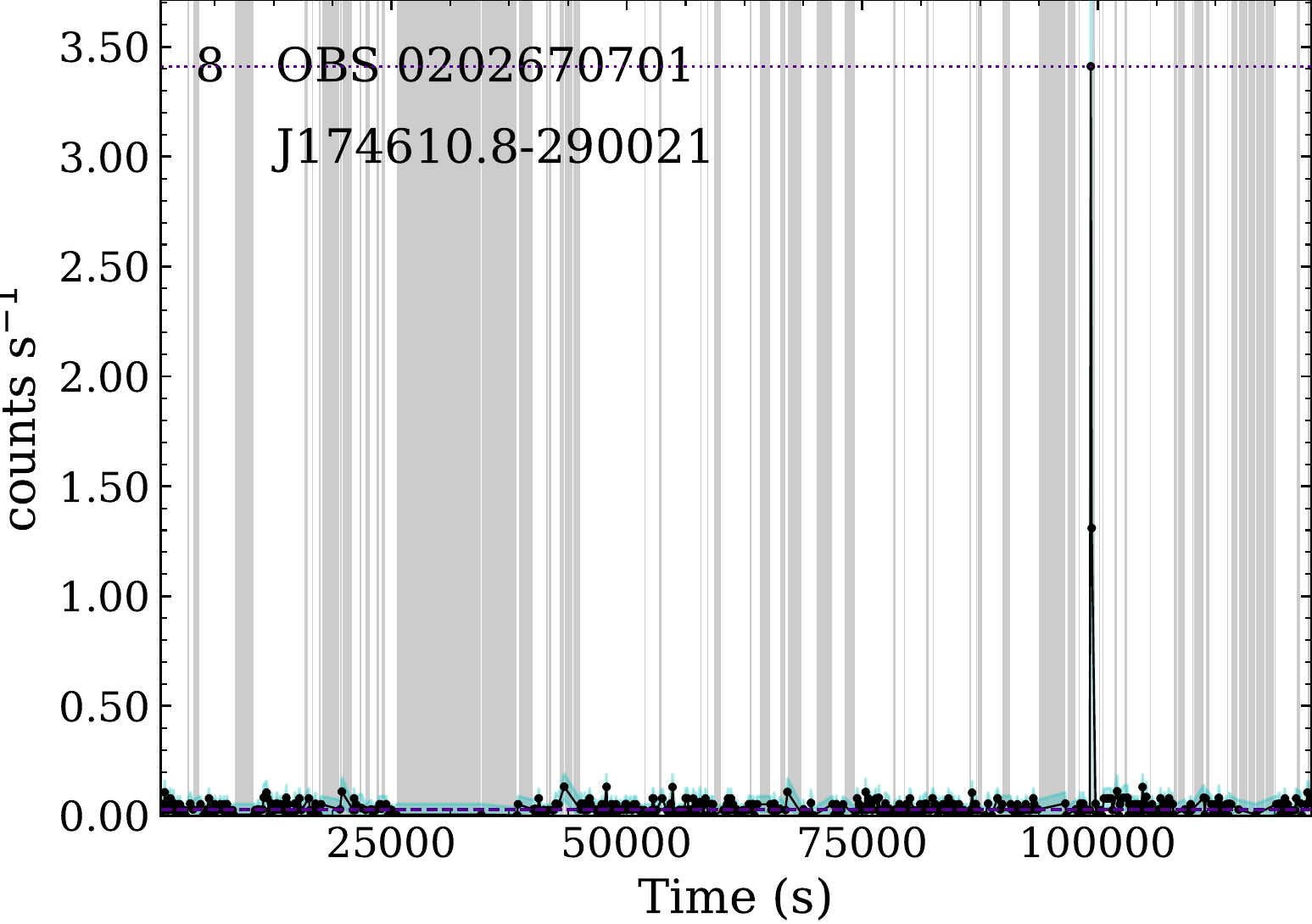}
	\includegraphics[width=0.325\textwidth]{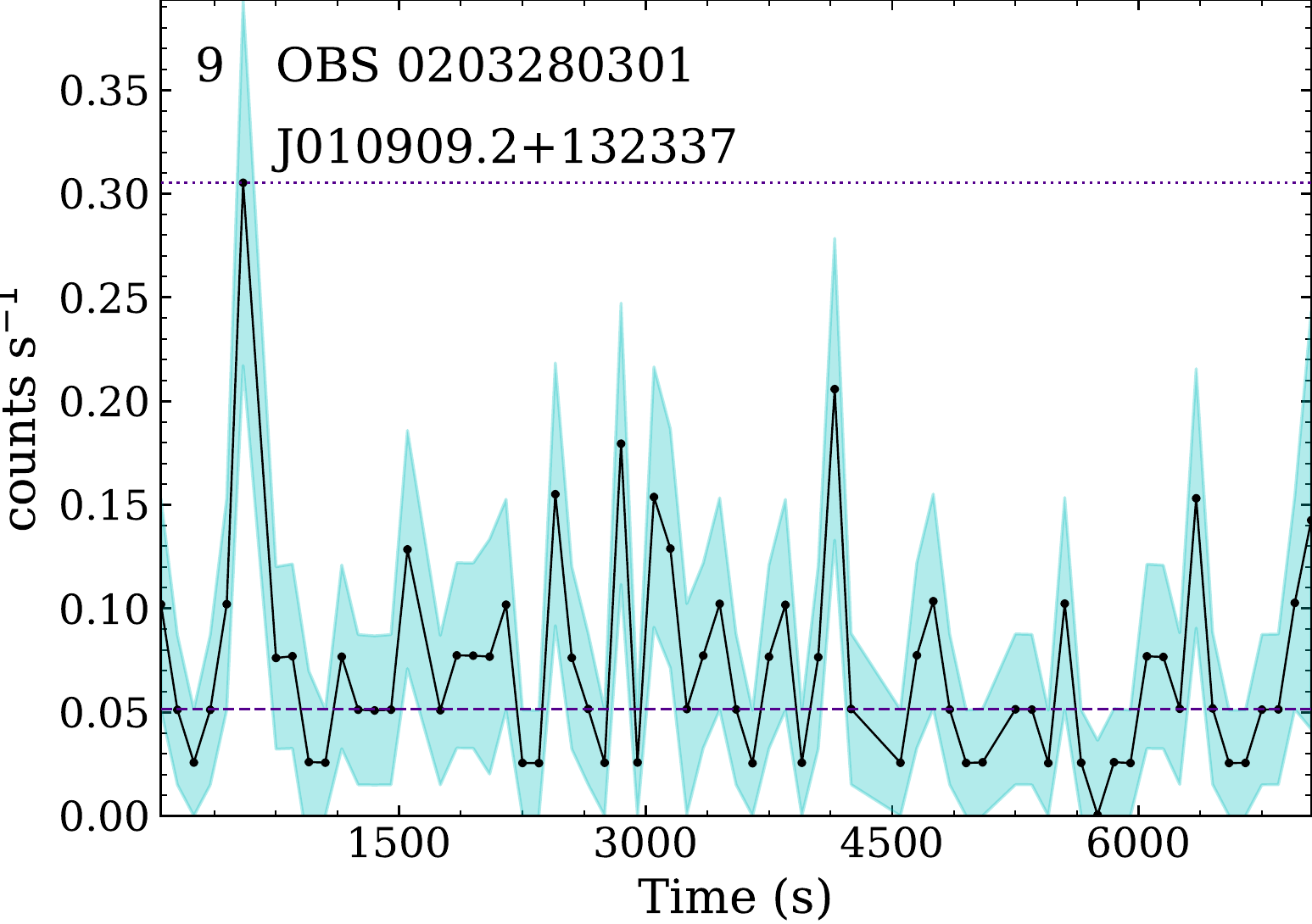}
	\includegraphics[width=0.325\textwidth]{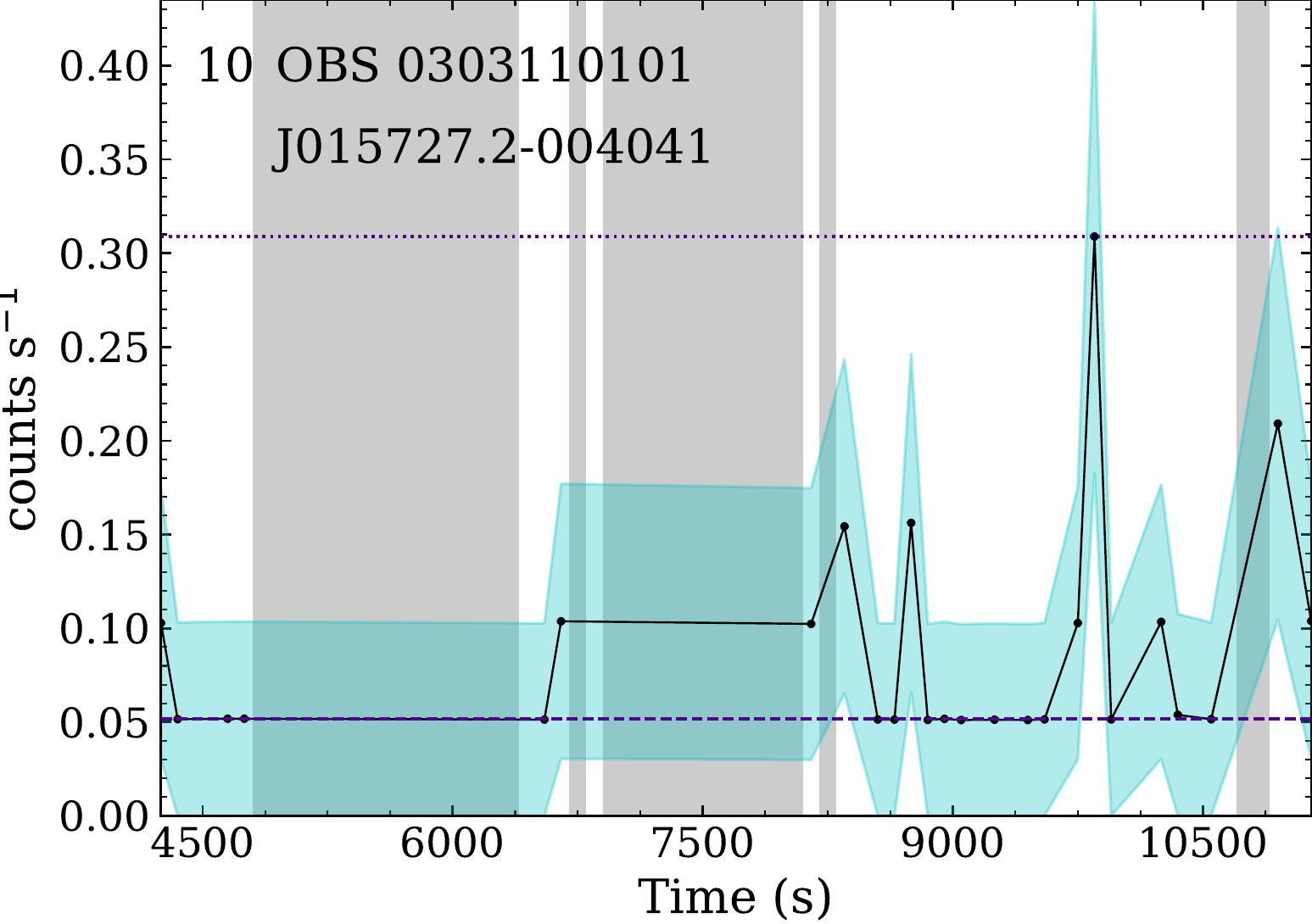}
	\includegraphics[width=0.325\textwidth]{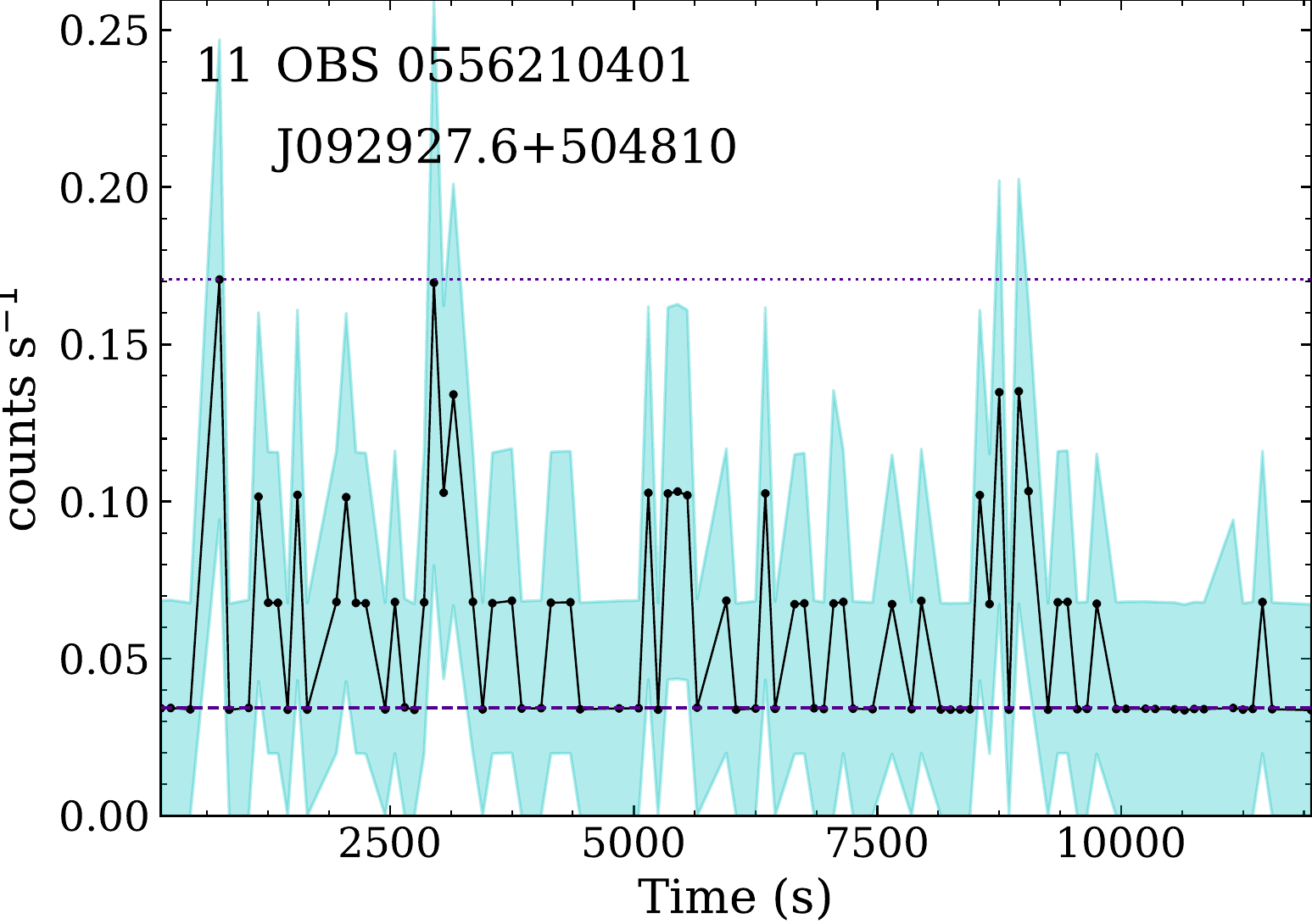}
	\includegraphics[width=0.325\textwidth]{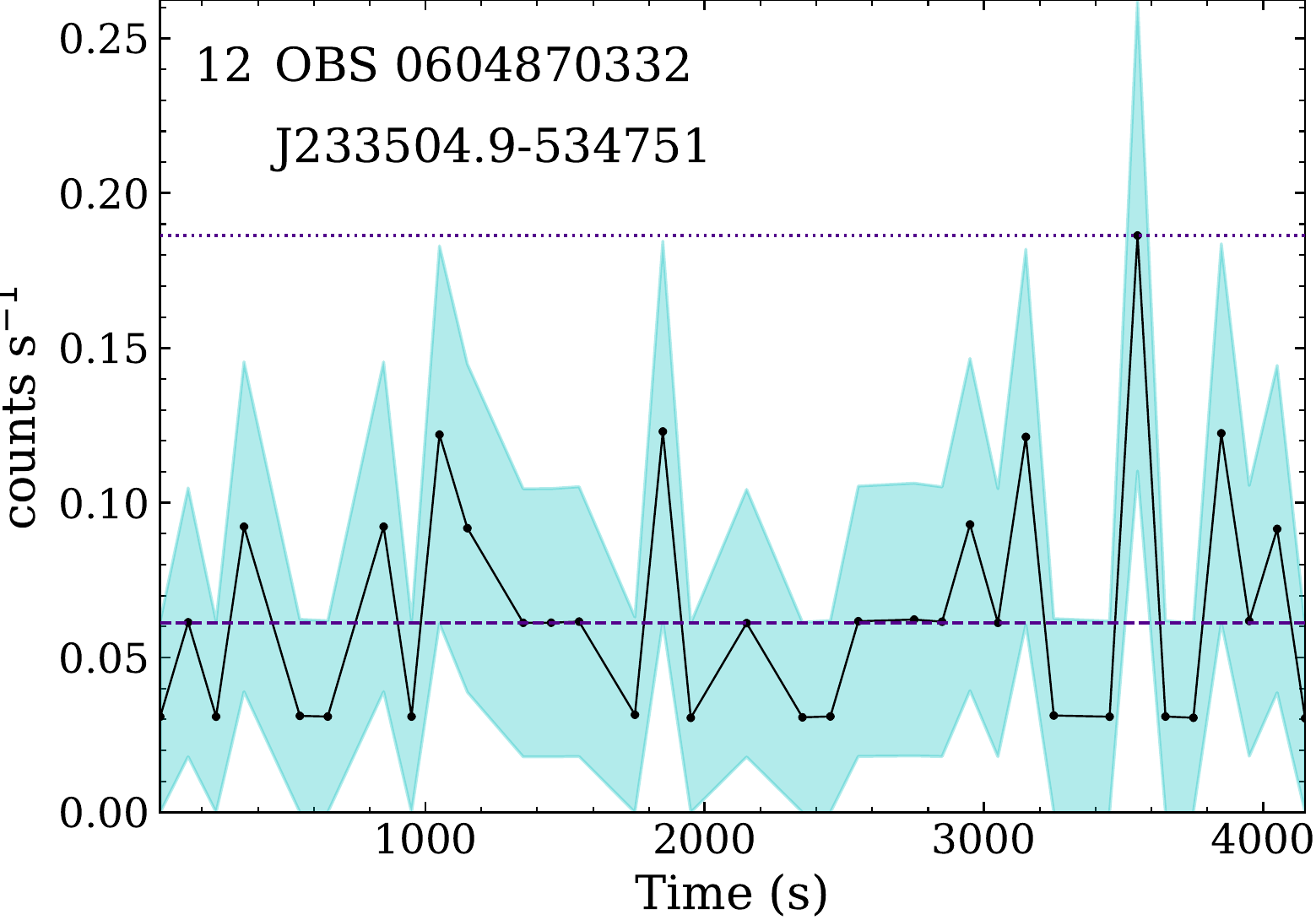}
	\includegraphics[width=0.325\textwidth]{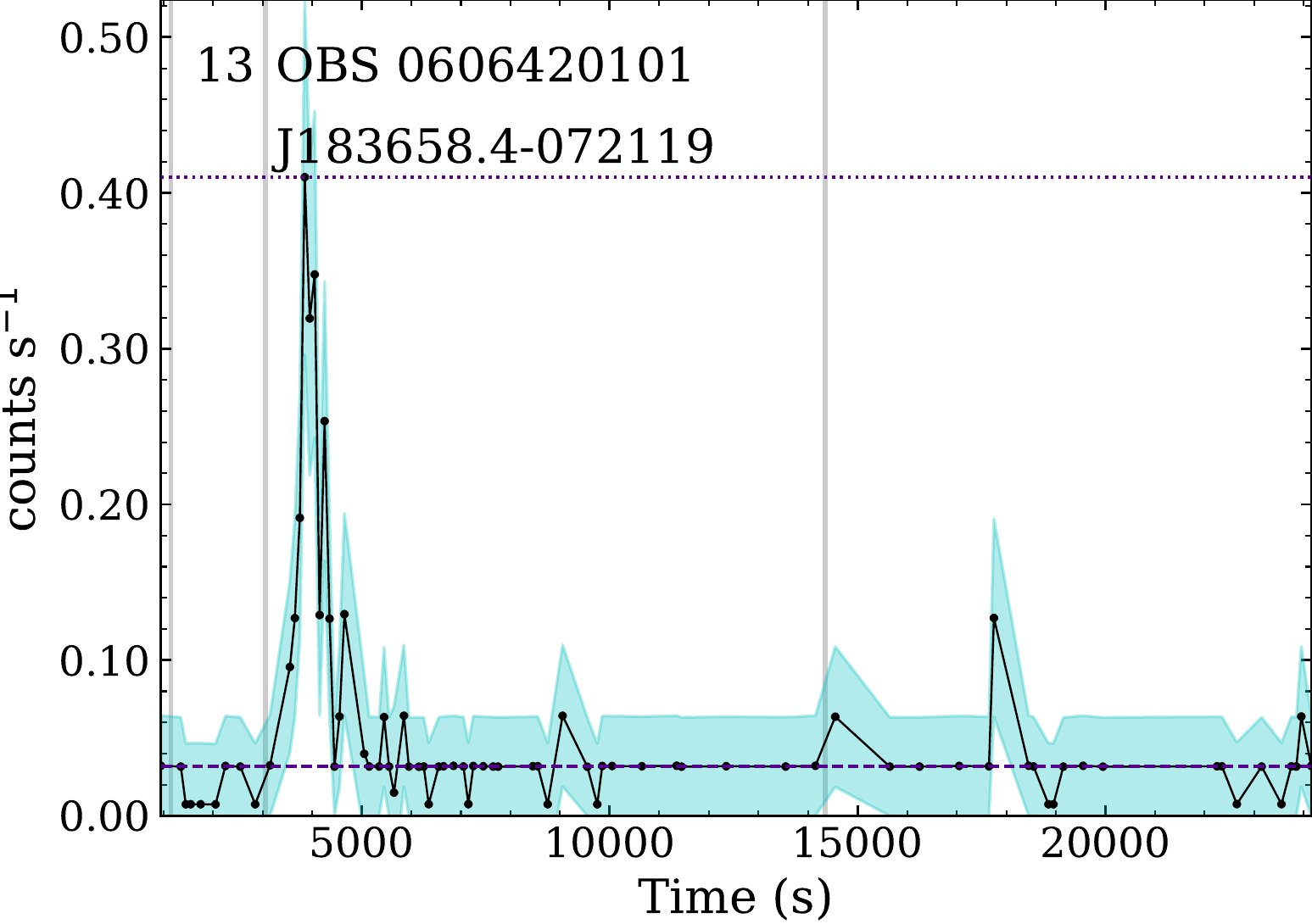}
	\caption{Light curves of sources detected with EXOD, present in 3XMM-DR8, but not classified as variable by the automatic pipeline. The light curves are plotted in black with \textit{cyan} shaded regions representing the 1$\sigma$ error bars. The dashed \textit{purple} line represents $\tilde{\mathcal{C}}$, the median number of counts. The dotted \textit{purple} line represents $\mathcal{C}_{max}$, the maximal number of counts. The gray vertical shaded regions represent the bad time intervals. Each plot gives the source ID, the OBSID in which it was detected and the 3XMM name of the source.}
	\label{fig:outbursts_1}
\end{figure*}

\begin{figure*}
	\centering
	\includegraphics[width=0.325\textwidth]{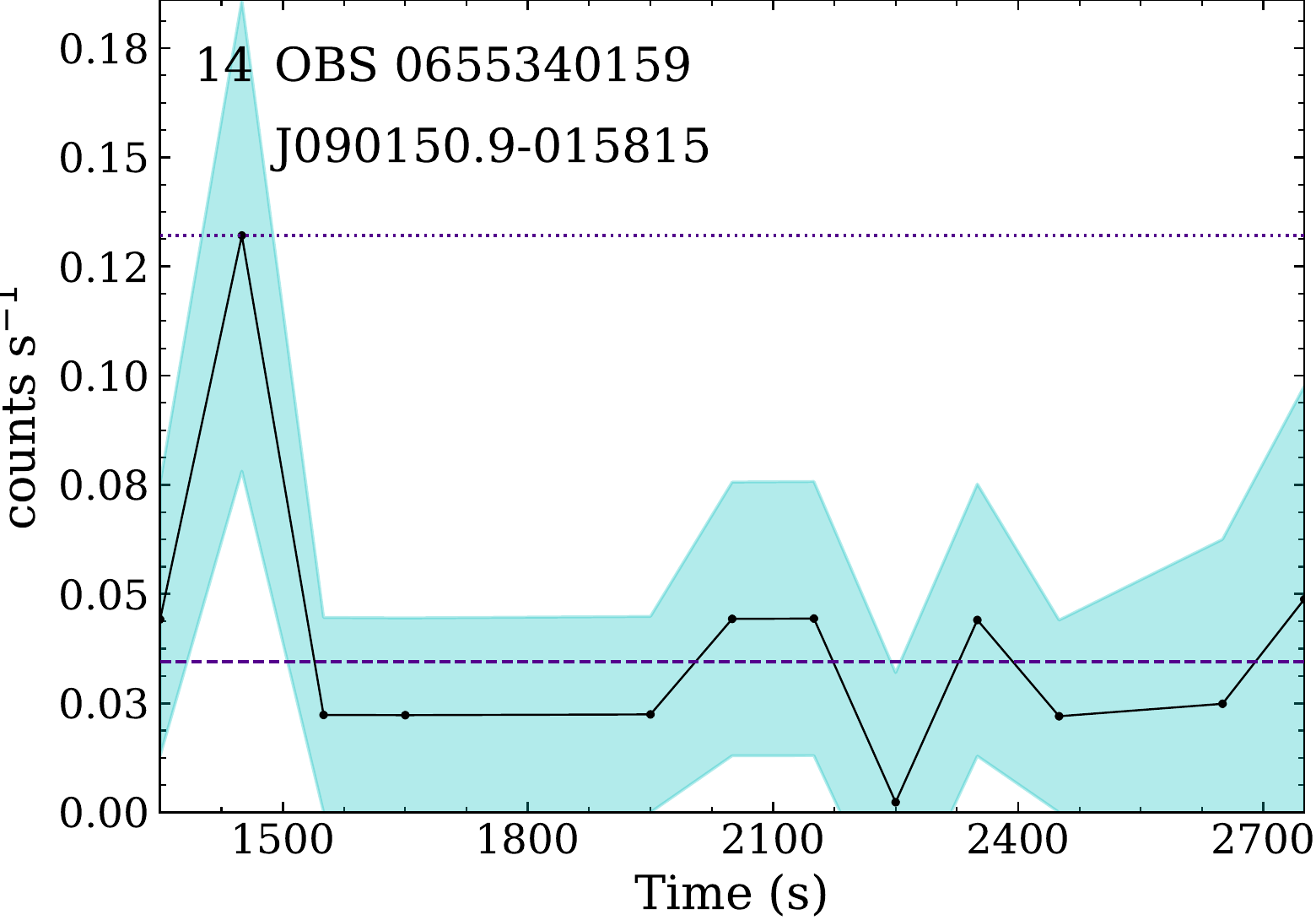}
	\includegraphics[width=0.325\textwidth]{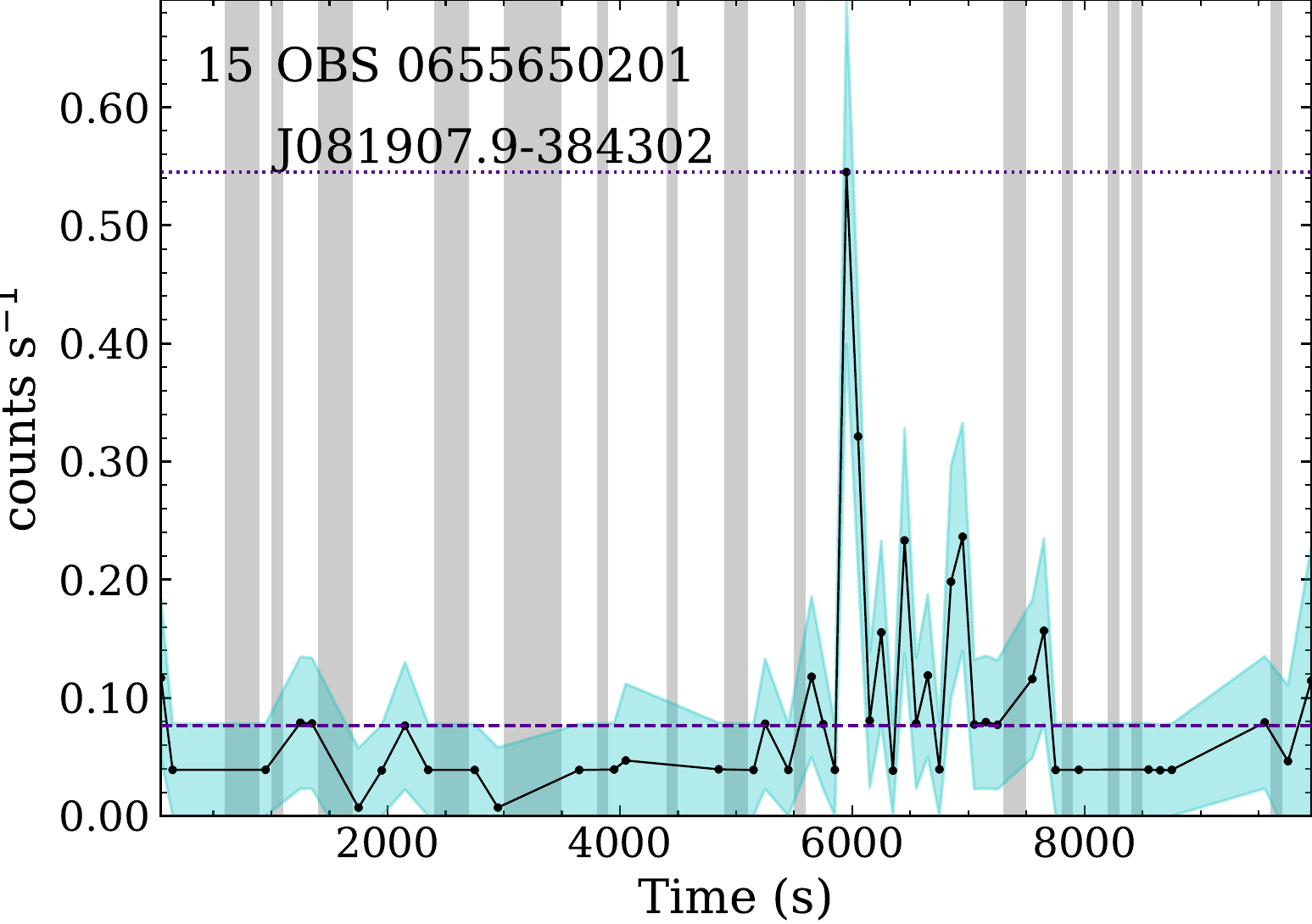}
	\includegraphics[width=0.325\textwidth]{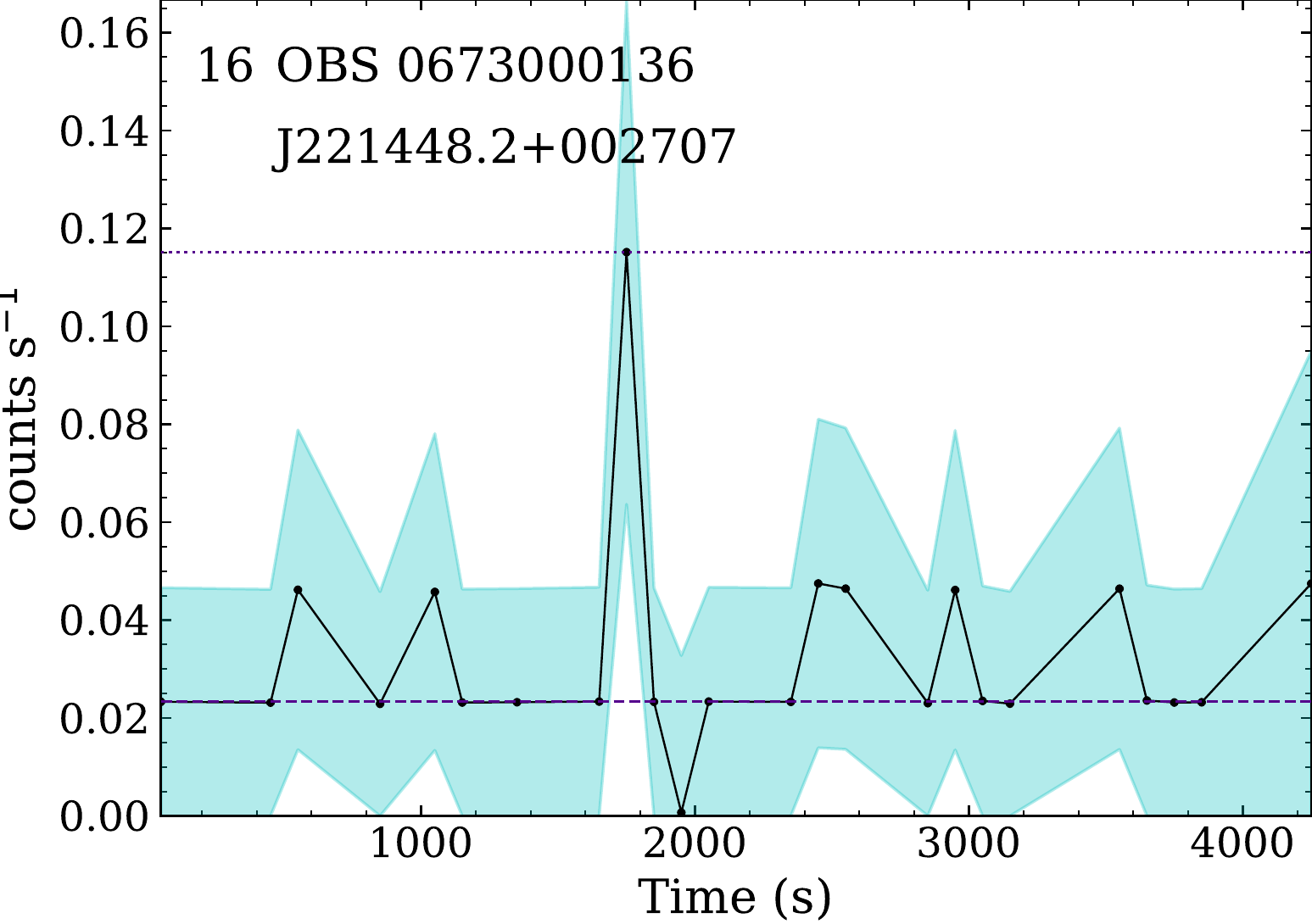}
	\includegraphics[width=0.325\textwidth]{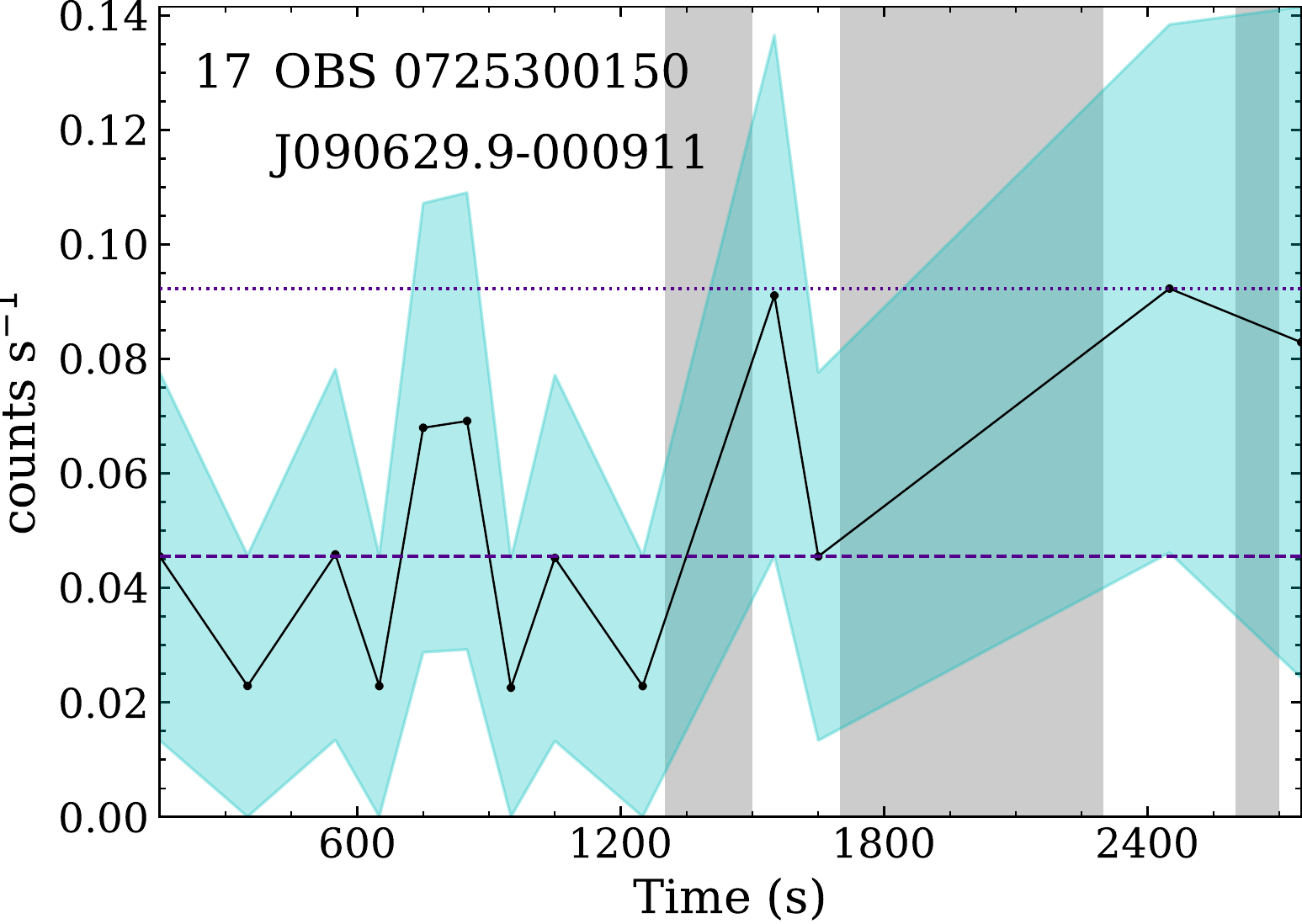}
	\includegraphics[width=0.325\textwidth]{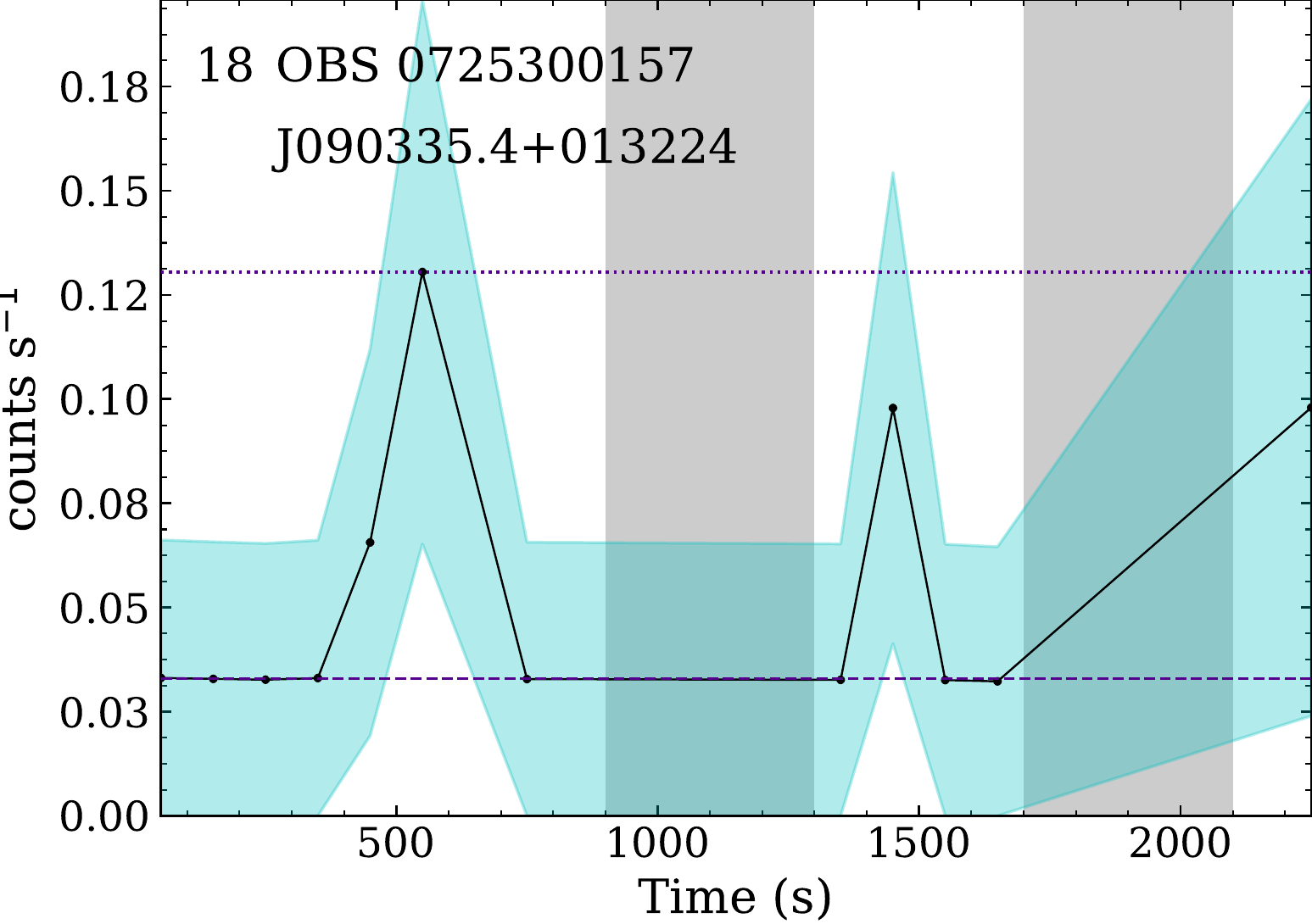}
	\includegraphics[width=0.325\textwidth]{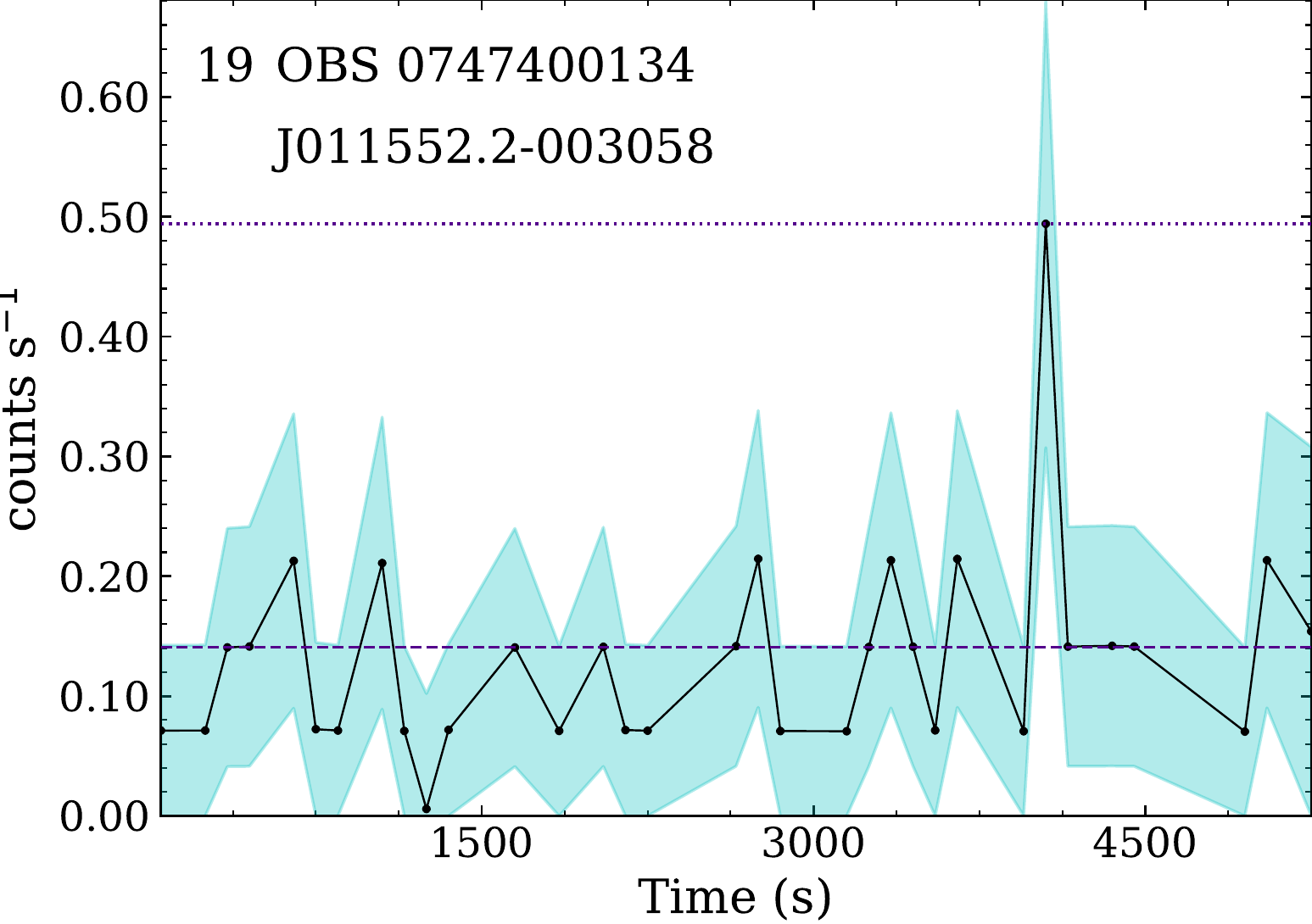}
	\includegraphics[width=0.325\textwidth]{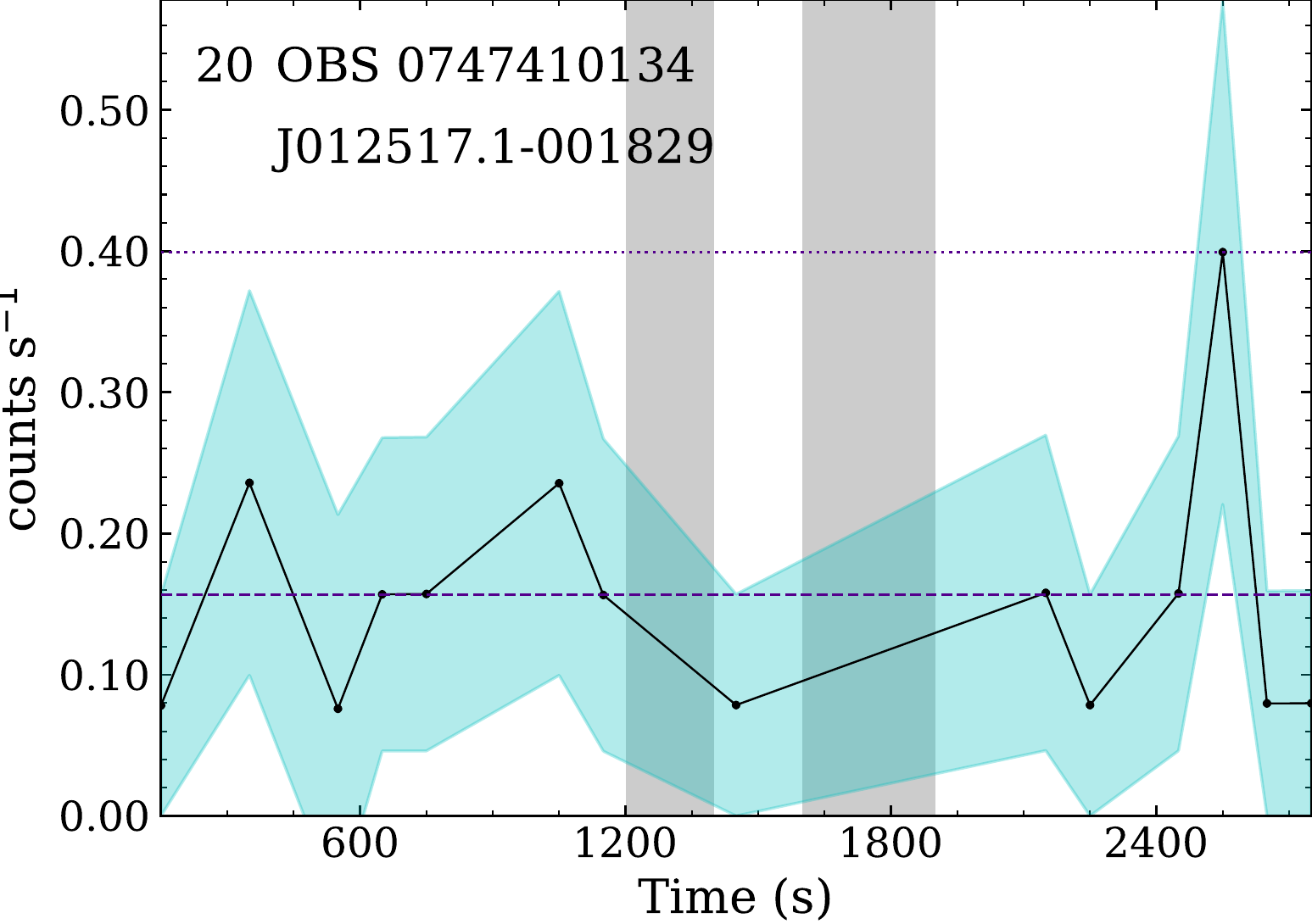}
	\includegraphics[width=0.325\textwidth]{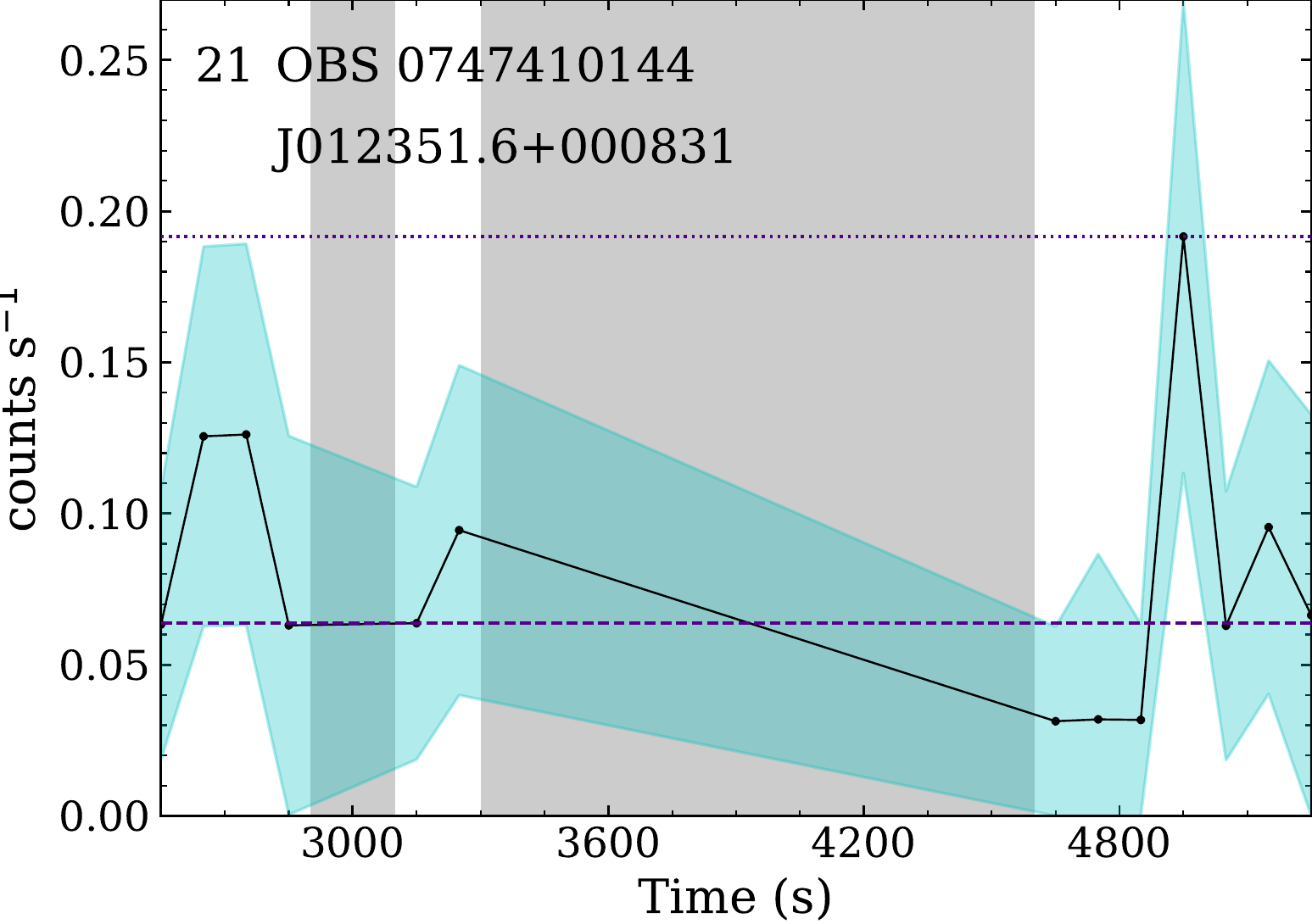}
	\includegraphics[width=0.325\textwidth]{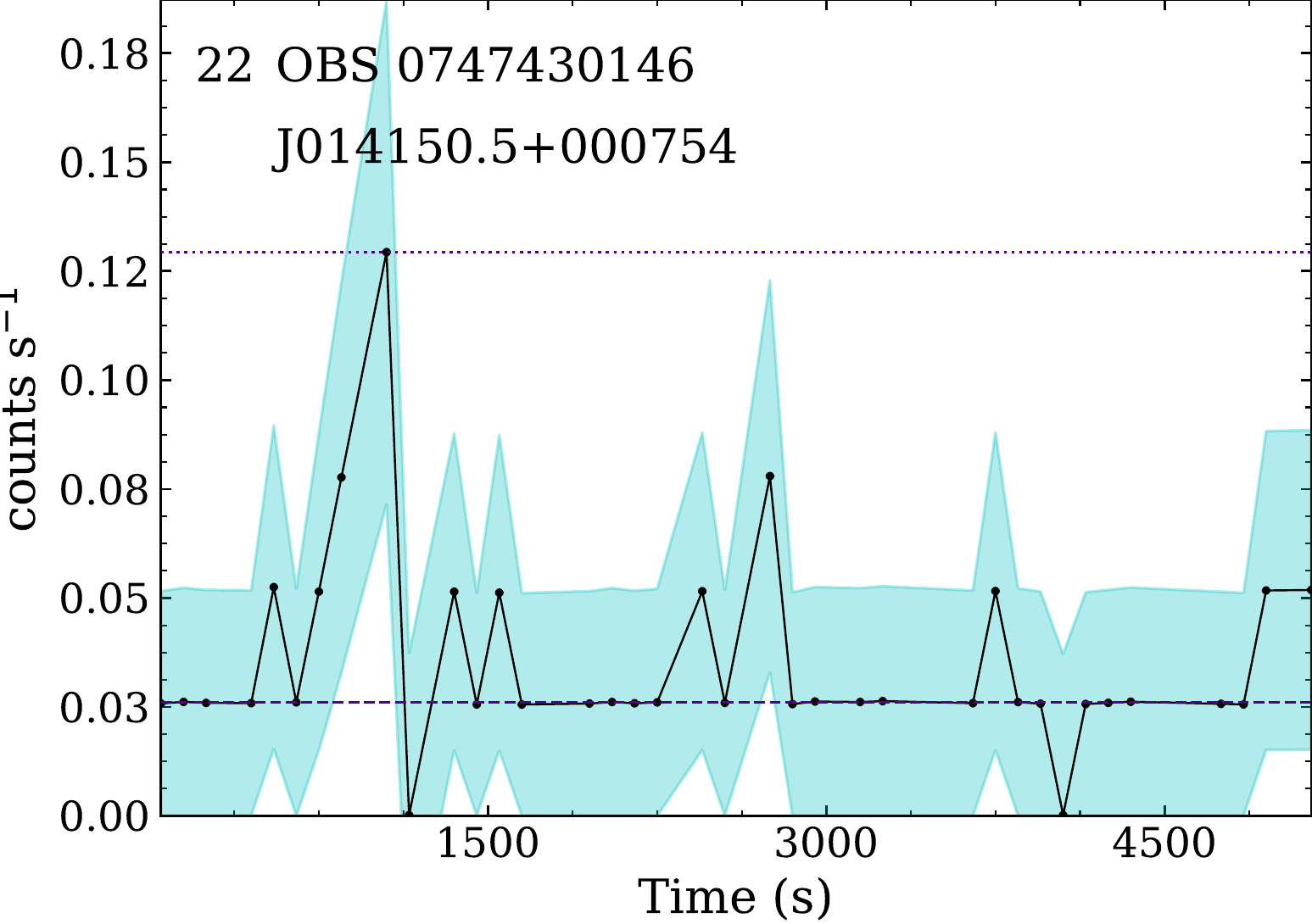}
	\includegraphics[width=0.325\textwidth]{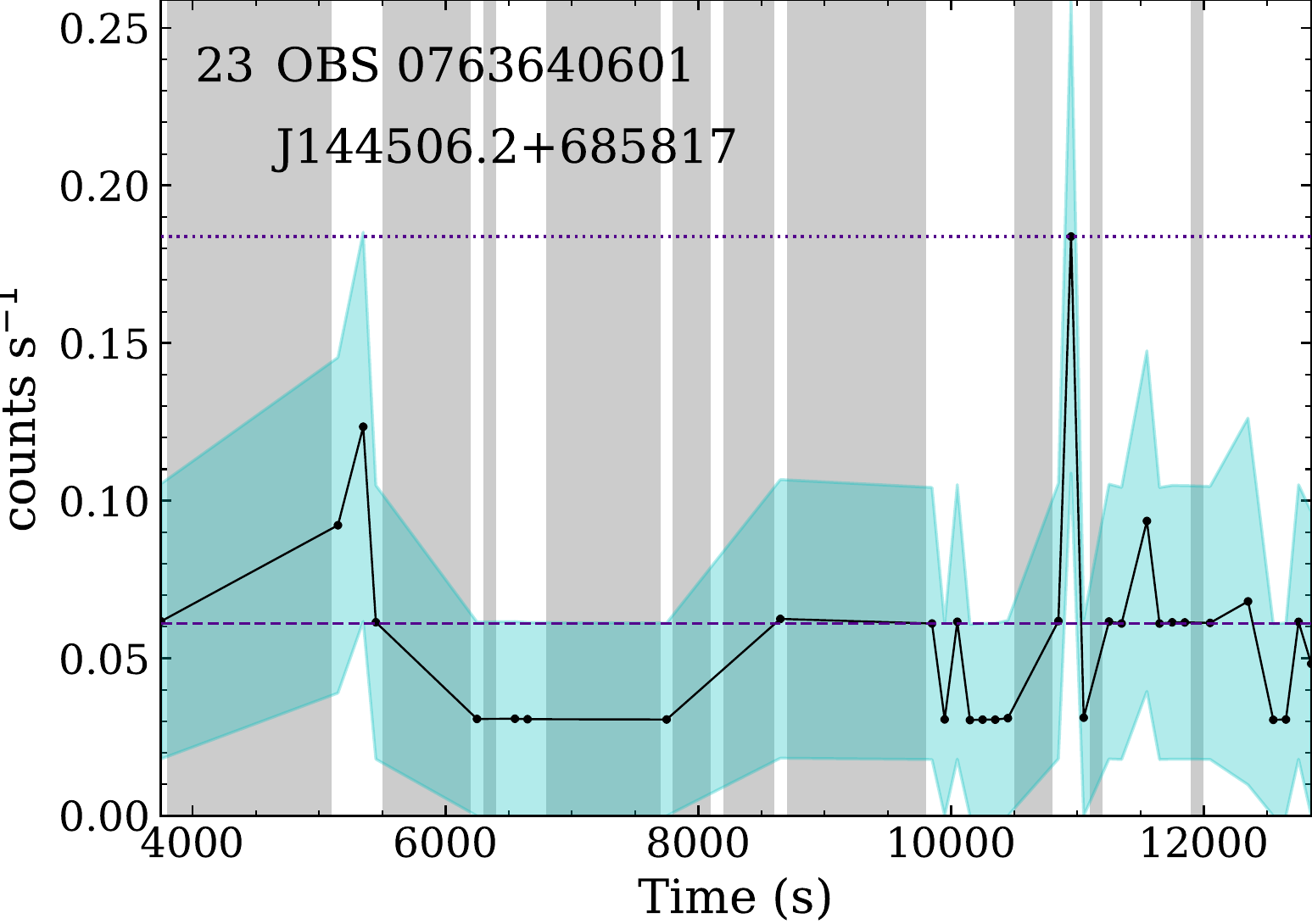}
	\includegraphics[width=0.325\textwidth]{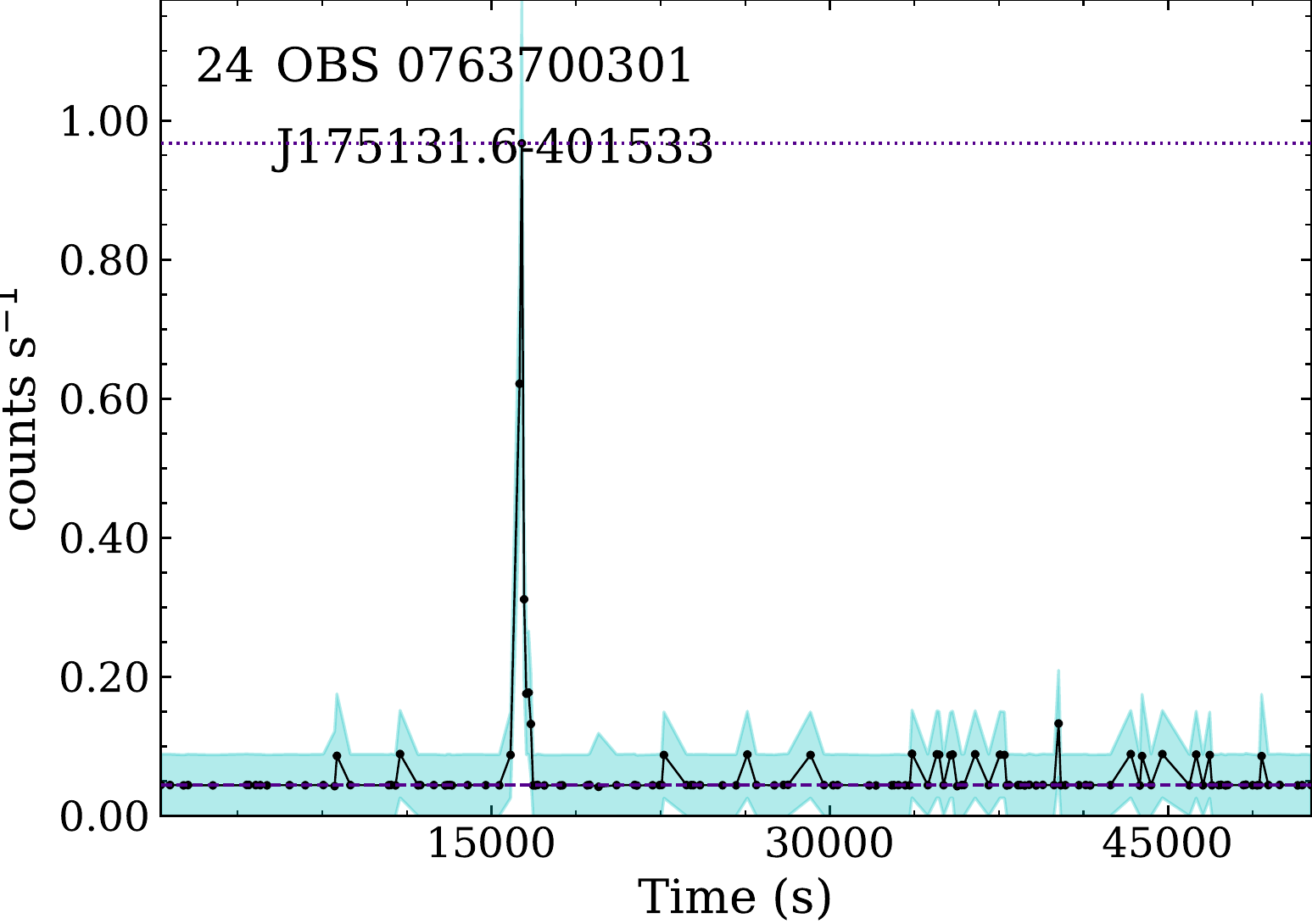}
	\includegraphics[width=0.325\textwidth]{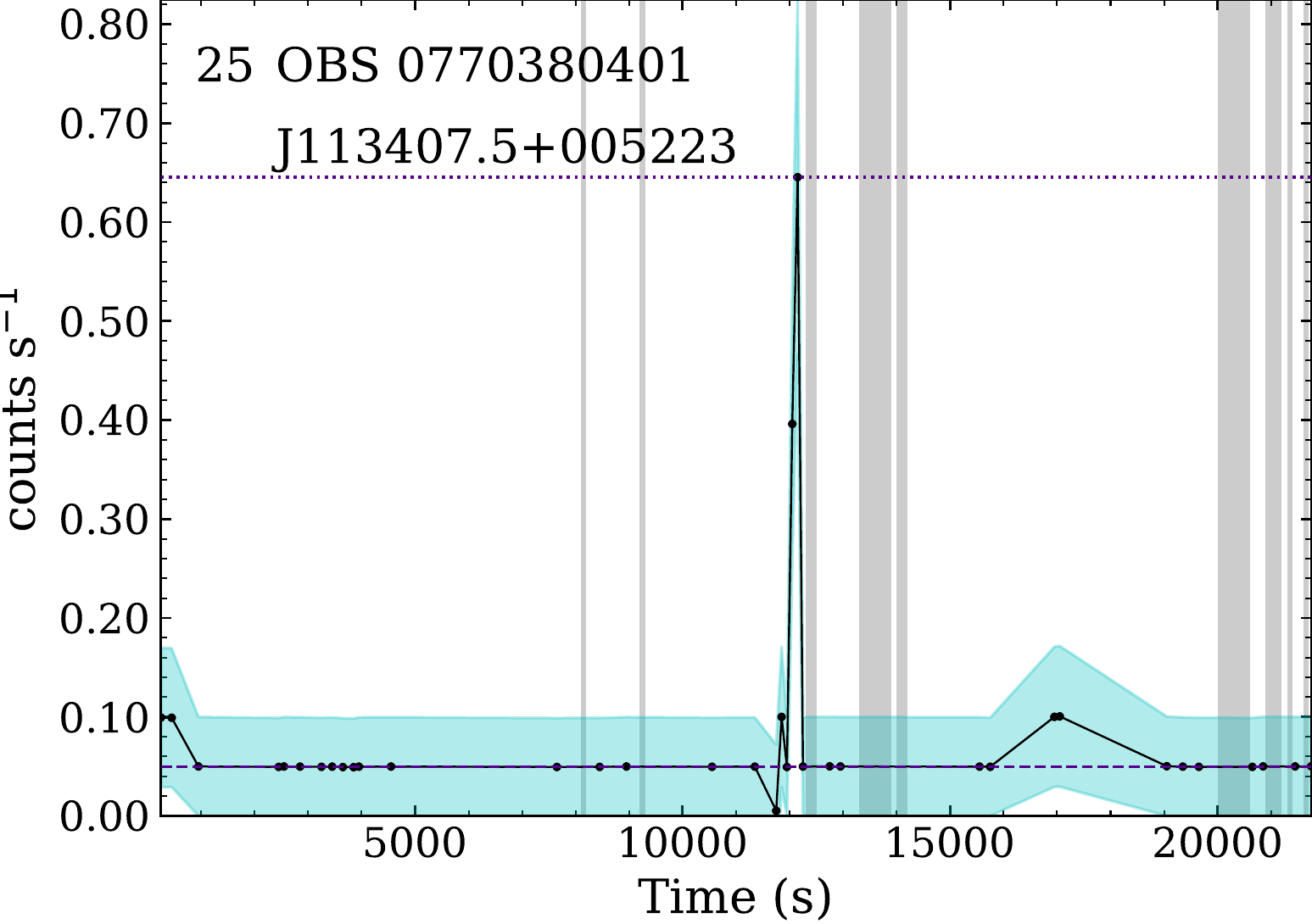}
	\includegraphics[width=0.325\textwidth]{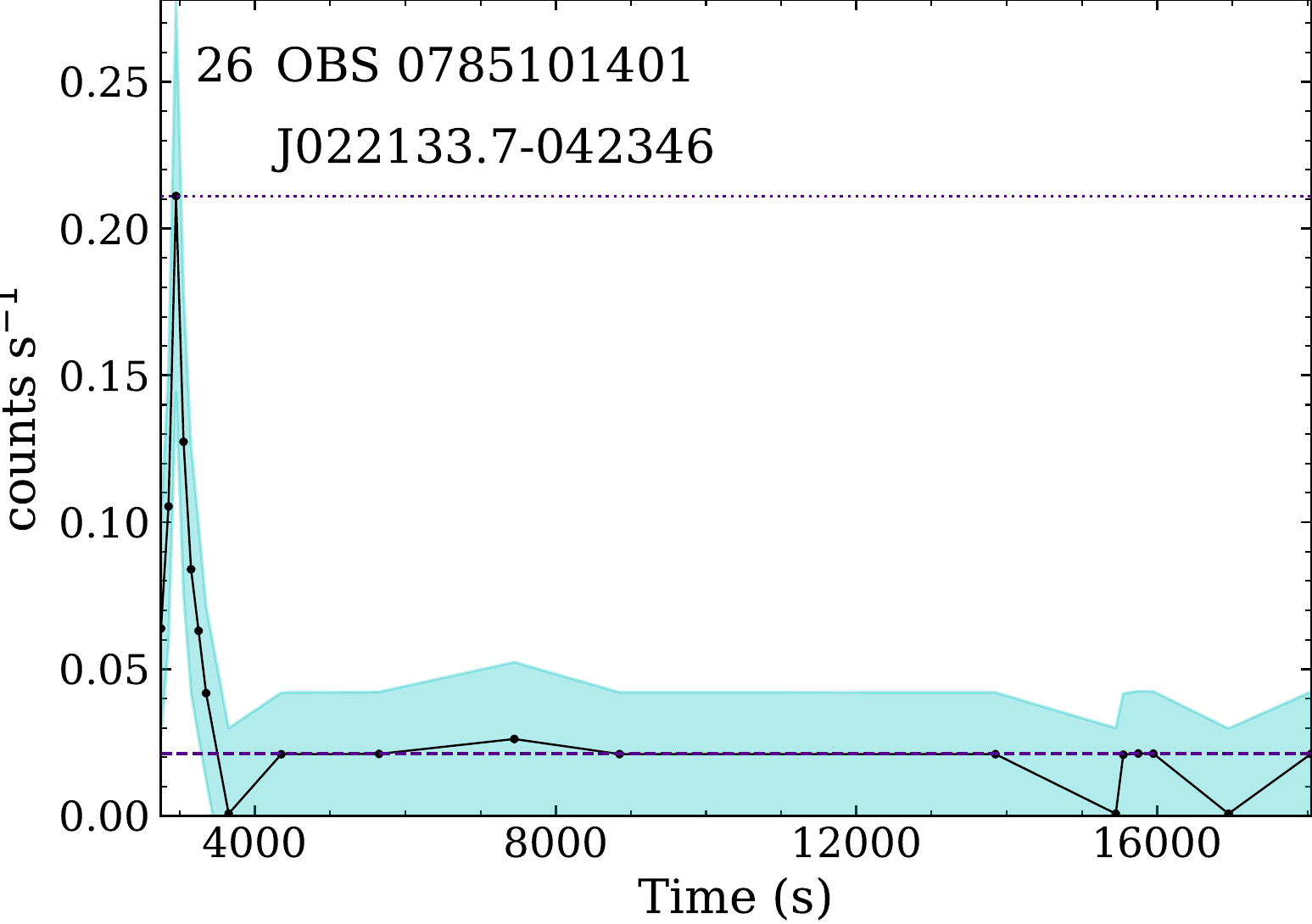}
	\caption{Same as Fig.~\ref{fig:outbursts_1}}
	\label{fig:outbursts_2}
\end{figure*}

\begin{figure*}
	\centering
	\includegraphics[width=0.325\textwidth]{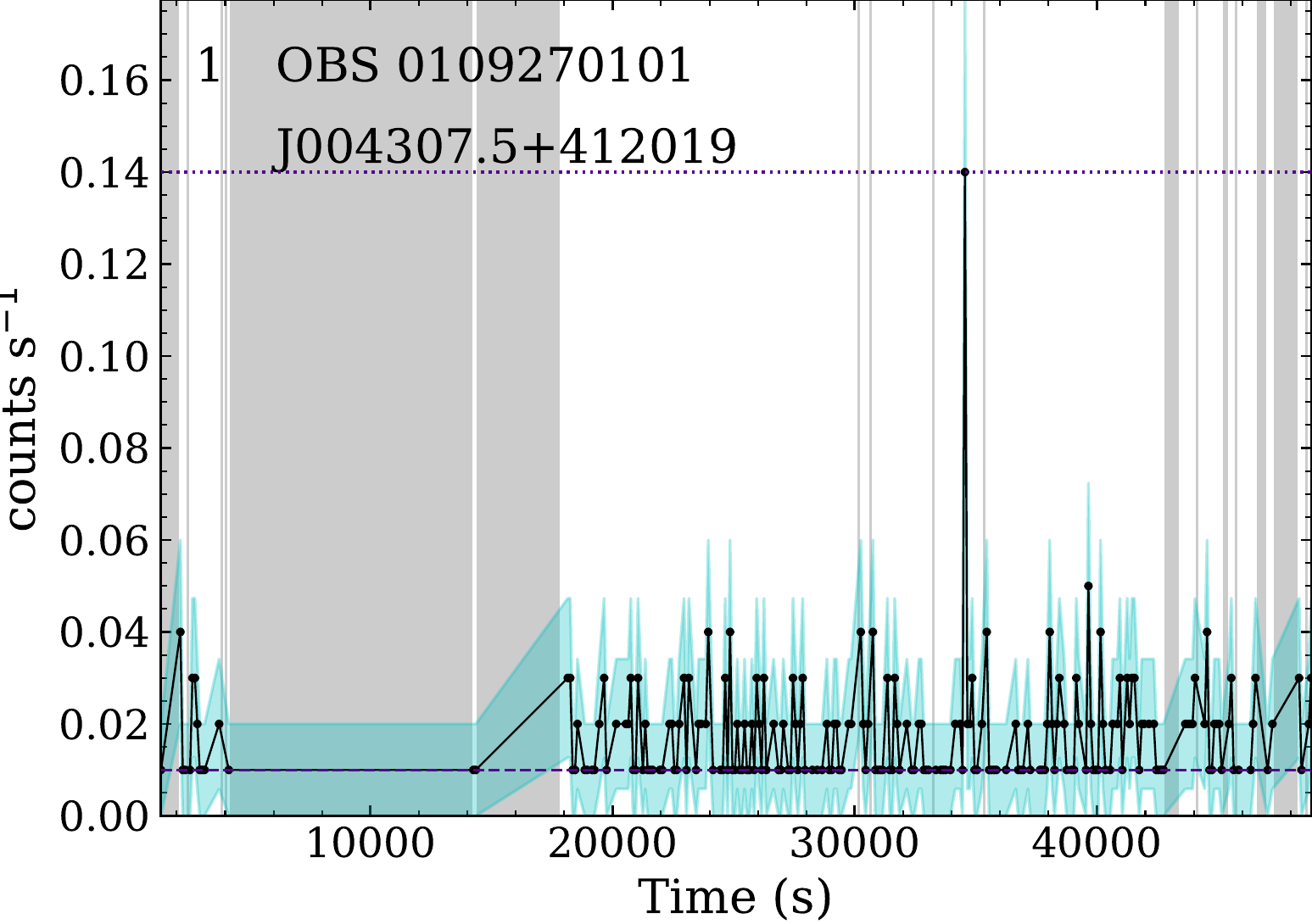}
	\includegraphics[width=0.325\textwidth]{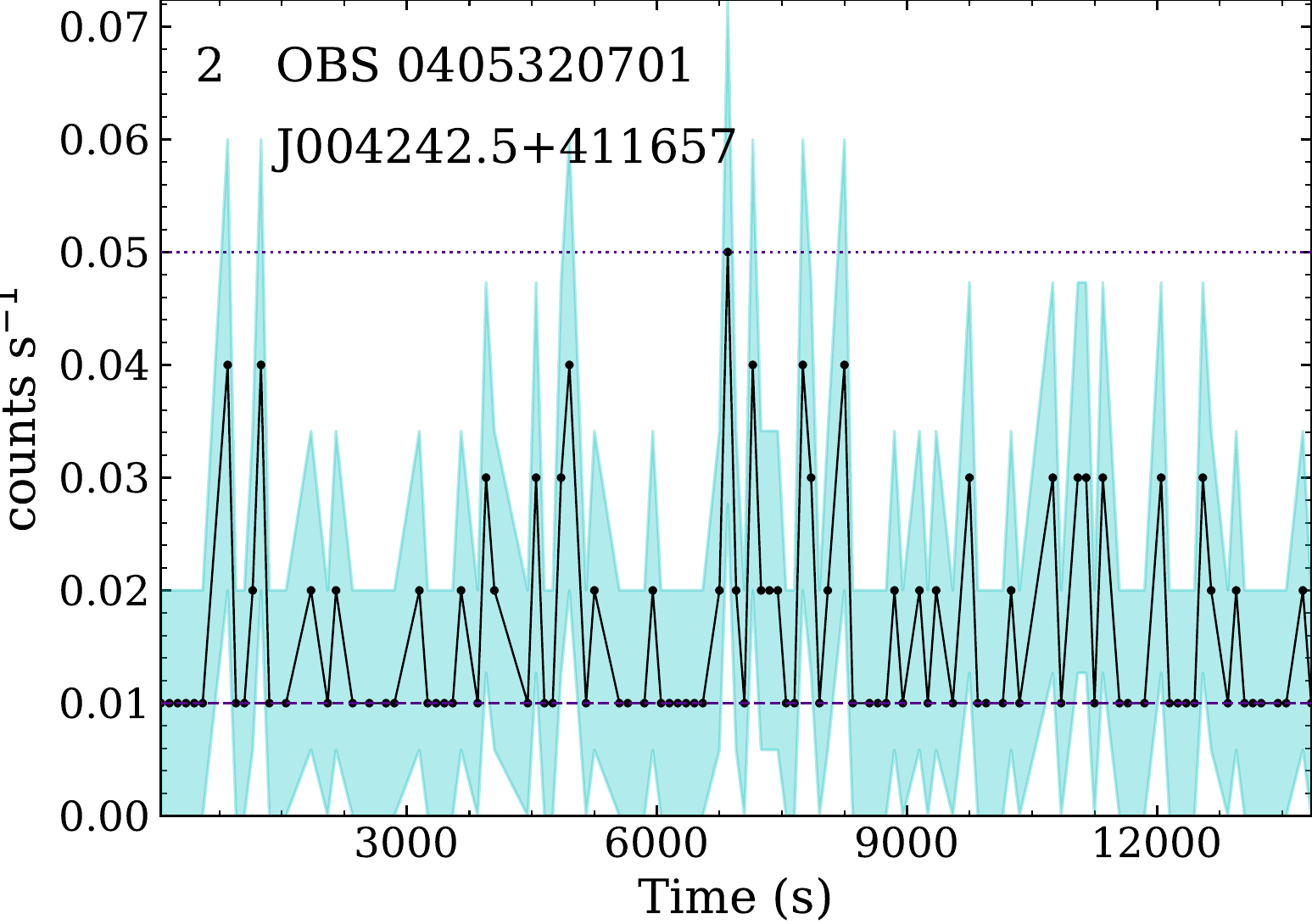}
	\includegraphics[width=0.325\textwidth]{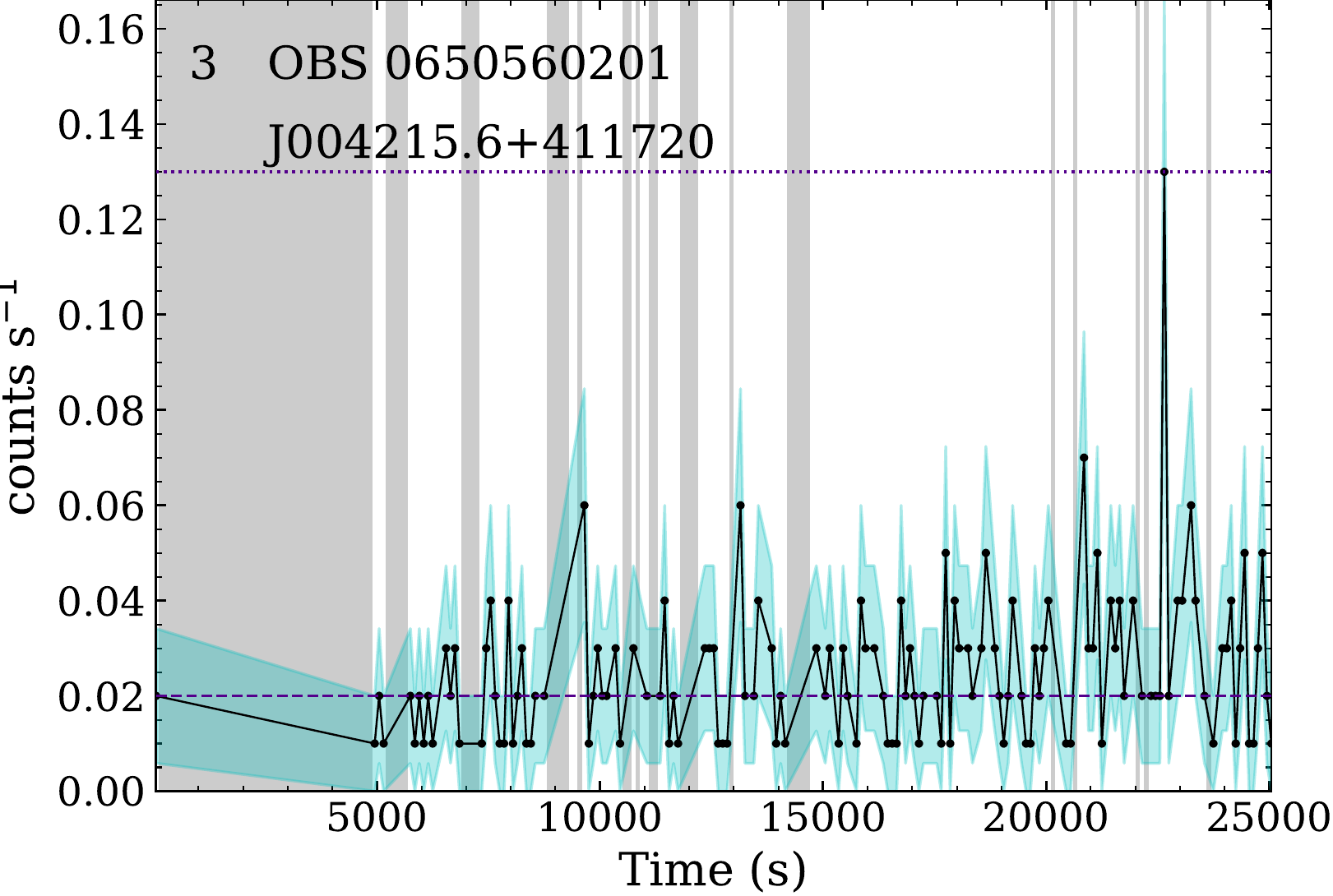}
	\includegraphics[width=0.325\textwidth]{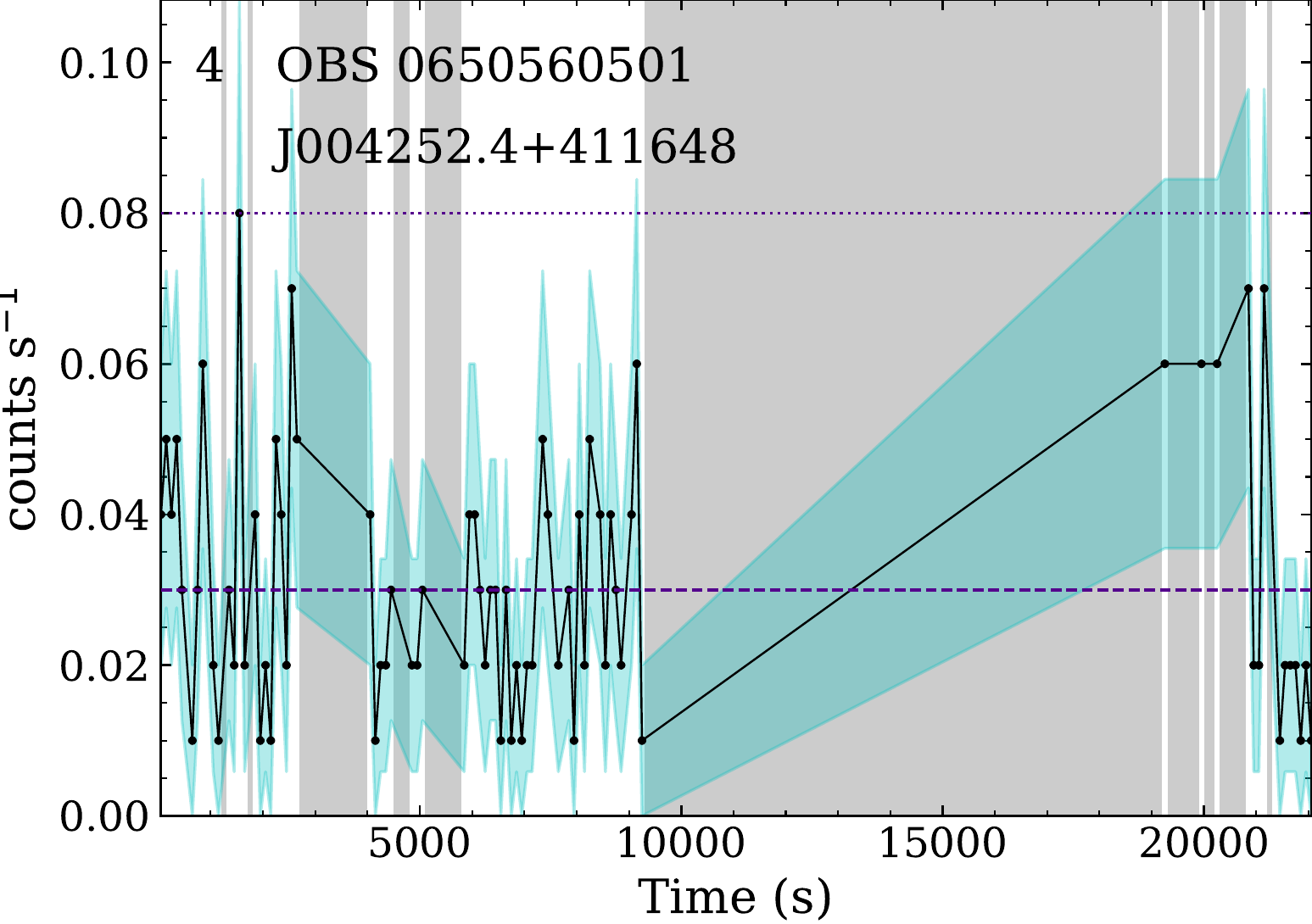}
	\includegraphics[width=0.325\textwidth]{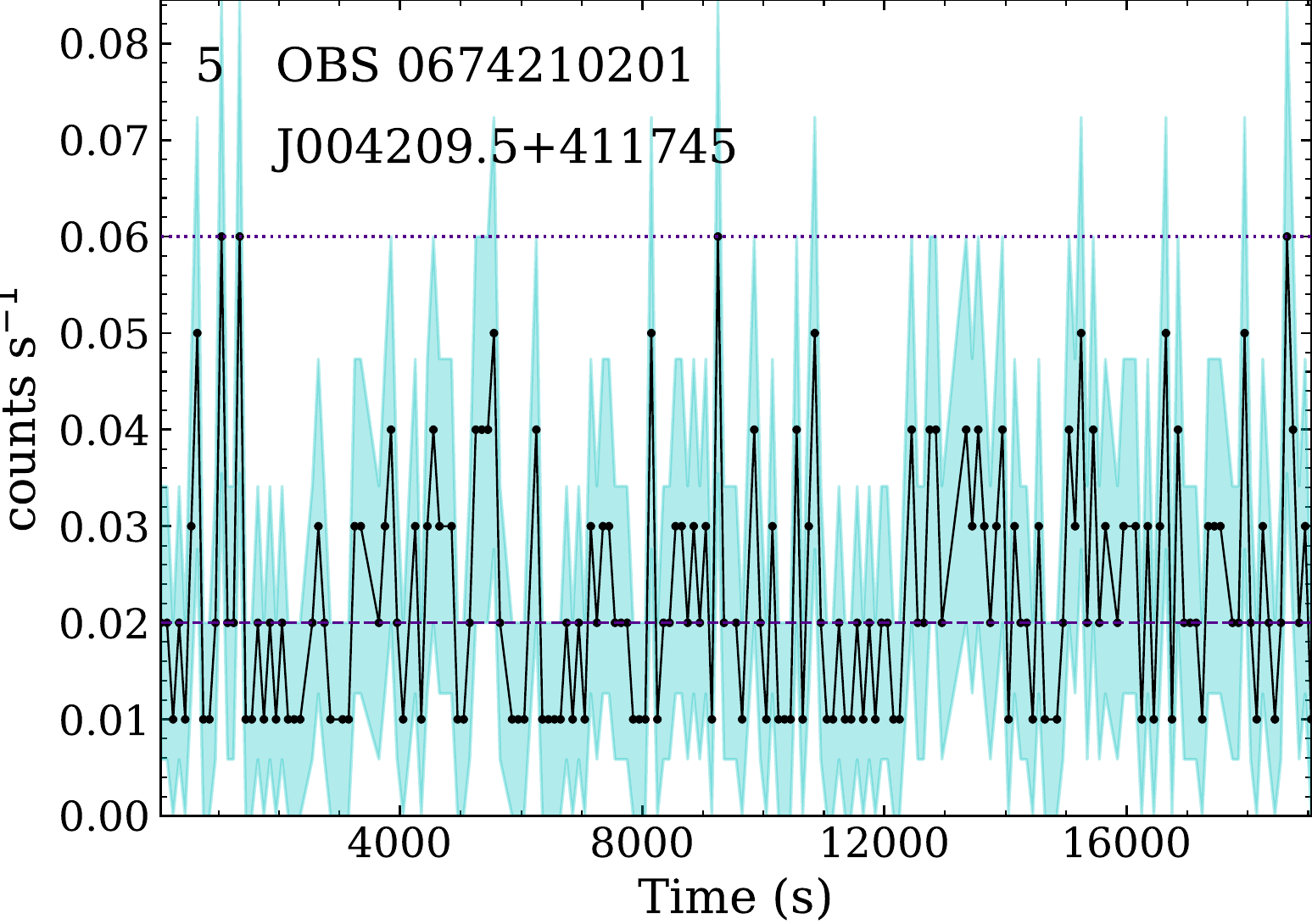}
	\includegraphics[width=0.325\textwidth]{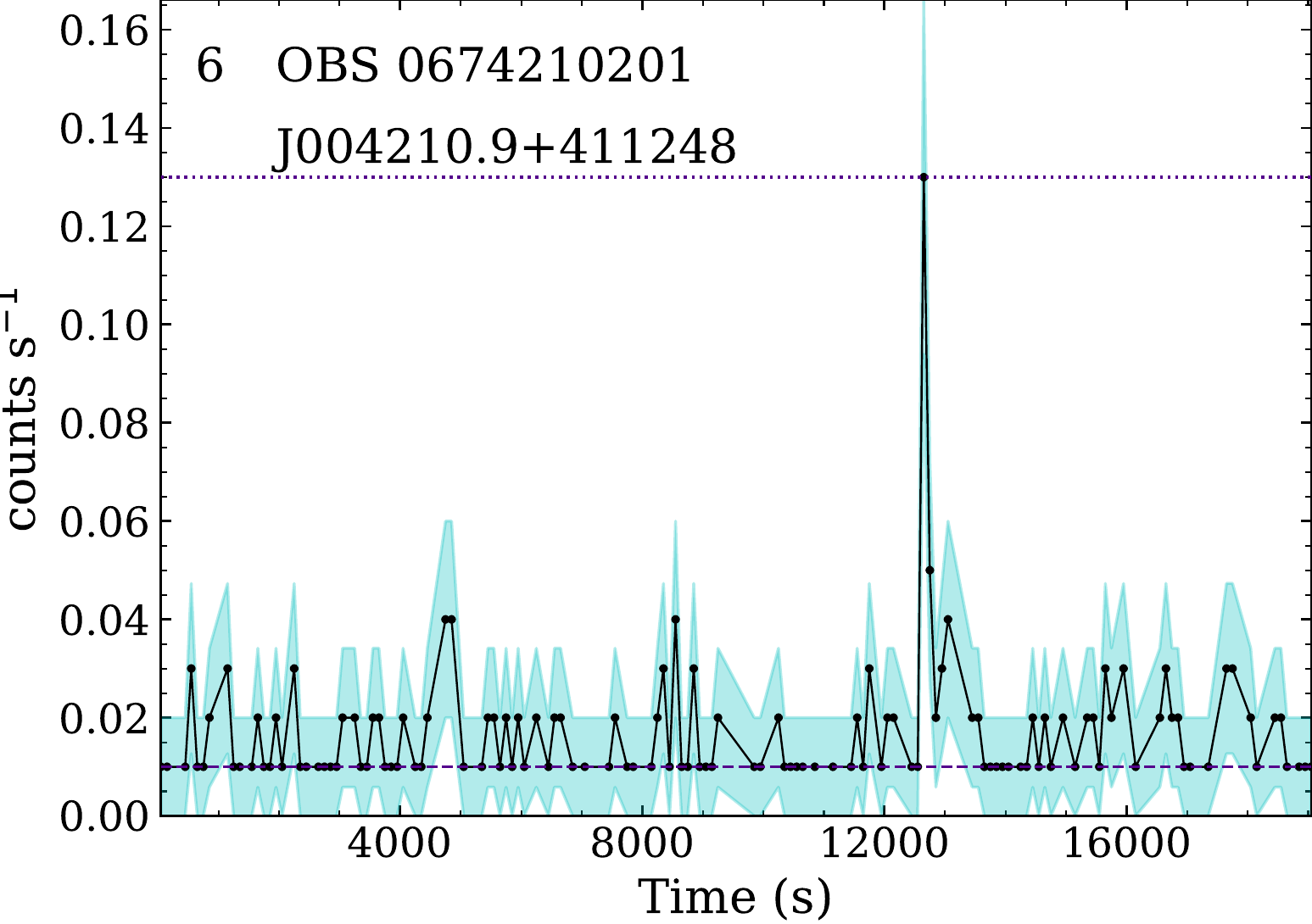}
	\includegraphics[width=0.325\textwidth]{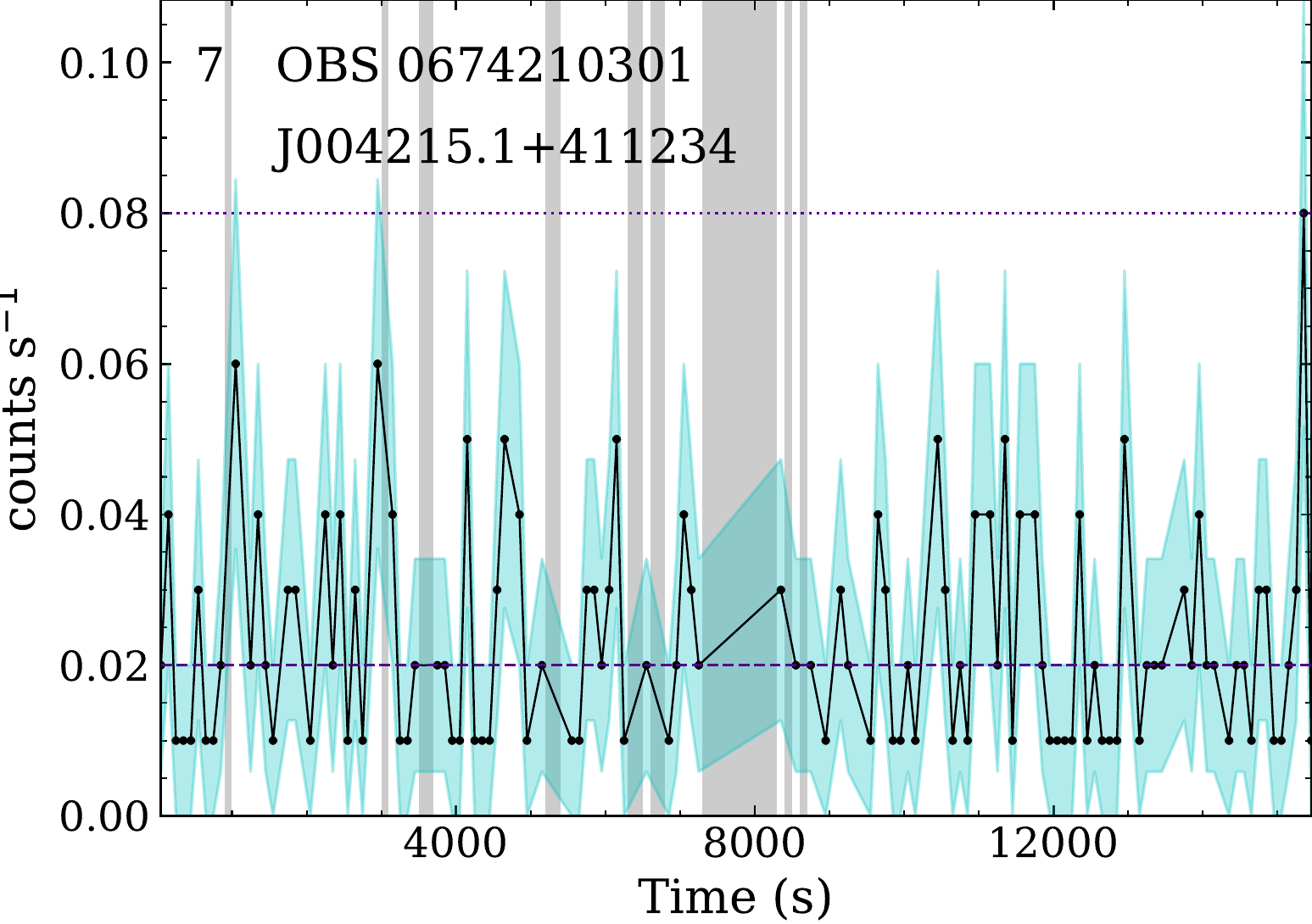}
	\includegraphics[width=0.325\textwidth]{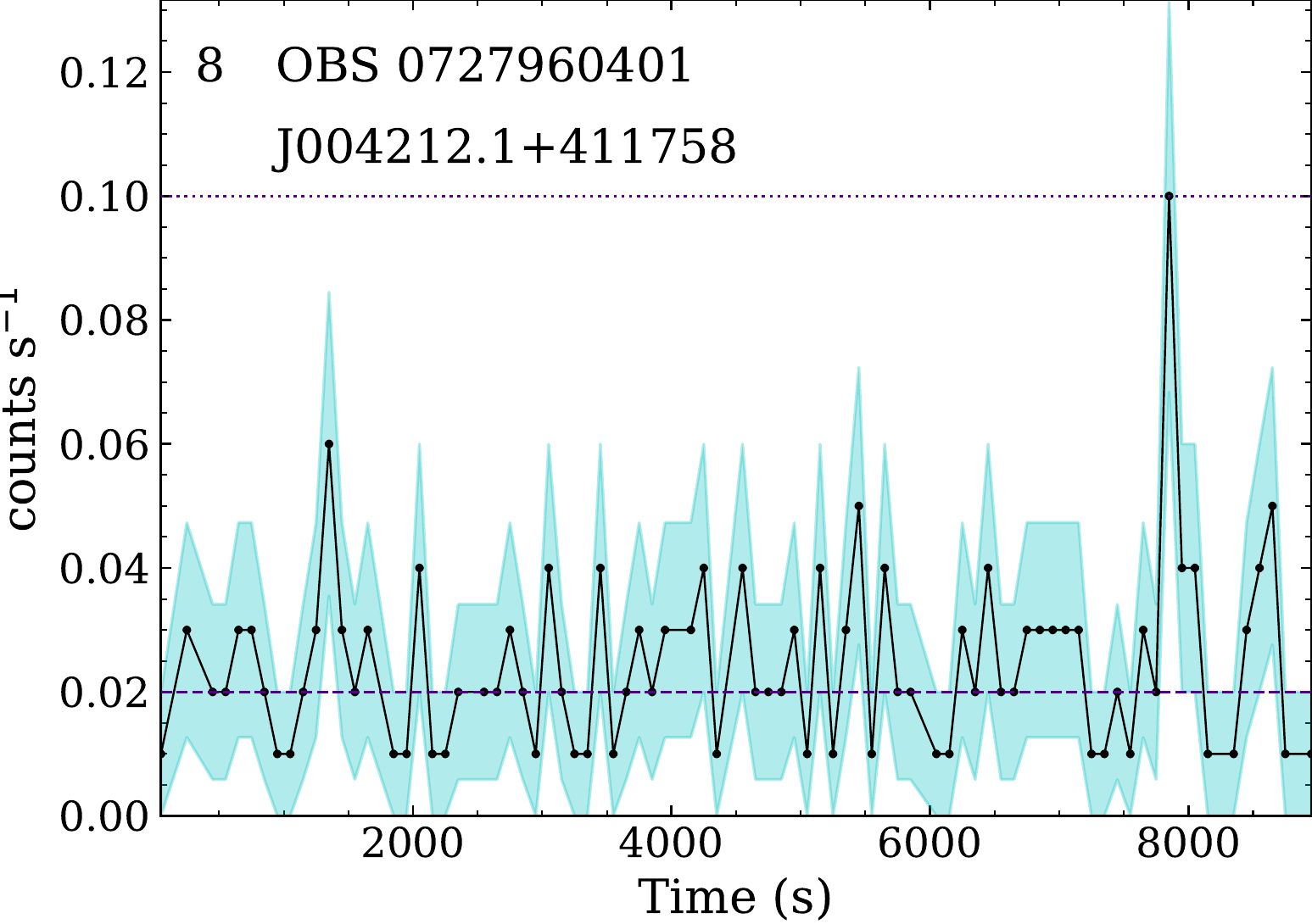}
	\includegraphics[width=0.325\textwidth]{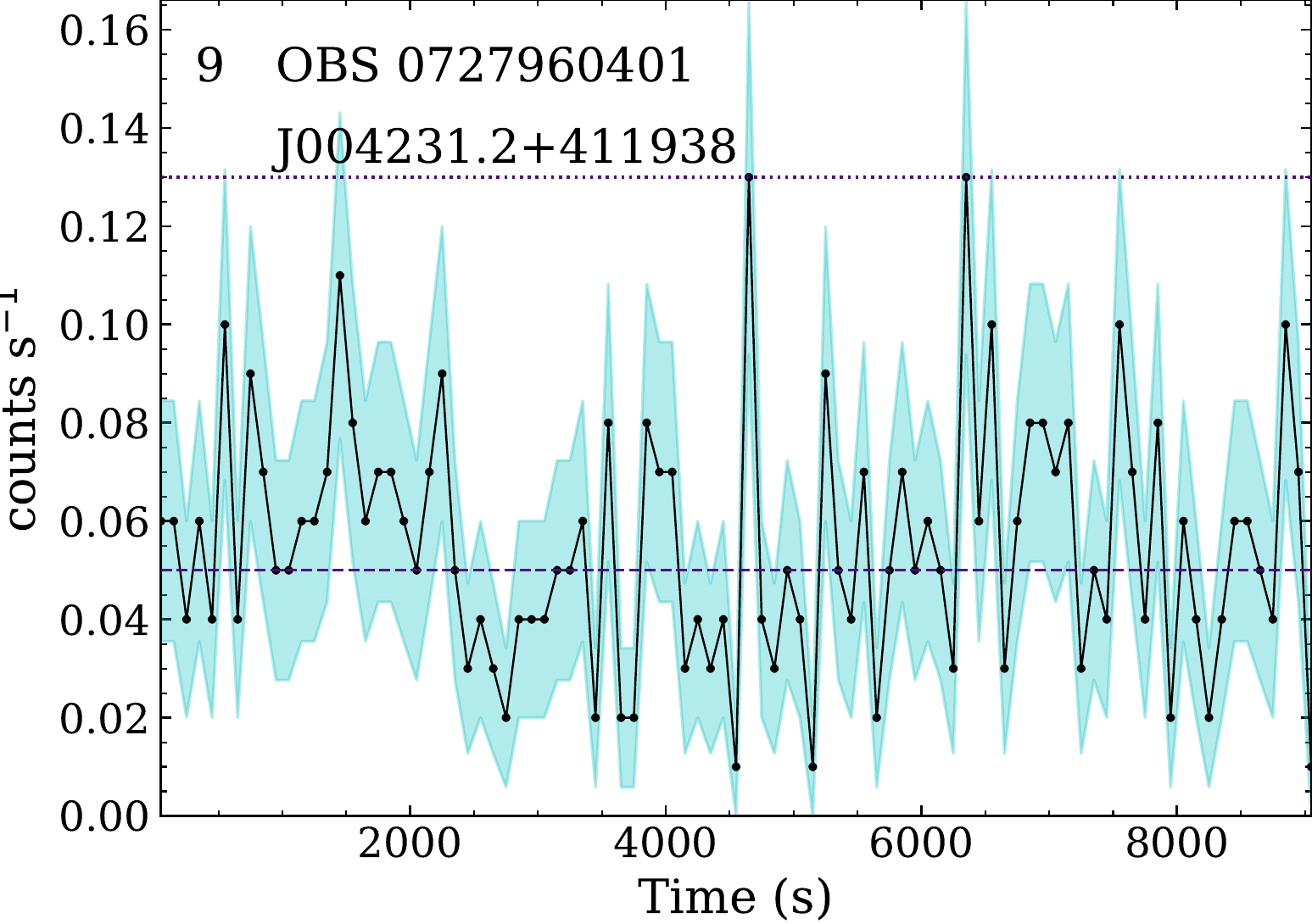}
	\caption{Same as Fig.~\ref{fig:outbursts_1}, but with sources found in M31.}
	\label{fig:outbursts_M31}
\end{figure*}

Since no completely new sources were found among the \textit{unknown source} class (Sect.~\ref{sub:crosscor}), we analysed the
\textit{no identification} sources in more detail.
We collected the sources not flagged as variable by \XMM's pipeline and visually inspected these. We removed the sources misidentified in the automatic procedure or affected by high background rates. Since the outburst selection was based on a visual inspection, we checked the P($\chi^2$) and P(KS) of the pn light curves. When these probabilities of constancy were $>10^{-4}$, we also computed them for the MOS1 and MOS2 light curves, and only kept the sources for which the P($\chi^2$) and P(KS) for pn, MOS1 or MOS2 were $<10^{-4}$. We ended up with a subset of 26 sources, whose light curves are presented in Fig.~\ref{fig:outbursts_1} and Fig.~\ref{fig:outbursts_2}.
Among the M31 observations for which we manually selected the light curve extraction regions, we found a further 9 sources previously detected but not flagged as variable, plotted in Fig.~\ref{fig:outbursts_M31}. The properties of these sources are summarized in Table~\ref{tab:new_src}.

All the new transients were detected with $TW = 100$\,s. Some of these transients were also detected by shorter $TW$, but no new sources were detected only with shorter time windows. If the duration of the transient is $\lesssim 100$\,s, most of the photons will fall within the same time bin if $TW = 100$\,s, increasing the signal to noise ratio. With a lower $TW$, the number of photons per time bin is lower, and thus has lower chances to reach the detection level. We found some interesting candidates that had been detected with $TW=3$\,s, but a closer look at these sources showed that they were damaged pixels. In particular, the pixel RAWX=1, RAWY=72, CCD=8 and the surrounding pixels were detected in $\sim20$ observations as what looked like very short, bright outbursts. In some cases, these had been included in 3XMM-DR8 as real detections.
The problem with the damaged pixels being detected as sources is addressed in 4XMM-DR9 (Webb et al., submitted), where there will be a flag to indicate that the source is likely to be spurious if it falls on or close to one of these pixels.
Most of the sources detected with EXOD that turned out to have a probability of constancy $>10^{-4}$ have in common that the net exposure time is less than 5000\,s, that as mentioned in Section~\ref{sec:variability}, cannot be ruled out as being due to statistical fluctuations.

While discussing the totality of the newly identified variable sources is beyond the scope of the paper, we analyse in more detail some particularly interesting candidates that show clear outbursts, some of them surprisingly undetected by \XMM's pipeline. In some cases this was done through the identification of a counterpart. In other cases, we extracted the spectra of the source and carried out spectral fitting using \textsc{Xspec} version 12.9.1p. We note however, that the low number of counts preclude us from performing a detailed spectral analysis, thus we use simple models like \texttt{powerlaw} and \texttt{bbody} to characterise the emission. We compute fluxes using the pseudo model \texttt{cflux} in \textsc{Xspec}. For the four transients found in M31 showing outbursts, we refine our analysis by extracting the spectra separately for the burst and the persistent emission, by visually selecting the times for each period. We fit each spectra separately and compute the persistent luminosity L$_{\text{pers}}$, peak luminosity L$_{\text{peak}}$ and luminosity ratio  (L$_{\text{peak}}$/L$_{\text{pers}}$) to study the properties of the emission in each case, in order to put constraints on their nature. Again, due to the low number of counts, we rely on \textsc{WebPIMMS}\footnote{\textsc{WebPIMMS}: \url{https://heasarc.gsfc.nasa.gov/cgi-bin/Tools/w3pimms/w3pimms.pl}} to compute this luminosity ratio. L$_{\text{pers}}$ is estimated using the model of the persistent emission and the median count rate of the light curve while L$_{\text{peak}}$ is computed with the burst model and the peak count rate of the light curve. 
Since the distance to M31 is known, we assume that the sources are located at the same distance, 0.78\,Mpc, to obtain the luminosity. 
Below we give a detailed analysis and discuss the nature of the selected sources. The detailed results and the plots of the spectral fitting can be found in Appendix~\ref{app:fit_spec}.\\

\Source{J173046.7+521846}{0021750201}{1}
This source, 2XMM J173046.8+521847 in \textsc{Simbad}, is classified as an Active Galactic Nucleus (AGN) candidate \citep{lin_classification_2012}. It shows an aperiodic variability within the duration of the observation ($\sim3500$ s). Long term X-ray variability (months/years) is common in AGNs, but short term variability (hours/days) has only been observed in Seyfert types 1.8 and 1.9 (e.g. \citealt{hernandez-garcia_x-ray_2017} and references therein). We thus propose that this source belongs to one of those classes.\\

\Source{J083941.3+192901}{0101440401}{4}
Although the automatic \textsc{Simbad} query classified this source in the \textit{no identification} group, we later found that the source is classified as a star in a cluster with a manual query of the \textsc{Simbad} database \citep{hambly_very_1995}.
This source, Cl* NGC 2632 HSHJ 283 in \textsc{Simbad}, shows an outburst lasting $\sim3000$\,s, with a main burst lasting $\sim500$\,s. The flare was not detected by the \XMM\ pipeline. Since we have not checked the totality of the source associations, we suspect there might be additional misidentifications like this one amongst the detected sources.\\

\Source{J015709.1+373739}{0149780101}{6}
We detected this $\sim 800$\,s long outburst in the direction of the open cluster NGC\,752, located at a distance of $\sim$\,430\,pc \citep{daniel_photometric_1994,giardino_x-ray_2008}. We identify a potential blue counterpart at a distance of 0.92$\arcsec$ with a magnitude of 22.83 in the Bj photographic band \citep[NBXA027004,][]{lasker_second-generation_2008}.
The counterpart is point like; if it is a star in this open cluster, the X-ray transient could be a stellar flare
that would correspond to a luminosity of $1.6\times10^{30}$\ergs. \\
While our paper was under review, \cite{alp_blasts_2020} also reported a discovery of this source, named XT\,030206 there. 
The authors identify the counterpart as a starburst galaxy and infer a photometric redshift of $z=1.17$ by fitting the SED of the optical spectrum. 
They conclude that the most likely explanation for this X-ray transient is a supernova shock breakout (SBO).
As the SED fitting makes the analysis of \cite{alp_blasts_2020} more thorough we defer to their interpretation of the transient as an SBO.
A new X-ray outburst of the source in the future would strengthen the stellar flare case. 
\\

\Source{J174610.8$-$290021}{0202670701}{8}
Although this observation is highly contaminated by soft proton flares, we detected a variable source with an outburst lasting ${\sim100}$\,s. Surprisingly, the light curve of this source was generated by \XMM's pipeline, and the burst is visible\footnote{Automatically generated light curve of Source 8: \url{http://xmm-catalog.irap.omp.eu/detection/102026707010084}}. However, the source was not classified as variable by the $\chi^2$ test, probably due to the short duration of the outburst.
This source, \mbox{CXOU J174610.8-290019} in \textsc{Simbad}, is classified as an \textit{X-ray source}. It is a perfect example of why EXOD is better adapted to detecting short transients than other variability tests.

This source is highly variable.  Its flux decreased by more than a factor 10 between 2000 and 2004 and then was never detected again by Swift or \XMM, even though these satellites returned to  the field on numerous occasions as the source is towards the Galactic center. It has a very absorbed spectrum, typical of sources in the Galactic center, $n_{\text{H}}\sim1.6\times 10^{23}$\,cm$^{-2}$, indicating a distance of $\sim8$\,kpc. 

The luminosity in the burst, assuming a distance of 8\,kpc, reaches $\sim 10^{34}$\ergs. Its non-flare spectrum shows indications of the presence of an emission line at $\sim6.66$\,keV, that we identify as an iron line or a cyclotron resonance line 
(See Appendix~\ref{app:fit_spec}).
The number of counts is too low to draw firm conclusions on the nature of this source, but it is most likely that it is an accreting neutron star in an X-ray binary. Whether it is an LMXB or an HMXB is not clear, since the search for an optical counterpart is complicated due to its location in a crowded region and the addition of the high extinction from the observed photoelectric absorption.
A 100\,s flare would be uncommon amongst HMXBs, but not unexpected due to the high variability that these sources present \citep[see e.g.][and references therein]{chaty_nature_2011}. A type I X-ray burst from an NS-LMXB, given the timing and spectral properties of the flare, could be an explanation.
\\

\Source{J183658.4$-$072119}{0606420101}{13}
The outburst of this source lasts $\sim 3000$\,s. 
It is located in a crowded region in the galactic plane and presents numerous optical/NIR counterparts within a $6\arcsec$ radius. We thus conclude that this source is most likely a star.
\\

\Source{J081907.9$-$384302}{0655650201}{15}
This source shows an outburst that lasts $\sim 300$\,s, but it is preceded and followed by higher flux periods before going back to its quiescent state. Including these periods, the outburst lasts $\sim 2500$\,s. This type of variability is also observed in stellar flares.
\\
	
\Source{J175131.6$-$401533}{0763700301}{24}
This outburst, lasting 1000\,s, is classified as a variable source in 3XMM-DR8, but it has no \textsc{Simbad} object associated. The source is however in a crowded region of the Galactic plane, indicating that this is most likely a star.\\

\Source{J113407.5+005223}{0770380401}{25}
This puzzling source has a burst that lasts only $\sim$ 200\,s, expected for type I X-ray burst, with some structure appearing in the burst. The burst spectrum appears to be fairly soft, although this is based on the joint fit of the persistent and burst emission due to the low number of counts.\\
\citet{alp_blasts_2020}  presented also this source 
(under name XT\,151219) 
as an SBO while our paper was under review. 
They associate the transient with a host galaxy located at redshift $z=0.62$. Although the presence of four galaxies within a 15\arcsec\  radius from the position of the X-ray source\footnote{NED query: \href{https://ned.ipac.caltech.edu/conesearch?search_type=Near\%20Position\%20Search&coordinates=11h34m07.5s\%20\%2B00d52m23s&radius=0.25&in_csys=Equatorial&in_equinox=J2000&out_csys=Equatorial&out_equinox=Same\%20as\%20Input&hconst=67.8&omegam=0.308&omegav=0.692&wmap=4&corr_z=1&iau_style=liberal&in_csys_IAU=Equatorial&in_equinox_IAU=B1950&z_constraint=Unconstrained&z_unit=z&ot_include=ANY&nmp_op=ANY&out_csys_nearname=Equatorial&out_equinox_nearname=J2000&obj_sort=Distance\%20to\%20search\%20center}{\nolinkurl{https://ned.ipac.caltech.edu/}}} casts some doubts on the identification of the host galaxy, \citet{alp_blasts_2020}  find the flux and spectral properties of the source to be in good agreement with blue supergiant SBO predictions.\\
In general, we too argue that this is likely an extragalactic source because it is located well outside the galactic plane and there is no clear bright counterpart other than background galaxies.
\\

\Source{J022133.7$-$042346}{0785101401}{26}
This source shows a $\sim 1000$\,s outburst, with a linear rise and an exponential decay. The source has a red-NIR counterpart at a distance of $\sim125$\,pc 
\citep{gaia_collaboration_gaia_2016,gaia_collaboration_gaia_2018,bailer-jones_estimating_2018}, indicating that it is a star.
\\

\Source{J004307.5+412019}{0109270101}{M31-1}
This outburst lasts $\sim 400$\,s. The source, [ZGV2011] 23 in \textsc{Simbad}, had been previously classified as a low-mass X-ray binary \citep{zhang_luminosity_2011}, but the nature of the accretor was unknown.
We extracted the spectrum of the burst and the persistent emission separately, and fitted each one with an absorbed black body (tbabs*bbody in \textsc{Xspec}) and with an absorbed power law (tbabs*pow). The low number of photons does not allow for a clear preference for one of these models, although a power law gives a slightly better fit for both burst and persistent emission.

From persistent to burst emission, the resulting black body temperature increases from $\sim0.2$ to $\sim0.3$\,keV, whereas the power law index $\Gamma$ goes from $\sim3.3$ to $\sim1.9$ (See Appendix~\ref{app:fit_spec}), indicating an apparent spectral hardening during the burst.
From the power law indices and a count rate going from 0.01\ctss\ to 0.14\ctss, we computed a L$_{\text{peak}}$/L$_{\text{pers}}\sim30$.
A persistent flux of $\sim0.3\times10^{-13}$\fcgs\ scales to a luminosity of $\sim2\times10^{36}$\ergs\ at the distance of M31, 0.78\,Mpc. This corresponds to a burst luminosity of $\sim6\times10^{37}$\ergs, around 30\% of the Eddington luminosity for an NS ($\sim1.8\times10^{38}$\ergs).
The presence of such a burst and the luminosity rise indicate that this is likely a type I X-ray burst, and would in this case identify the accretor as an NS.
Such a source shows the power of EXOD, since it allows one to identify the nature of a compact object.
\\

\Source{J004215.6+411720}{0650560201}{M31-3}
This outburst lasts $\sim 300$\,s. The source is classified as an X-ray binary candidate \citep{lin_classification_2012}. 
The spectral fitting seems to indicate a hardening of the source during the burst, with $\Gamma$ going from $\sim1.6$ in the persistent emission to $\sim0.7$ during the burst, although still consistent within the error bars. These values are nevertheless hard for type I X-ray bursts, making it difficult to identify the nature of the compact object in the system. 
By using the aforementioned power law indices and the count rates from the persistent (0.02\ctss) and burst (0.13\ctss) emission, we calculate L$_{\text{peak}}$/L$_{\text{pers}}\sim16$.
The presence of such a burst makes the type I X-ray burster the most likely scenario, although with the present data we cannot robustly confirm its nature.
\\

\Source{J004210.9+411248}{0674210201}{M31-6}
The outburst lasts $\sim 200$\,s, and it is followed by a higher flux period before going back to the quiescent state. Including this, the outburst lasts $\sim 500$\,s. This type of outburst is consistent with a type I X-ray burst.
The source, XMMM31 J004211.0+411247 in \textsc{Simbad}, where it is catalogued as an \textit{X-ray source}, has been detected in 44 \XMM\ observations. It is so faint that its spectrum and time series have only been extracted in 3 out of 44 observations. The observation where we detected its variability was not included\footnote{Source M31-6: \url{http://xmm-catalog.irap.omp.eu/source/201125704010111}}, and the source is thus not catalogued as variable.

We fitted the spectrum with an absorbed black body (tbabs*bbody) and an absorbed power law (tbabs*pow).
The fitted power law gives $\Gamma\sim2.4$ for the persistent emission and $\Gamma\sim1.1$ at peak. However, there are only 30 available counts during the flare and the source was additionally located next to a chip gap in the pn detector, reducing the number of available photons for the spectral fitting. We obtain L$_{\text{peak}}$/L$_{\text{pers}}\sim40$ for a count rate going from  0.01\ctss\ to 0.13\ctss. The flare spectrum as well as the duration of the burst are all comprised within the typical values of type I X-ray bursts, and it makes this source an NS-LMXB candidate.\\

\Source{J004212.1+411758}{0727960401}{M31-8}
This outburst lasts $\sim 500$\,s. 
Previously identified as an LMXB named [ZGV2011] 27 \citep{zhang_luminosity_2011}, it has been detected in 47 \XMM\ observations. In order to identify the nature of the accretor, we fitted its persistent and burst spectrum with an absorbed black body (tbabs*bbody) and with an absorbed power law (tbabs*pow). 
This source was also located near a chip gap in the pn detector, considerably reducing the number of photons used for the spectral fitting, with only 20 available photons during the burst, and thus not enough to detect spectral shape deviations between the persistent and burst emission.
However, the presence of a short burst reinforces the previous identification with an LMXB and makes an accreting NS the most likely explanation.
\\

\begin{table*}
\caption{New variable EXOD sources.}
\label{tab:new_src}
\centering
\begin{tabular}{r c c c c r c p{5.5cm} }
\hline\hline
	ID & Name & OBSID & RA & Dec  & \multicolumn{1}{c}{Burst} & Var. & Comments\\ 
	& & & (J2000) & (J2000) & \multicolumn{1}{c}{dur. (s)} & & \\
\hline
	1 & J173046.7+521846 & 0021750201 & 17:30:46.7 & +52:18:46 &  & PN & AGN candidate.\\ 
	2 & J160502.1+430401 & 0025740101 & 16:05:02.1 & +43:04:01 & & PN & Short observation.\\ 
	3 & J191515.1+044348 & 0075140501 & 19:15:15.1 & +04:43:48 & & PN & No \textsc{Simbad} object associated.\\ 
	4 & J083941.3+192901 & 0101440401 & 08:39:41.3 & +19:29:01 & 3000 & PN & Star in cluster.\\ 
	5 & J182929.4$-$092530 & 0135744801 & 18:29:29.4 & $-$09:25:30 & & PN & Stellar flare. Off-axis detection.\\ 
	6 & J015709.1+373739 & 0149780101 & 01:57:09.1 & +37:37:39 & 800 & PN & Stellar flare.\\ 
	7 & J070509.8$-$112940 & 0201390201 & 07:05:09.8 & $-$11:29:40 & & PN & Short observation. Star.\\ 
	8 & J174610.8$-$290021 & 0202670701 & 17:46:10.8 & $-$29:00:21 & 100 & PN & Type I X-ray burst candidate.\\ 
	9 & J010909.2+132337 & 0203280301 & 01:09:09.2 & +13:23:37 & & PN & AGN candidate.\\ 
	10 & J015727.2$-$004041 & 0303110101 & 01:57:27.2 & $-$00:40:41 & 200 & PN & Short observation, off-axis detection.\\ 
	11 & J092927.6+504810 & 0556210401 & 09:29:27.6 & +50:48:10 & & PN & Quasar candidate.\\ 
	12 & J233504.9$-$534751 & 0604870332 & 23:35:04.9 & $-$53:47:51 & & PN & AGN. \\ 
	13 & J183658.4$-$072119 & 0606420101 & 18:36:58.4 & $-$07:21:19 & 3000 & PN & Close to CCD gap. Star.\\ 
	14 & J090150.9$-$015815 & 0655340159 & 09:01:50.9 & $-$01:58:15 & 600 & PN & Short observation. Blue ctp.\\ 
	15 & J081907.9$-$384302 & 0655650201 & 08:19:07.9 & $-$38:43:02 & 2500 & PN & IR-optical ctp. Star.\\ 
	16 & J221448.2+002707 & 0673000136 & 22:14:48.2 & +00:27:07 & 400 & MOS & QSO. \\ 
	17 & J090629.9$-$000911 & 0725300150 & 09:06:29.9 & $-$00:09:11 & & MOS & Short observation. Star.\\ 
	18 & J090335.4+013224 & 0725300157 & 09:03:35.4 & +01:32:24 & & PN & Short observation. Seyfert 1.\\ 
	19 & J011552.2$-$003058 & 0747400134 & 01:15:52.2 & $-$00:30:58 & 300 & PN & No \textsc{Simbad} object associated.\\ 
	20 & J012517.1$-$001829 & 0747410134 & 01:25:17.1 & $-$00:18:29 & 300 & PN & QSO.\\ 
	21 & J012351.6+000831 & 0747410144 & 01:23:51.6 & +00:08:31 & & MOS & Short observation, highly contamined.\\ 
	22 & J014150.5+000754 & 0747430146 & 01:41:50.5 & +00:07:54 & 400 & PN & Optical-UV ctp. \\ 
	23 & J144506.2+685817 & 0763640601 & 14:45:06.2 & +68:58:17 & 200 & PN & Contamined observation. Blue ctp. \\ 
	24 & J175131.6$-$401533 & 0763700301 & 17:51:31.6 & $-$40:15:33 & 1000 & PN & var\_flag=True. Optical ctp. Star.\\ 
	25 & J113407.5+005223 & 0770380401 & 11:34:07.5 & +00:52:23 & 200 & PN & No \textsc{Simbad} ctp.\\ 
	26 & J022133.7$-$042346 & 0785101401 & 02:21:33.7 & $-$04:23:46 & 1000 & PN & NIR-blue ctp. Star.\\ 

\hline
\multicolumn{8}{c}{M31}\\
\hline
	1 & J004307.5+412019 & 0109270101 & 00:43:07.5 & +41:20:19 & 400 & PN & Type I X-ray burst. NS-LMXB.\\ 
	2 & J004242.5+411657 & 0405320701 & 00:42:42.5 & +41:16:57 & & PN & \\ 
	3 & J004215.6+411720 & 0650560201 & 00:42:15.6 & +41:17:20 & 300 & PN & Type I X-ray burst. NS-LMXB.\\ 
	4 & J004252.4+411648 & 0650560501 & 00:42:52.4 & +41:16:48 & & PN & \\ 
	5 & J004209.5+411745 & 0674210201 & 00:42:09.5 & +41:17:45 & & PN & \\ 
	6 & J004210.9+411248 & 0674210201 & 00:42:10.9 & +41:12:48 & 500 & PN & Type I X-ray burst. NS-LMXB. \\ 
	7 & J004215.1+411234 & 0674210301 & 00:42:15.1 & +41:12:34 & & PN & \\ 
	8 & J004212.1+411758 & 0727960401 & 00:42:12.1 & +41:17:58 & 500 & MOS & Type-I X-ray burst. NS-LMXB.\\ 
	9 & J004231.2+411938 & 0727960401 & 00:42:31.2 & +41:19:38 & & PN & \\ 
\hline
\end{tabular}
\tablefoot{Newly variable sources detected by EXOD with $TW=100$\,s. Column 1 gives the source ID in this paper. Column 2 the name of the source in 3XMM-DR8. Column 3 the observation ID in which the source was detected, columns 4 and 5 the RA and Dec. respectively. Column 6 the duration of the burst if there is one, column 7 in which instrument the light curve is variable (PN or MOS), and column 8 some comments, including the presence of a counterpart (ctp.). The top 30 are the general
survey, the bottom 9 are in the direction of M31.}
\end{table*}


\section{Discussion}\label{sec:discussion}

\subsection{Effectiveness and speed}
We present a new algorithm that can be used as a tool to promptly detect interesting sources variable on short timescales in \XMM's EPIC-pn observations. EXOD is computationally inexpensive when compared to other variability tests, like $\chi^2$ and KS. This makes EXOD faster than generating the light curve of all the sources in one observation.

The combination of the parameters $TW = 100$\,s, $DL = 8$, $r_{GT} = 1$ and $b = 3$ pixels detects a higher number of variable sources according to the $\chi^2$ and KS variability tests with a lower proportion of non-variable sources. We chose a low $DL$ in order to detect a higher number of new variable sources, accepting the consequent increase in false positives. The variability of the whole observation with these parameters is computed faster than the generation of a single light curve with the SAS. This makes it optimal to find new variable sources. However, a visual inspection of the detected sources becomes necessary since the amount of detections in bright or extended sources is higher than with higher $DLs$, and the $TW = 100$\,s can lead to a worse localization than with other $TW$.

With the optimal parameters applied to different time windows, the percentage of variable sources according to the $\chi^2$ and KS tests varies between 64.3\% and 83.7\%. However, only 22.7\% of the sources classified as variable in 3XMM-DR8 are detected. The high rate of false negatives is partially compensated by the fact that we get variability information from faint sources for which the light curves were not previously automatically generated (27.1\% of the EXOD sources). 

\subsection{Contamination}
The source detection with the sliding box technique gives the same problems when applied here as when applied to the source detection in \XMM's pipeline. It is a simple technique that usually gives good results, except in complicated cases where extended sources, bright sources, crowded regions or OoT events are present, since the spatial background can vary rapidly and give rise to spurious variability detections. These problems have been listed in \cite{watson_xmm-newton_2009}, and such sources have been flagged as spurious in the 3XMM-DR8 catalogue thanks to visual screening. Fortunately, one can easily distinguish real sources from false detections with a visual inspection.

We are confident that the variability computation algorithm gives the desired results. Nevertheless, the number of spurious detections ($\sim$1/8 of the total) remains quite large due to the source detection procedure.
A more sophisticated source detection algorithm could get rid of this problem, but it would probably be more computationally expensive. A comparison of different source detection procedures for \XMM\ can be found in \cite{valtchanov_comparison_2001}.

We have noted an additional issue for bright sources:
The spikes of the PSF are instrumental features with considerable stochastic variability, and they give rise to spurious variable detections around  sources with a high number of counts. In most of these cases, the center of the PSF is detected as a low variability region.
This can be visualised in Fig.~\ref{fig:example}.

\subsection{Discoveries}
EXOD computes the variability of the whole field of view, which uncovers the variability of faint sources with <100 counts. It also allows the identification of variable sources that had been classified as non variable by \XMM's pipeline due to a non-optimal time binning or to the short duration of the outburst, helping to unravel the nature of these sources.

Applying EXOD to 5,751 EPIC-pn observations in the full frame mode with different parameters led to the detection of a total of 2,961 sources. Of these, 2,536 were previously known sources in 3XMM-DR8 or \textsc{Simbad}, and we consider the remaining ones to be spurious detections. These variable sources belong to a wide variety of categories, including stellar flares, cataclysmic variables, type I X-ray bursts, supergiant fast X-ray transients, supernova shock breakouts, AGNs and more.

Finally we discuss 35 sources previously known, but not known to be variable among those classified as \textit{no identification} with \textsc{Simbad} and not classified as variable in the 3XMM-DR8 catalogue. Some of the sources that were detected in observations with a net exposure time lower than 5000\,s are not classified as variable by other variability tests. We thus recommend the EXOD user to be cautious when analysing such short observations.

Variability is a prominent feature that can be usdiscusseded for classification purposes. Earlier works aiming to identify unidentified sources in the \XMM\ catalogue applied a position cross-match to catalogs at other wavelengths. Other known parameters, such as optical to X-ray flux ratio, are used afterwards to discriminate between source classes \citep{pineau_cross-correlation_2011,lin_classification_2012}. More recent works have applied machine learning algorithms where timing parameters are the major classifying feature \citep{lo_automatic_2014,farrell_autoclassification_2015}. 
Around 27.1\% of the EXOD detected sources did not have a previously generated light curve and thus no variability classification. Additionally, many EXOD detections that were bright enough for a light curve generation were classified as non-variable by \XMM's pipeline, where the light curve showed a clearly visible outburst. EXOD thus provides an additional component that can be used to improve the classification of faint sources.

\subsection{M31}

We identified four transients in M31 that are likely to be neutron-star low mass X-ray binaries given their variability and spectra. These double the population of known NSs in M31. 

Previous neutron star searches in M31 have used various  techniques.
\cite{pietsch_xmm-newton_2005} found two NSs in LMXBs from \XMM\ data, through their bursting nature.
\cite{esposito_extras_2016} and \cite{zolotukhin_slowest_2017} found a 1.2\,s X-ray pulsar in M31, while \citet{rodriguez_castillo_discovery_2018} detected a 3\,s X-ray pulsar.

Radio pulsar searches have also been carried out. \cite{rubio-herrera_search_2013} used the Westerbork telescope at 328\,MHz to detect six bursts at a similar
dispersion measure (DM), suggesting a neutron star emitting as a  rotating radio transient, but the periodicity searches
have been unsuccessful, and the source has not been confirmed.
Deep radio-pulsar searches with LOFAR at 150\,MHz have been reported by \citet{mikhailov_radio_2018} and \citet{van_leeuwen_andromeda_2019}, without detections.

That combines to a total number of known neutron stars in M31 of only 4. In this galaxy more massive than the Milky Way, many more than these four must clearly exist. 

Given the increased number of \XMM\ observations of M31 since the detections of e.g. \cite{pietsch_xmm-newton_2005}, and the sensitivity of EXOD to outburst detections demonstrated in Section~\ref{sec:vary}, additional type I X-ray bursters could be expected to be found through EXOD.

And indeed, our four new type I X-ray burster candidates J004307.5+412019, J004215.6+411720, J004210.9+411248, and J004212.1+4111758 significantly increase the limited available population of known extragalactic neutron stars.

\subsection{Further potential for discovery}

EXOD has the potential to discover further examples similar to the tidal disruption event recently discovered with unexplained quasi periodic eruptions as it declines to quiescence \citep{miniutti_nine-hour_2019}. Whilst the data showing the quasi periodic eruptions was not in the 3XMM-DR8 catalogue, another similar source was included,  RX J1301.9+2747 \citep{giustini_x-ray_2020}. This source was not found during the initial study as the observation was taken in the Extended prime full window mode, a mode that was not included in our study. However, the source was detected highly significantly by a run of EXOD on this data. It was identified as a galaxy and therefore there may be more of these rare and unexplained objects in that category. Identifying them should provide data that will help us understand the nature of the quasi periodic eruptions.
Our goal is to apply EXOD to \XMM\ observations performed in the Extended Full Frame mode and the Large Window mode in the future, where we expect to find a high number of unknown transients.

The two sources referred as source 6 and 25 in this paper that were independently detected by \cite{alp_blasts_2020} and identified as SBOs while this paper was under review, represent exceptional examples of EXOD's extragalactic transient discovery potential.

It is very likely that there are other variable sources that have an inaccurate identification following the automatic \textsc{Simbad} query. Some could also have been eliminated by the 0.5\ctss\ GTI rate threshold. Finding these would require manual perusal of the associations, and a visual inspection of all the detected sources. This will be pursued in future work.

EXOD can further be adapted to other existing X-ray observatories with similar detector properties, such as \textit{Chandra}, \textit{NuSTAR}, \textit{Swift} or \textit{eROSITA}. It could also be applied to future missions like \textit{Athena}.

\section{Conclusions}\label{sec:conclusion}
In this work we have presented  EXOD, a new algorithm able to detect sources that vary within the duration of an
\XMM\ observation. It applies an imaging technique proven in optical and radio transient surveys.
Its main strength is to detect the variability of faint sources for which no light curve has been generated by \XMM's automatic pipeline.

We tested its performance by applying it to 5751 observations and subsequently implementing the $\chi^2$ and Kolmogorov-Smirnov variability tests on the detected sources.
With this technique, we were able to find a net count of 2,536 variable sources such as stellar flares, type-I X-ray bursts, supergiant X-ray transients, cataclysmic variables, AGNs and QSOs.

Thirty-five of the sources detected in \XMM\ archival data were unknown transients\footnote{Although two were identified with an independent transient search algorithm while this paper was under review.}. 
In spite of the low number of counts of these sources, usually <100, we performed spectral fitting to get a hint on their nature, and we searched for
archival counterparts at other wavelengths.

We find that four of these sources are extragalactic type I X-ray bursters, located in M31. Other sources, in spite of having analysed their spectra and looked for counterparts at other wavelengths, continue to be unidentified since the number of photons of these sources remains too low to draw any firmer conclusions.

EXOD is a computationally inexpensive algorithm, making it particularly advantageous to rapidly find sources of interest in \XMM\ observations and to provide additional information suitable to identify the nature of a transient source. It can be applied to future \XMM\ observations, and we envision to later adapt it to other X-ray observatories with similar detection techniques. This could yield discoveries of new transients and ease their multi-wavelength, even multi-messenger follow-up.

\begin{acknowledgements}
IPM thanks A. G\'urpide and the anonymous referee for their input and discussions.
IPM, NAW \& DTW are grateful to the CNES for financing this study.
IPM further acknowledges funding from the Netherlands Research School for Astronomy (NOVA5-NW3-10.3.5.14).
JvL received funding from the European Research Council under the European Union's Seventh Framework Programme (FP/2007-2013) / ERC Grant Agreement n. 617199 ('ALERT'), and from Vici research programme 'ARGO' with project number 639.043.815, financed by the Netherlands Organisation for Scientific Research (NWO).
This research has made use of data obtained from the 3XMM \XMM\ serendipitous source catalogue compiled by the 10 institutes of the \XMM\ Survey Science Centre selected by ESA;
of the \textsc{Simbad} database, operated at CDS, Strasbourg, France;
and of data from the European Space Agency (ESA) mission
{\it Gaia} (\url{https://www.cosmos.esa.int/gaia}), processed by the {\it Gaia}
Data Processing and Analysis Consortium (DPAC,
\url{https://www.cosmos.esa.int/web/gaia/dpac/consortium}). Funding for the DPAC
has been provided by national institutions, in particular the institutions
participating in the {\it Gaia} Multilateral Agreement.

\end{acknowledgements}


\bibliographystyle{yahapj}
\bibliography{biblio} 

\begin{appendix}

\section{EXOD output}

\begin{figure*}
\centering
\includegraphics[width=\textwidth]{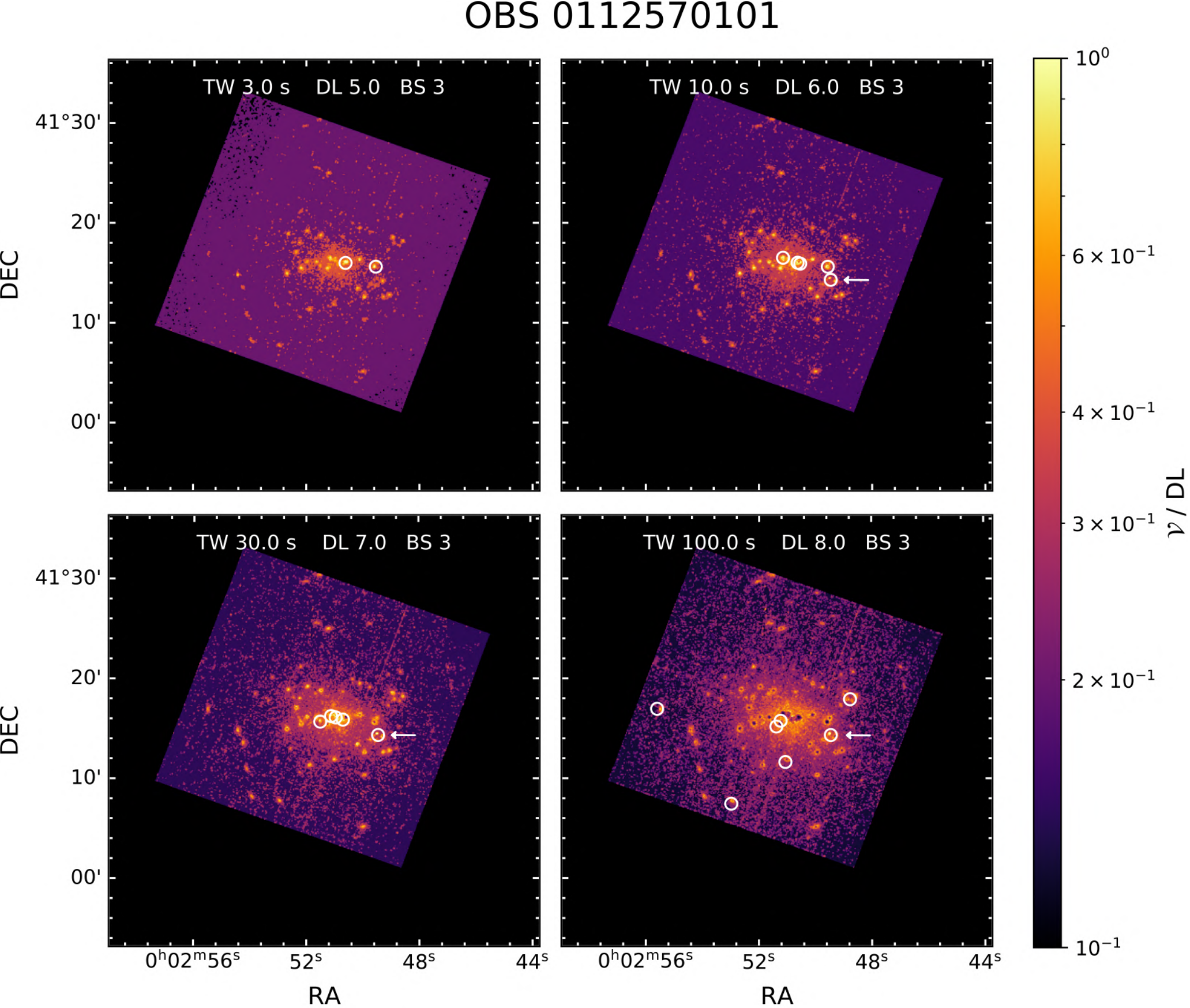}
\caption{Variability of Observation 0112570101 of M31's central region. Computed with $TW$ = 3\,s (top left), 10\,s (top right), 30\,s (bottom left) and 100\,s (bottom right). The colorbar is the same for all the plots and represents the variability divided by the detection level used for each time window, from less variable ($\mathcal{V}/DL = 0.1$, darker) to more variable ($\mathcal{V}/DL = 1$, lighter).
The detected variable sources are marked by white circles.
One of the sources that is detected with $TW=100$\,s, $TW=30$\,s and $TW=10$\,s in this observation was reported in \cite{pietsch_xmm-newton_2005} as a type I X-ray burst, and is it marked here with white arrows.}
\label{fig:example}
\end{figure*}

In Fig.~\ref{fig:example}, we present an example of one of the M31 observations analysed with the four sets of parameters given in Table~\ref{tab:final_param} to illustrate the output of \textit{EXOD} and how the chosen time window affects the value of the variability.

In this particular observation, \cite{pietsch_xmm-newton_2005} reported the detection of a type I X-ray burst in M31. Other sources are detected in the same field of view as being variable, up to seven with $TW=100$\,s.

One can see that the sources detected with different $TW$ can change. A remarkable fact is that some of the brightest sources that are variable on short timescales (3, 10 s) are seen as a low-variability centre embedded in a variable surrounding area at $TW = 100$\,s.

\clearpage
\section{Algorithm}\label{app:algo}
Algorithm~\ref{algo:exod_full} shows the variability computation explained in Section~\ref{sec:var-comp} with an algorithmic notation.

\begin{algorithm*}
    \caption{EXOD variability computation}
    \label{algo:exod_full}

    \KwData{The events list $\mathcal{E}\; =\; \{<x,\ y,\ t>\}$, the good time intervals $GTI\; =\; \{<t_{n},\ t_{m}>\}$, the list of TWs $T\; =\; \{<t_{n},\ t_{n+TW}>\}$ and the good time ratio $r_{GT}$}

    \KwResult{The variability per pixel, in \ctss}

    \Begin{
        \tcc{Event count per pixel per time window}

        $\mathcal{C} \longleftarrow ()_{64,\ 254,\ |T|}\ $ \;

        \ForEach{$x \in [0..64]$}{
            \ForEach{$y \in [0..254]$}{
                \ForEach{$tw \in T$}{

                    $\mathcal{C}_{x,\ y,\ tw} \longleftarrow |\{\bigcup_{i\ =\ x-1}^{x+1} \bigcup_{j\ =\ y-1}^{y+1} \bigcup_{t\ \in\ tw} e_{i,\ j,\ t}\ \in\ \mathcal{E}\}|$ \;

                }
            }
        }

        \tcc{Good time ratio computation and TWs filtering}

        $tw \longleftarrow T_{0}$ \;

        \ForEach{$g \in GTI$}{
            \tcc{Deletion of the TWs outside GTIs}
            \While{$ tw \cap g = \emptyset$} {
                \lIf{$r_{GT} \neq 0$}{remove $tw$ in $T$, $\mathcal{C}$ and $\tilde{\mathcal{C}}$\ }
                $tw \longleftarrow next\ TW \in T$ \;
            }

            \tcc{Application of $r_{GT}$ for each TW into a GTI}
            \While{$tw \cap g \neq \emptyset$}{
                $r_{GT}^{(tw)} \longleftarrow \frac{|tw\ \cap\ g|}{|tw|}$ \;
                \If{$r_{GT}^{(tw)} \geq r_{GT}$}{
                    $\mathcal{C}_{\_,\ \_,\ tw} \longleftarrow (x_{i,\ j,\ tw} \times r_{GT}^{(tw)}\ :\ \forall\; i\ \in\ [0..64]\  \forall\; j\ \in\ [0..254])$ \;
                }
                $tw \longleftarrow next\ TW \in T$ \;
            }
        }

        \tcc{Variability computation}

        $\mathcal{V} \longleftarrow (0)_{64,\ 254}$\ ;

        \ForEach{$x \in [0..64]$}{
            \ForEach{$y \in [0..254]$}{
                $\tilde{\mathcal{C}} \longleftarrow median({\mathcal{C}_{x,\ y,\ tw} : \forall\; tw \in T})$ \;
                $\mathcal{C}_{max} \longleftarrow max({\mathcal{C}_{x,\ y,\ tw} : \forall\; tw \in T})$ \;
                $\mathcal{C}_{min} \longleftarrow min({\mathcal{C}_{x,\ y,\ tw} : \forall\; tw \in T})$ \;

                \eIf {$\tilde{\mathcal{C}} \neq 0$}{
                    $\mathcal{V}_{x,\ y} \longleftarrow max(\mathcal{C}_{max} - \tilde{\mathcal{C}},\ |\mathcal{C}_{min} - \tilde{\mathcal{C}}|)$ \;
                }{
                 $\mathcal{V}_{x,\ y} \longleftarrow \mathcal{C}_{max}$ \;
                }
            }
        }
        \Return{$\mathcal{V}$}\;
    }

\end{algorithm*}
\clearpage

\section{Spectral fitting}\label{app:fit_spec}
\begin{figure*}
    \centering
    \includegraphics[width=0.325\textwidth]{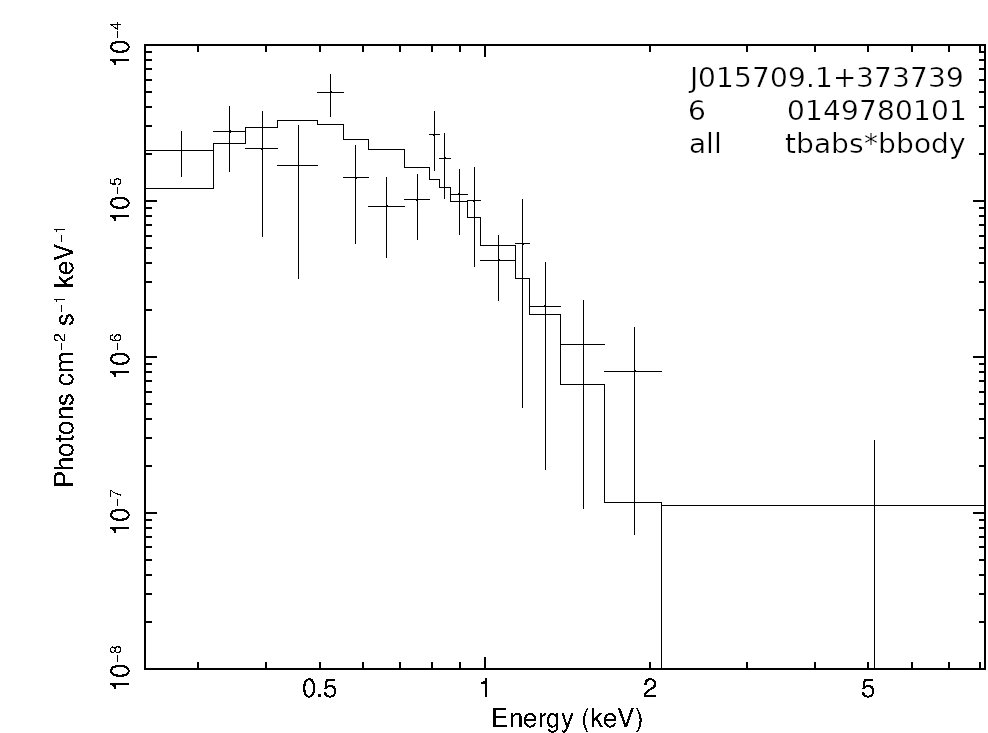}
    \includegraphics[width=0.325\textwidth]{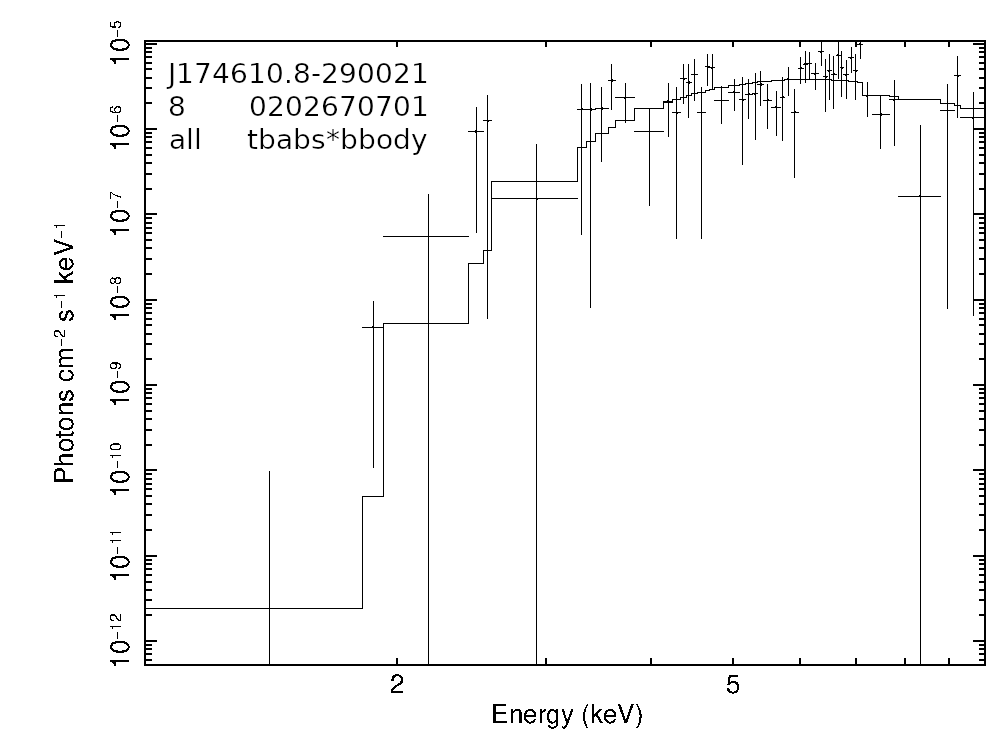}
    \includegraphics[width=0.325\textwidth]{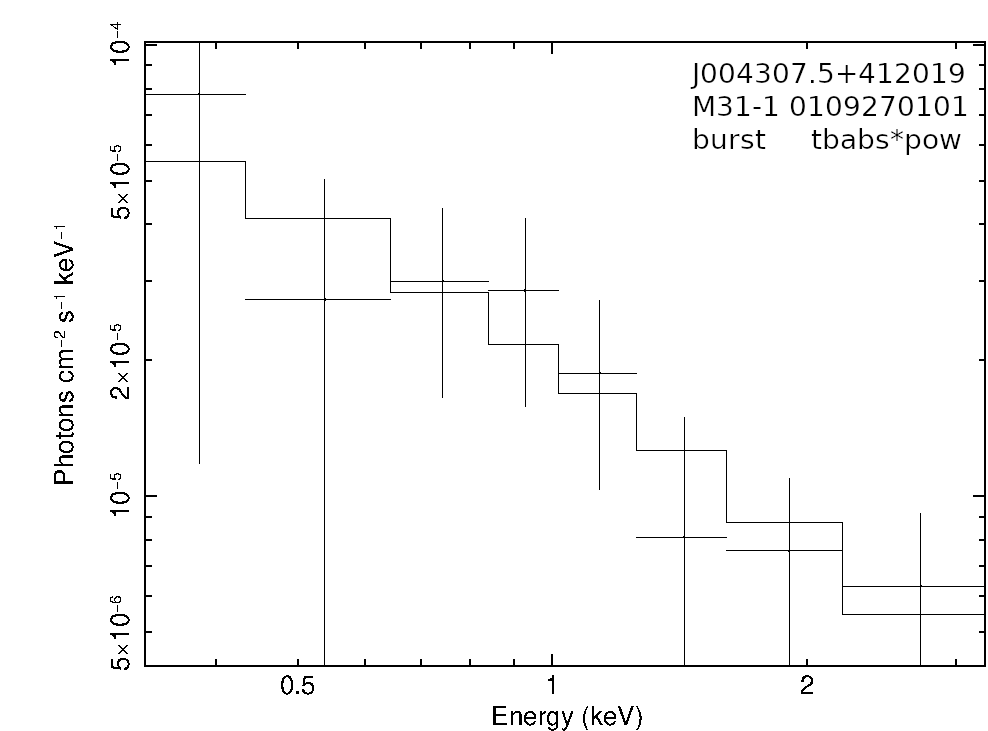}
    
    \includegraphics[width=0.325\textwidth]{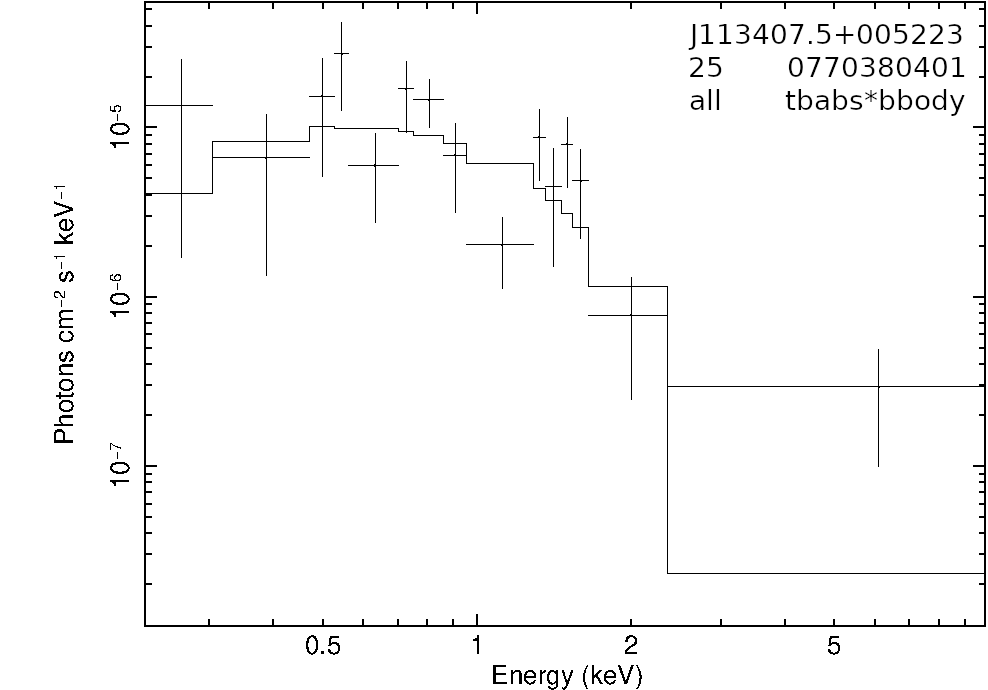}
    \includegraphics[width=0.325\textwidth]{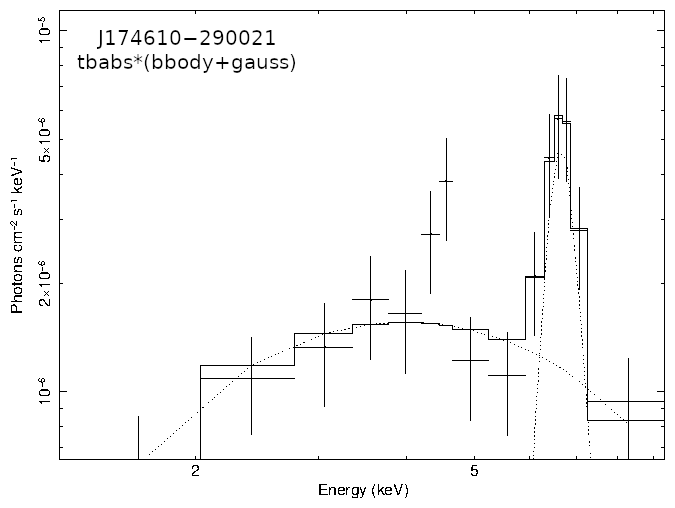}
    \includegraphics[width=0.325\textwidth]{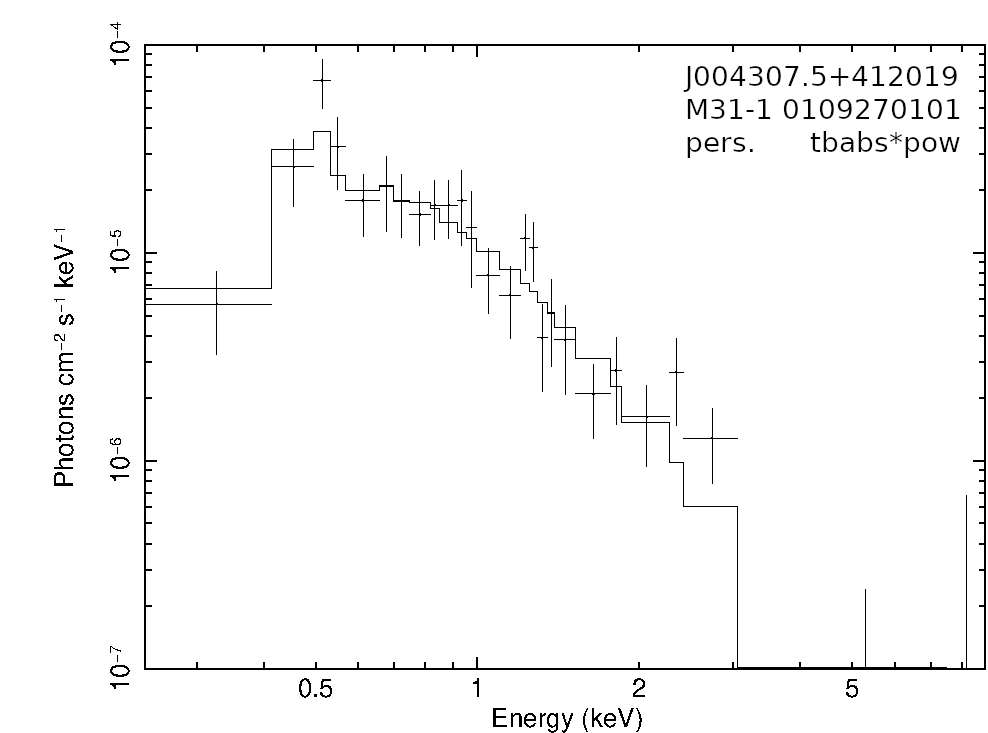}
    \vspace{1cm}
    
    \includegraphics[width=0.325\textwidth]{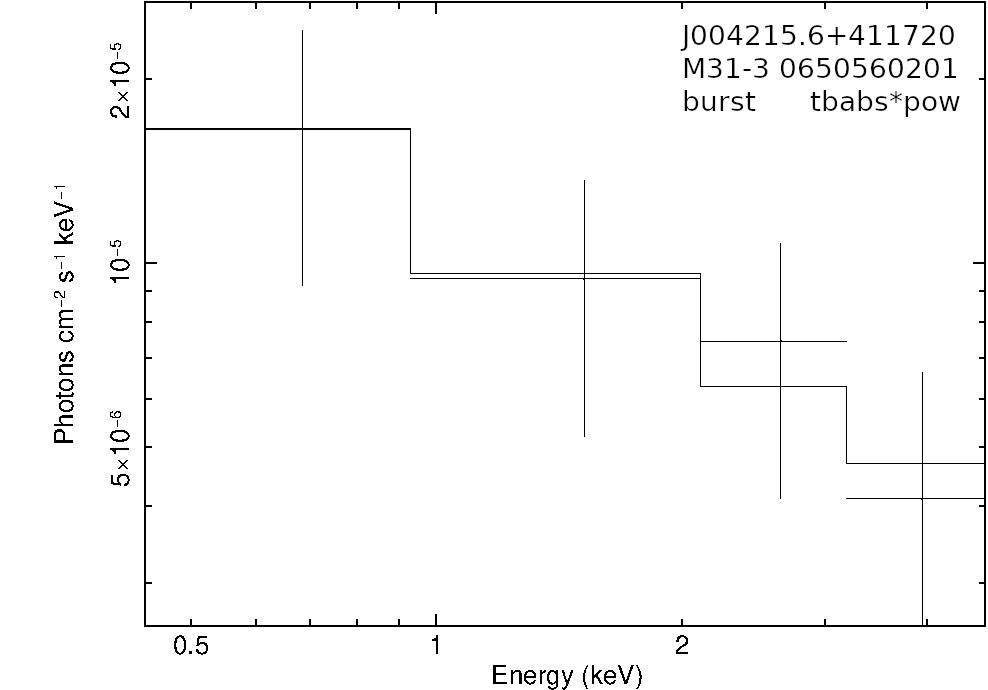}
    \includegraphics[width=0.325\textwidth]{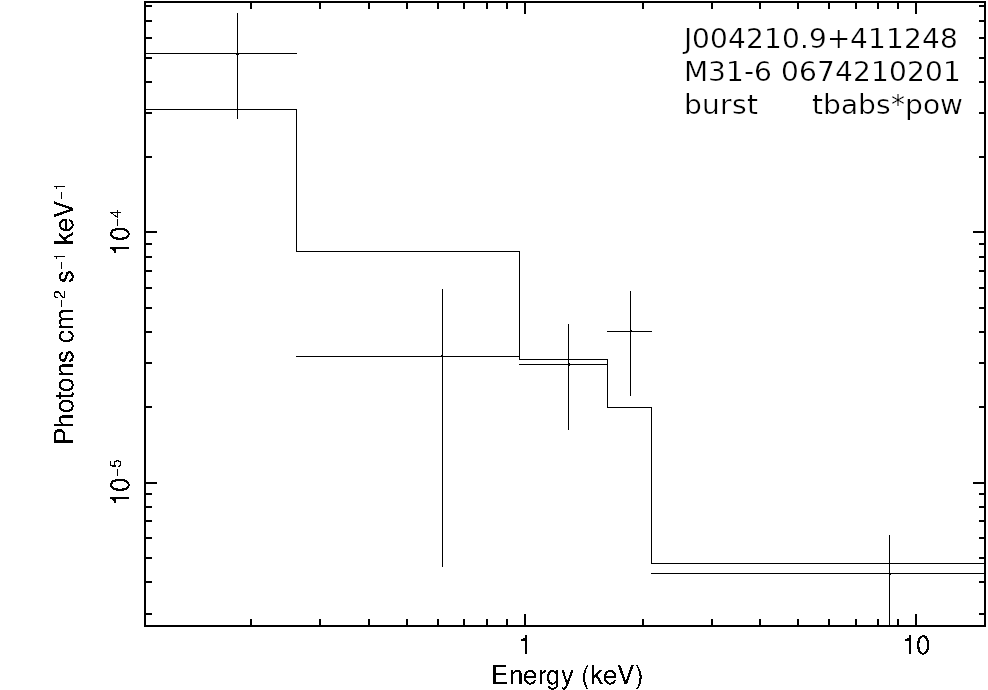}
    \includegraphics[width=0.325\textwidth]{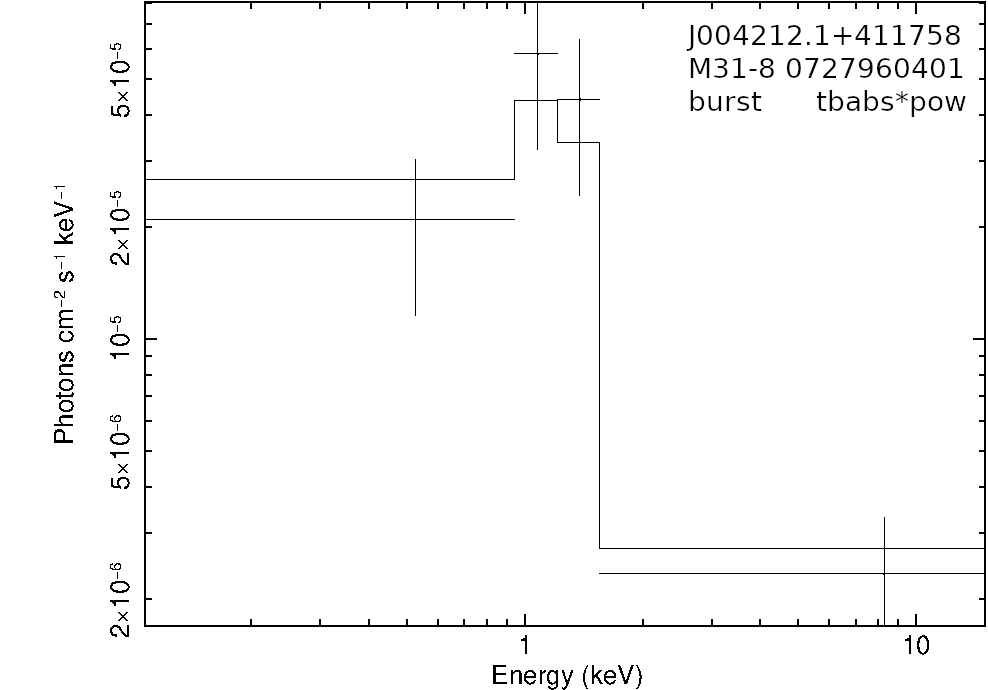}
    
    \includegraphics[width=0.325\textwidth]{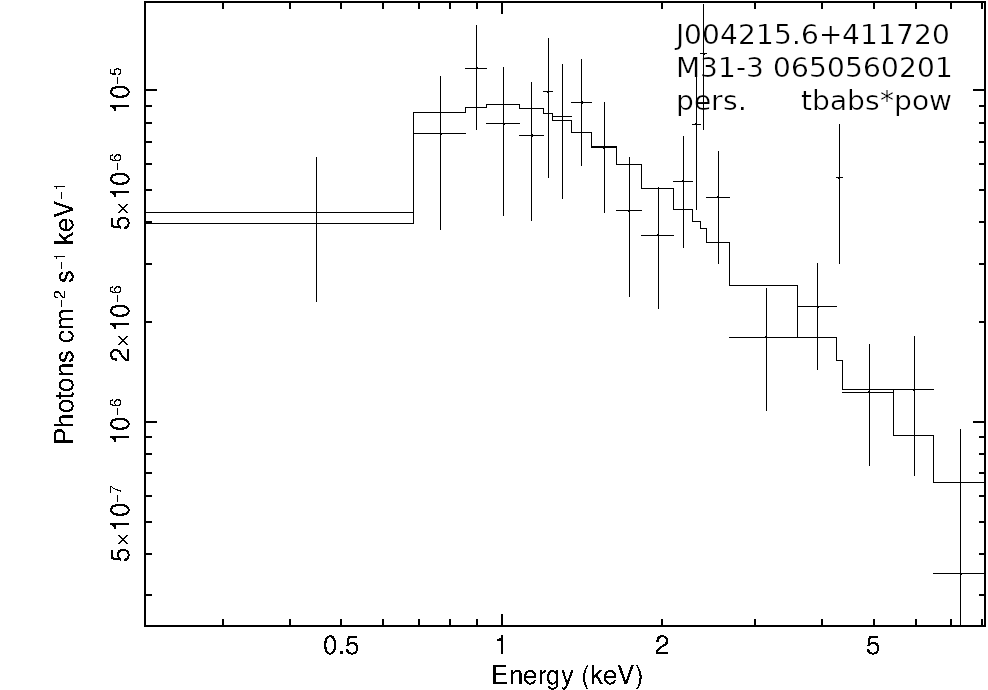}
    \includegraphics[width=0.325\textwidth]{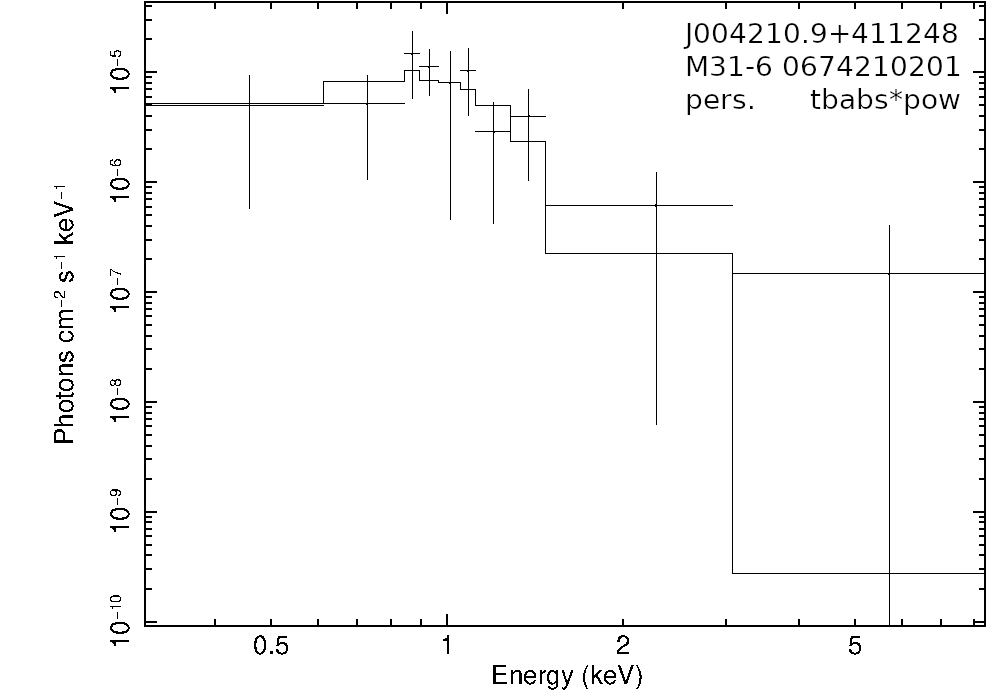}
    \includegraphics[width=0.325\textwidth]{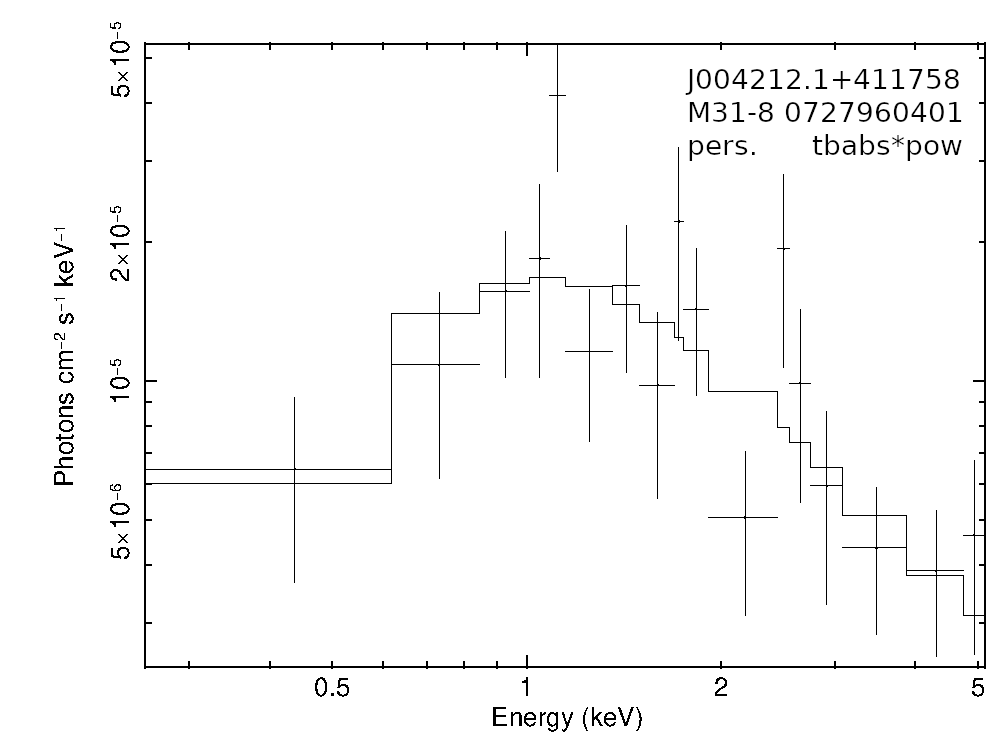}
    \caption{EPIC-pn fitted spectra in the 0.2--10\,keV band. Beginning by first pair of rows and then second pair of rows, from left to right: 1) J015709.1+373739 fitted with a tbabs*bbody model. 2) J113407.5+005223 fitted with a tbabs*bbody model. 3) and 4) respectively: J174610.8$-$290021 fitted with a tbabs*bbody model and persistent emission fitted to a tbabs*(bbody + gauss) model. 5) and 6) J004307.5+412019 burst and persistent emission fitted with a tbabs*pow model. 7) and 8) J004215.6+411720 burst and persistent emission fitted with a tbabs*pow model. 9) and 10) J004210.9+411248 burst and persistent emission fitted with a tbabs*pow model. 11) and 12) J004212.1+411758 burst and persistent emission fitted with a tbabs*pow model.}
    \label{fig:fit_spec}
\end{figure*}

In this section, we give the detailed results of the spectral fittings performed with Xspec. The fitted spectra are shown in Fig.~\ref{fig:fit_spec}, and Table~\ref{tab:fit_spec} shows the C-stat fit of the sources to an absorbed black body or an absorbed power law model.
\\

The spectra were generated with the SAS task \texttt{evselect} using standard filters as recommended. The spectra were rebinned to have at least 5 counts per spectral bin.

\begin{table*} \label{tab:fit_spec}
    \centering
    \caption{Results from fitting different models to the pn spectra.}
    \renewcommand{\arraystretch}{1.5}
    \begin{tabular}{rccccccc}
    \hline\hline
    \multicolumn{1}{c}{(1)} & (2) & (3) & (4) & (5) & (6) & (7) & (8) \\
    \multicolumn{1}{c}{Object} & Time & $n_{\text{H}}$ & kT & $\Gamma$ & C-stat (dof) & F$_{\text{abs}}$ & F$_{\text{unabs}}$ \\
    & selection & ($\times 10^{22}$ at.\,cm$^{-2}$) & (keV) &   &   & \multicolumn{2}{c}{($10^{-13}$\fcgs)} \\
    \hline
    \multirow{2}{*}{6} J015709.1+373739     
    & A & $<0.12$ & 0.16\errors{0.03}{0.03} & ... & 19.79 (19) & 0.16\errors{0.03}{0.03} & 0.26\errors{0.19}{0.05} \\
    & & $<1.14$ & ... & 2.95\errors{7.05}{0.76} & 14.83 (19) & 0.20\errors{0.12}{0.09} & 0.41\errors{4208}{0.19} \\
    \hline
    
    \multirow{2}{*}{8} J174610.8$-$290021   
    & A & 22.71\errors{44.14}{10.52} & 1.92\errors{2.90}{1.37} & ... & 88.88 (90) & 1.81\errors{0.36}{0.33} & 4.10\errors{3.88}{1.34} \\
    & & 30.38\errors{57.93}{15.97} & ... & 1.74\errors{3.28}{0.72} & 88.69 (90) & 1.86\errors{0.36}{0.33} & 9.17\errors{47.8}{5.22} \\
    \hline
    
    \multirow{2}{*}{25} J113407.5+005223     
    & A & $<0.18$ & 0.30\errors{0.07}{0.08} & ... & 36.30 (20) & 0.15\errors{0.04}{0.04} & 0.17\errors{0.04}{0.04} \\
    & & 0.31\errors{0.39}{0.23} & ... & 3.55\errors{6.04}{2.12} & 35.56 (20) & 1.89\errors{75.8}{1.51} & 0.18\errors{0.12}{0.06} \\
    \hline
    \hline
    
    \multirow{4}{*}{M31-1 J004307.5+412019} 
    & B & <0.50  & 0.32\errors{0.18}{0.17} & ... & 1.72 (2) & 0.83\errors{0.54}{0.39} & 1.15\errors{1.90}{0.42} \\
    & & <0.12 & ... & 1.91\errors{1.02}{1.04} & 0.68 (2) & 1.91\errors{4.45}{1.09} & 2.68\errors{4.07}{1.02} \\
    
    & P & <0.16 & 0.22\errors{0.03}{0.03} & ... & 73.23 (51) & 0.23\errors{0.03}{0.03} & 0.39\errors{0.11}{0.05} \\
    & & 0.23\errors{0.16}{0.11} & ... & 3.31\errors{1.11}{0.72} & 56.71 (51) & 0.32\errors{0.08}{0.06} & 2.03 \errors{8.91}{1.16} \\
    \hline
    
    \multirow{4}{*}{M31-3 J004215.6+411720} 
    & B & <0.39 & 0.68\errors{0.56}{0.23} & ... & 2.72 (3) & 1.33\errors{1.36}{0.63} & 1.47\errors{1.32}{0.66} \\
    & & <0.54 & ... & 0.74\errors{1.67}{0.67} & 0.48 (3) & 3.55\errors{3.86}{2.31} & 3.60\errors{3.85}{1.94} \\
    
    & P & <0.19 & 0.65\errors{0.20}{-0.13} & ... & 50.42 (50) & 0.56\errors{0.21}{0.14} & 0.61\errors{0.21}{0.13} \\
    & & 0.25\errors{0.20}{0.14} & ... & 1.64\errors{0.50}{0.41} & 36.78 (50) & 0.96\errors{0.27}{0.23} & 1.32\errors{0.42}{0.25} \\
    \hline
    
    \multirow{4}{*}{M31-6 J004210.9+411248} 
    & B & <3.86 & 0.71\errors{0.36}{0.45} & ... & 7.23 (2) & 3.22\errors{2.29}{1.00} & 3.50\errors{2.76}{1.49} \\
    & & <3.55 & ... & 1.15\errors{0.42}{0.47} & 5.87 (2) & 5.18\errors{3.34}{2.56} & 5.18\errors{3.34}{2.56} \\
    
    & P & 0.12* & 0.23\errors{0.09}{0.06} & ... & 18.59 (23) & 0.13\errors{0.05}{0.04} & 0.23\errors{0.08}{0.07} \\
    & & 0.12* & ... & 2.38\errors{0.67}{0.59} & 21.39 (23) & 0.25\errors{0.16}{0.06} & 0.47\errors{0.17}{0.17} \\
    \hline
    
    \multirow{4}{*}{M31-8 J004212.1+411758} 
    & B & 0.12* & 0.43\errors{0.17}{0.12} & ... & 0.28 (2) & 1.75\errors{1.04}{0.69} & 2.15\errors{1.06}{0.76} \\
    & & 0.32* & ... & 2.34\errors{0.72}{0.64} & 1.18 (2) & 2.51\errors{1.84}{1.10} & 6.15\errors{4.27}{2.11} \\
    
    & P & <0.19 & 0.73\errors{0.11}{0.11} & ... & 31.21 (23) & 1.37\errors{0.38}{0.30} & 1.48\errors{0.39}{0.30} \\
    & & 0.32\errors{0.22}{0.17} & ... & 1.55\errors{0.62}{0.54} & 24.69 (23) & 2.27\errors{0.95}{0.69} & 3.06\errors{0.99}{0.59} \\
    \hline
    
    \end{tabular}
    \renewcommand{\arraystretch}{1}
    \tablefoot{Column (1) gives the name of the object in this paper and in 3XMM, (2) Time selection of the fit refers to all photons of the observation (A), the burst (B) or persistent (P) emission. Column (3)  gives the interstellar absorption; columns (4) and (5) give the black-body temperature (kT) or the power law index ($\Gamma$) respectively; column (6) the goodness of fit measured using the C-statistic and the number of degrees of freedom; columns (7) and (8) give an estimate of the absorbed (F$_{\text{abs}}$) and unabsorbed (F$_{\text{unabs}}$) fluxes in the 0.2--10.0\,keV band, respectively. All the errors are given for 90\% confidence for one interesting parameter. For the fluxes, the errors are at 68\% confidence.\\
    * Values frozen for the fit.}
\end{table*}

J174610$-$290021 without the photons emitted during the flare in observation 0202670701 was fitted with an absorbed black body with a Gaussian emission line (tbabs*(bbody+gaus)). The fitted parameters are the following: \\

\noindent
$n_{\text{H}} = 1.08^{+1.87}_{-1.08}\times10^{22}$ atom\,cm$^{-2}$\\
kT $= 2.12^{+1.33}_{-0.78}$\,keV\\
Gaussian energy $= 6.66^{+0.17}_{-0.13}$\,keV\\
Gaussian $\sigma = 0.28^{+0.22}_{-0.12}$\\
Flux $= 1.175 \times10^{-13}$\fcgs\\
C-stat = 0.8 (9 dof)\\

One should note that the $n_{\text{H}}$ value obtained by fitting the spectrum during the flare is an order of magnitude smaller than the value of the persistent emission. The large errors are due to the small number of data points. Both values are compatible at $3\sigma$.

\end{appendix}
\end{document}